# Aeroacoustic signatures reveal fast transient dynamics of vapor-jet-driven cavity oscillations in metallic additive manufacturing

Haolin Liu[1,2]*†, S. Kiana Naghibzadeh[3]†, Zhongshu Ren[4,5,6], Yanming Zhang[7], Jiayun Shao[4], Samuel J. Clark[8], Kamel Fezzaa[8], Xuzhe Zeng[1], Lin Gao[9], Wentao Yan[7], Noel Walkington[10], Kaushik Dayal[1,10,11], Tao Sun[4,8], Anthony D. Rollett[2]*, Levent Burak Kara[1]*

Affiliations

[1]Department of Mechanical Engineering, Carnegie Mellon University; Pittsburgh, PA, USA.

[2]Department of Materials Science and Engineering, Carnegie Mellon University; Pittsburgh, PA, USA.

[3]Department of Civil and Environmental Engineering, Massachusetts Institute of Technology; Cambridge, MA, USA.

[4]Department of Mechanical Engineering, Northwestern University; Evanston, IL, USA.

[5]Department of Materials Science and Engineering, University of Virginia; Charlottesville, VA, USA

[6]Physical Sciences and Research Operations Division, National Synchrotron Light Source II, Brookhaven National Laboratory; Uptown, NY, USA

[7]Department of Mechanical Engineering, National University of Singapore; Singapore.

[8]X-ray Science Division, Advanced Photon Source, Argonne National Laboratory; Lemont, IL, USA.

[9]Department of Mechanical Engineering, The University of Alabama; Tuscaloosa, AL, USA.

[10]Center for Nonlinear Analysis, Department of Mathematical Sciences, Carnegie Mellon University; Pittsburgh, PA, USA.

[11]Department of Civil and Environmental Engineering, Carnegie Mellon University; Pittsburgh, PA, USA.

*Corresponding authors. Email: haolinl@andrew.cmu.edu; rollett@andrew.cmu.edu; lkara@andrew.cmu.edu.

†These authors contribute equally to this work.

## Summary paragraph

**Aeroacoustic emissions from intense evaporation are widely measured yet often treated as noisy byproducts and used mainly in empirical monitoring[1–8]. Here, we show that airborne sound encodes physics-governed sub-millisecond fingerprints of vapor-jet dynamics in excessive vaporization, exemplified by vapor keyholes in laser metal processing[9–16]. From first principles, we develop a vapor-jet-cavity oscillation framework and incorporate it into an aeroacoustic formulation, thereby coupling measured sound to transient cavity depth and oscillation frequency[17–30]. Reconciled with synchronized multimodal in-situ data, airborne acoustics enable accurate tracking of vapor-cavity properties within tens to hundreds of microseconds. Combined with newly discovered correlations, cavity-jet-acoustic theory recasts the transition from steady, pore-free to pore-shedding vaporizations as a critical-frequency event. Aeroacoustic emissions thus become scalable, physics-guided, and cost-efficient probes of rapidly evolving liquid–vapor systems.**

Airborne sound accompanies nearly every instance of rapid vaporization, a common physical process that occurs from boiling and cavitation to laser ablation and metal welding[1–5]. Yet, even when acoustic emissions from excessive evaporation exhibit repeatable tonal features, the mechanisms governing how those features evolve, as well as how they map quantitatively onto momentary vaporization dynamics, remain only partially understood[1,2,4–6]. Arising from mass and momentum exchange between vapor jets and the surrounding medium, these ubiquitous emissions are often treated as noisy, broadband, and configuration-sensitive, hence hindering mechanistic interpretation[2,5–7]. Accordingly, despite the widespread empirical usage of acoustic sensing, a deterministic and physics-based understanding of how transient liquid–vapor motion relates to measurable airborne sound has remained elusive[4,6–8]. This gap raises a fundamental question: *can the ubiquitous yet seemingly noisy sound of excessive vaporization encode transient physics-consistent signatures of the process itself?*

More broadly, this question concerns *observability and mechanistic attribution* in non-equilibrium intense vaporization across a wide range of scientific and engineering topics[4,5]. Notably, laser-based metal processing (LMP), where concentrated laser energy deposition melts and resolidifies metal feedstock to produce dense components, offers a particularly tangible testbed for addressing this gap. In laser-based welding and powder bed fusion (PBF), excessive laser heating frequently occurs and drives the melt pool to temperatures near or above the alloy's boiling point[9]. The ensuing rapid vaporization generates airborne acoustic emissions and produces a recoil pressure that depresses the liquid surface, forming a vapor-filled cavity known as a *keyhole*[10]. Under vigorous and highly transient evaporation, the vapor jet sustains a deep keyhole geometry while emitting sound[8]; oscillations of the deep keyhole, sometimes accompanied by intermittent collapse and gas entrapment that shed near-spherical pores and seed defects known as keyhole porosity[10], can in turn modulate the vapor jet and its associated acoustic emissions[9–11]. Albeit a specific case, LMP couples acoustic emissions with driven vapor jets, compliant liquid cavities, and strongly time-dependent interfacial dynamics, making it representative of a broader class of vaporization-driven dynamical systems and a useful context for developing transferable observability principles.

Another crucial and impactful aspect of the aforementioned observability and interpretability hinges on *time-resolved accessibility*, namely, the extent to which they can be achieved rapidly using easily accessible, cost-effective sensing methods across diverse practical settings. In recent years, monitoring approaches spanning from high-fidelity *in situ* and *ex situ* measurements to computational and data-centric predictive models have been developed to advance evaporation-associated physical understanding and enable quantitative inference of process behaviours[12–16]. These approaches have been applied to resolve phase-transition dynamics, defect formation, and manufacturing outcomes in systems involving ablative evaporation[14–16]. However, many diagnostic modalities remain costly and resource intensive, limiting their ability to capture the fast transient vaporization dynamics that often govern process quality[12,14–16]. By contrast, airborne acoustics are commonly viewed as an inexpensive indicator of process stability[12–14], but aeroacoustic signatures from excessive vaporization are typically treated as indirect, low-signal-to-noise-ratio (SNR) observables and interpreted primarily through empirical rules or data-driven correlations[16–24]. This limitation reflects both the sensitivity of liquid–vapor behaviors to coupled laser-driven vaporization and interfacial forces and the lack of analytical, mechanistic descriptions linking transient vapor-cavity dynamics to measurable sound generation[10,17–20]. As a result, airborne acoustic monitoring often requires extensive condition-specific datasets[14,16], hindering scalable, physics-informed monitoring strategies and efficient models for this inherently complex system[12,19,25–28].

In this work, we address these long-standing obstacles to deepen mechanistic understanding and broaden the practical utility of airborne acoustic emissions from rapid and excessive vaporizations. We discover and demonstrate that, despite appearing noisy and ambiguous, aeroacoustic signatures from intense evaporation encode informative, physics-governed fingerprints of vapor-jet dynamics and cavity oscillations. As illustrated in Fig. 1, we first construct a laser-PBF experimental prototype that integrates synchronized multimodal in situ characterization with a machine-learning-enabled data-processing pipeline to observe and analyze laser-induced excessive vaporization in Ti-6Al-4V alloy (Fig. 1A). Inspired by newly revealed cross-modality correlations, we develop a perturbation-based linear framework for vapor-jet–cavity oscillations and couple it with aeroacoustic formulations[29,30] to derive a *Vapor-jet–Cavity (keyhole) Aeroacoustic Equation (VCAE)* that links aeroacoustic signatures to vapor-jet and cavity-depth accelerations, showing that measurable airborne sound encodes the underlying vapor-jet–cavity physics (Fig. 1B). Next, we extend the analysis to the nonlinear regime by deriving the *Vapor-jet–Cavity (keyhole) Oscillation Dynamics Equation (VCODE)*, a new and generalizable ordinary differential equation from continuum conservation laws that governs the transient nonlinear coupling between deep cavity-depth fluctuations and vapor-jet evolution (Fig. 1C). We reconcile these predictions and theoretical inferences with synchronized measurements, achieving qualitative agreement and enabling, for the first time, high-fidelity quantitative inference of key vaporization variables over sub-millisecond windows from airborne sound. Finally, in the context of laser-PBF, we identify in acoustic data a threshold frequency associated with the transition from steady, porosity-free keyholes (N-KH) to pore-shedding keyholes (P-KH) in Ti-6Al-4V. Combined with our theoretical framework, this finding motivates a mechanistic reinterpretation of the keyhole defect boundary (KH-bound) on the laser power–scan speed ($P$–$V$) map[19] as a critical iso-frequency contour in the same parameter space. Together, these results recast aeroacoustic signatures from vapor-jet-driven cavities as physics-informed observables, establishing a practical and cost-effective route to interrogate rapidly evolving excessive-vaporization systems using airborne acoustic emissions. By enabling accurate, rapid transient inference over time windows of a few hundred microseconds, our work leverages airborne acoustics to facilitate vaporization characterization by two to three orders of magnitude, aided by its broad generalizability, cost-effectiveness, fast response, and ease of deployment.

### ***In situ*** **multimodal prototype for characterization of laser-induced vaporization**

Our laser-induced vaporization prototype, *i.e.*, a laser-PBF platform, is equipped with synchronized *in situ* characterization modalities, including synchrotron X-ray imaging and ultrasonic airborne acoustic sensing (Fig. 1A). We performed single-track laser-melting experiments on Ti-6Al-4V using a focused laser beam across a range of laser power–scan speed ($P$–$V$) combinations to systematically probe vapor-cavity behavior. All modalities were registered to a common timeline to enable direct correlation between acoustic emissions and X-ray-resolved cavity geometry. An ultrasonic microphone was positioned near the top of the fusion zone to capture near-source acoustic signals from vapor jets exiting the cavity with minimal turbulence contamination. During processing, acoustic and other time-series signals were analyzed in the time domain and, when needed, via continuous wavelet transforms (CWT) to obtain spectro-temporal scalograms and extract spectral components. In parallel, we quantified side-view cavity profiles using a high-accuracy U-Net[31] segmentation model trained on only a handful of precisely manually labeled images, enabling automated cavity segmentation, contour extraction, and geometric feature quantification across the full X-ray dataset[32]. Together with automated cross-modality synchronization, this pipeline eliminated labor-intensive manual measurements and manual data pairing (Methods).

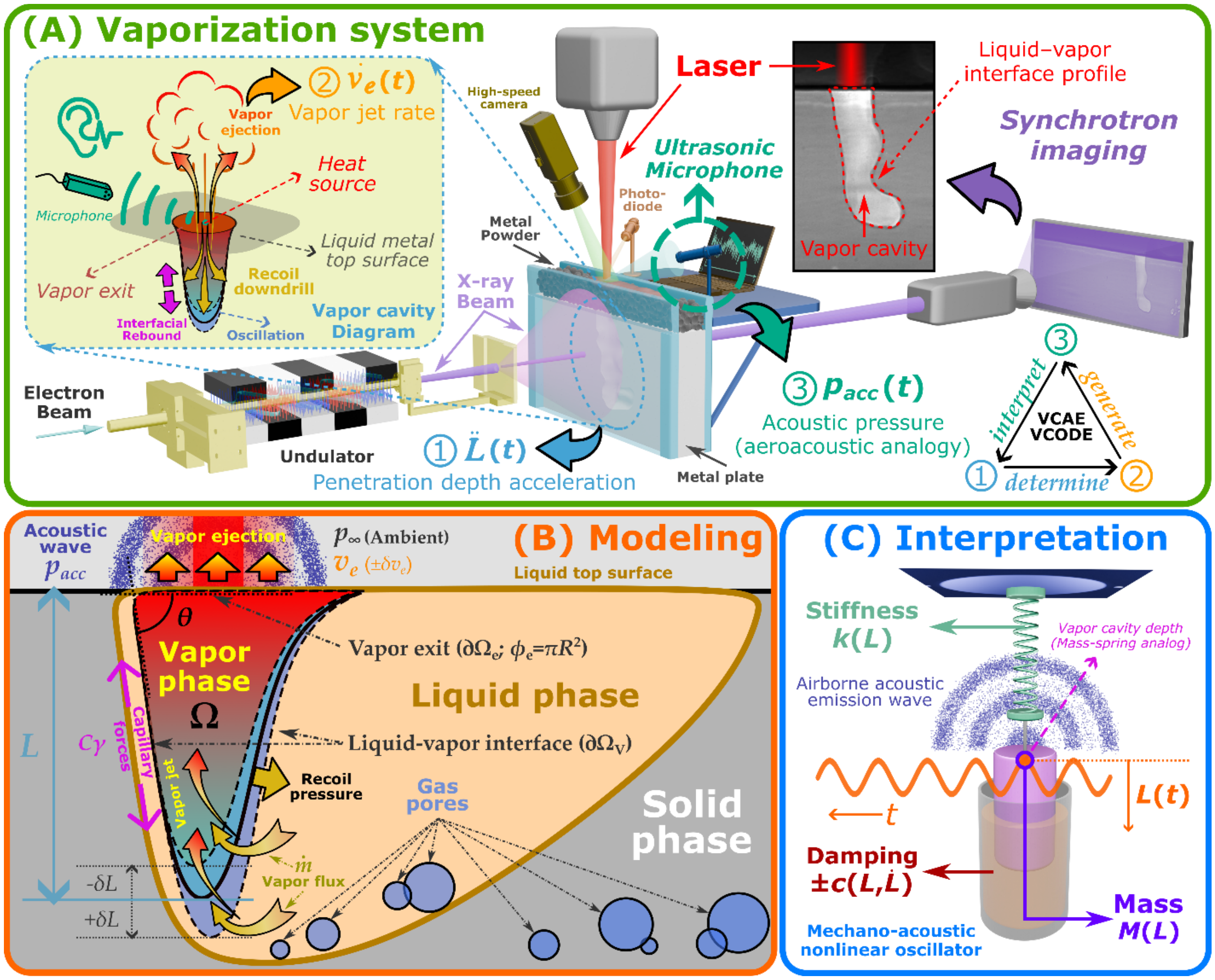

**Fig. 1. Multimodal prototype and mechanistic interpretation of laser-driven excessive vaporization. (A)** Vaporization paradigm and synchrotron-compatible experimental prototype. Side-view synchrotron X-ray imaging resolves transient vapor-cavity profiles, while an ultrasonic microphone records airborne pressure fluctuations emitted at the vapor exit; auxiliary sensors (e.g., infrared camera and photodiode) monitor thermal/optical signatures. The coupled process links cavity-depth acceleration ( $\ddot{L}(t)$ ), rate of vertical vapor-jet velocity ( $\dot{v}_e(t)$ ), and radiated acoustic pressure ( $p_{acc}(t)$ ; interpreted via acoustic formulations), enabling inference between cavity dynamics and acoustics. **(B)** Modeling schematic of the vapor cavity side-view highlighting recoil pressure, interfacial capillarity, and vaporization mass flux that shape the liquid–vapor interface, modulate cavity depth $L$ , and drive oscillatory vapor ejection that produces airborne acoustic waves and can seed porosity. These considerations lead to a theoretical model

that describes transient depth evolution, constrained by time-resolved X-ray measurements of interface geometry and exit conditions. **(C)** Vapor-cavity depth is interpreted as a nonlinear mechano-acoustic oscillator that strongly modulates vapor ejection, the resulting sound, and its stability. The extracted oscillator model has effective mass $M(L)$, effective stiffness $k(L)$, and effective damping $c(L,\dot{L})$ with either positive or negative values, all of which are derived from vapor jet parameters and cavity's geometric properties.

After synchronization, the registered measurements were partitioned into short 400 or 800 $\mu s$ snippets for subsequent correlation analysis. This synchronized snippet dataset provides a time-resolved view of excessive vaporization in which cavity geometry and airborne acoustics are measured simultaneously under systematically varied conditions. Within these snippets, cavity motion and acoustic signatures co-evolve in a very structured manner (Fig. 2A): changes in cavity depth accelerations and oscillation phase track repeatable shifts in the emitted ultrasound, indicating that the airborne signal is not merely incidental noise but an informative projection of the instantaneous cavity state. These discoveries motivate the theoretical development that follows: we use the X-ray-derived cavity evolution to formulate and constrain a first-principles, fluid-dynamics-based description of vapor-jet-driven cavity oscillations, and we use the collocated acoustic measurements to test and quantify how those dynamics are encoded in scalable airborne signatures. More details are referred to the Supplementary Information.

## Coupling between vapor jet and cavity oscillations

To interpret the multimodal measurements and observations, we start from developing a rough and simple mechanistic description of steady vapor-cavity oscillation dynamics and their associated aeroacoustic emissions. Figure 1B shows the side-view profile of the vapor cavity and the depiction of the modeling framework for the mechano-acoustic system of interest. Assuming a near-cylindrical shape ($\Omega$, also used to denote the volume) of the vapor-cavity with its depth ($L$) and vapor-jet velocity ($v_e$) at the cavity opening ($\partial\Omega_e$) as the predominant variables, we obtain the following balance of mass equation within $\Omega$:

$$\dot{m} = \rho_V \frac{d\Omega}{dt} + \rho_V \phi_e v_e = \rho_V \phi_e \left( \dot{L} + v_e \right) \qquad \text{(Eq. 1)}$$

where $\rho_V$ and $\phi_e$ denote the vapor density (assumed constant for now) and area of $\partial\Omega_e$, respectively, and $\dot{m}$ denotes the total mass vaporization rate across the liquid–vapor interface ($\partial\Omega_V$) of the vapor cavity. Equation 1 essentially states that the rate of change of mass ($\rho_V \dot{\Omega}$) within $\Omega$ equals the mass influx ($\dot{m}$) minus the mass outflux ($\rho_V \phi_e v_e$). During steady and pore-free oscillations observed in experiments, we have both $L$ and $v_e$ fluctuate with small amplitudes. Therefore, we further assume that under small geometric perturbations, $\dot{m}$ —predominantly associated with the total laser absorption determined by the laser absorptivity and cavity's unperturbed size and shape—remains approximately constant, *i.e.*, $\ddot{m} = 0$[18–20,23,33]. Guided by this interpretation and introducing small perturbations to $v_e$ and $L$, *i.e.*, $\delta v_e$ and $\delta L$, about their steady state $v_{e0}$ and $L_0$, we differentiate both sides of Eq. 1 with respect to time and yield $\dot{\delta v_e} = -\ddot{\delta L}$ (or $\dot{v}_e = -\ddot{L}$), revealing that under small changes of $v_e$ and $L$, the time derivative of $v_e$ is proportional to the negative of the acceleration of $L$. This relationship is fundamental to understanding the airborne acoustic emissions driven by the vapor jet around $\partial\Omega_e$, as subsequent analyses demonstrate that $\dot{v}_e$ is directly proportional to the measured acoustic amplitude.

**Connecting acoustic signatures with transient vaporization dynamics**

During intense laser-induced vaporization, mechanical acoustic waves are generated by fluctuating jets around the vapor exit and propagate outward until they reach the ultrasonic microphone, where they are recorded as time-series signals. To investigate further the mechanism by which the airborne acoustics are generated near $\partial\Omega_e$, we use *Lighthill's aeroacoustic formulations*[29,30] to formulize emitted acoustic signatures, which can be further connected with vapor-jet dynamics and cavity oscillations using Eq. 1. By characterizing source terms in *the Lighthill equation for aeroacoustics*[29] with $v_e$, we obtain the following expression consisting of leading terms for the resultant acoustic pressure amplitude ( $p_{acc}$ ) originating from the cavity exit $\partial\Omega_e$:

$$p_{acc} - p_\infty \sim \frac{\rho_V \phi_e}{4\pi r}\left(2M+1\right)\frac{dv_e}{dt} \qquad \text{(Eq. 2)}$$

where $p_\infty$ represents the ambient atmospheric pressure, and $r$ and $M$ denote the distance between the microphone and $\partial\Omega_e$ and the Mach number of the vapor-jet, respectively. It is worth noting that under perturbations, Eq. 2 can be combined with Eq. 1 and obtain $p_{acc} - p_\infty = -\frac{\rho_V \phi_e}{4\pi r}(2M+1)\frac{d^2L}{dt^2}$. We call Eq. 2 the *Vapor-jet–Cavity (keyhole) Aeroacoustic Equation (VCAE)*, more derivation details of which are referred to the Supplementary Materials. Essentially, VCAE tells us that the measured acoustic signal near the source, whose amplitude depends nonlinearly on $\dot{v}_e$ and $M$ ( $v_{e0}/c_0 \approx v_e/c_0$ ) and decays proportionally to $1/r$, can be interpreted as emission yielded effectively by a monopole and a dipole directed along the vapor ejection axis. For most vapor-cavity cases where $M \ll 1$ (*i.e.*, low $v_e$ regime and no shock-wave involved), the acoustic emission amplitude is dominated by those of the monopole component and thus becomes proportional to $\dot{v}_e$, which, under steady oscillations, matches the evolution of $-\ddot{L}$. VCAE also serves as a confirmation and extension of the previously investigated laser welding acoustic mechanism[8], as well as an alternative interpretation of the *Ffowcs Williams–Hawkings (FW–H) equation*, used in aeroacoustics to predict the noise generated by moving rigid surfaces[36]. Finally, we emphasize that our analysis treats acoustic emission dominated by vapor-jet unsteadiness. Unless strong vortex shedding occurs near $\partial\Omega_e$, we for now neglect far-field turbulence- and eddy-driven sound better described by classical "jet-noise" models. This distinction clarifies what our airborne acoustics measures near the vapor-jet exit[20].

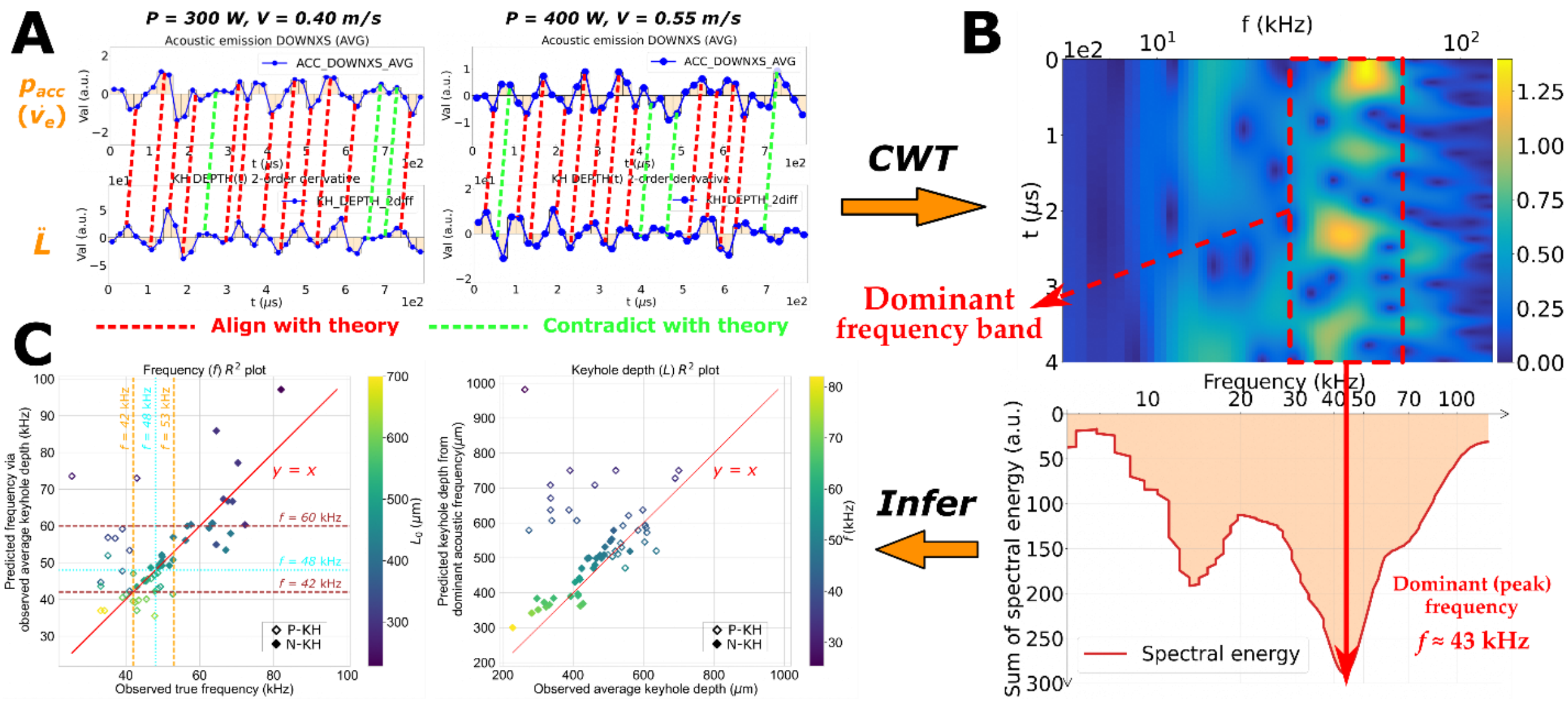

**Fig. 2. Discovered correlations between cavity-depth acceleration and measured aeroacoustic signatures enable interpretable inference of vapor-cavity oscillation properties on sub-millisecond timescales.** (**A**) Correlation between the acoustic data ( $p_{acc}$ , top row) and the second time-derivative of the measured vapor-cavity depth ( $\ddot{L}$ , bottom row) of two example cases. The red dashed lines connecting signals from two independent modalities reveal clear inverse correlation implied by Eq. 1 with a constant phase shift, illustrated by the slope of the dashed lines due to inherent synchronization offset of the data collection system (*i.e.*, around 20–40 $\mu s$); while a few point pairs labeled by the green dashed lines contradict the theory due to inevitable noise and arbitrary factors. (**B**) Data processing that converts time-series data to spectro-temporal scalograms via continuous wavelet transform. From the resultant scalograms, we extracted the highest peak of the energy summation value of each frequency band to represent the dominant acoustic frequency of the corresponding snippet. (**C**) Predictions of vapor-cavity oscillation characteristics, including oscillation frequency and average cavity depth, across process conditions within 400 $\mu s$ time window. For each 400 $\mu s$ snippet, VCODE: (1) uses vapor-cavity geometric features to estimate the oscillation frequency (left), and (2) uses the dominant acoustic frequency to infer the mean cavity depth (right). Solid markers denote N-KHs, and hollow markers denote P-KHs. In the left plot, we observe two critical threshold frequencies ( $f_c$ ) along each axis (orange and brown dashed lines) that roughly partition the map into three regimes—predominantly P-KHs, predominantly N-KHs, and a mixed region—with a few outliers. Another frequency ( $f_c = 48$ kHz, cyan dashed lines) emerges as a critical intersection between N-KHs and P-KHs along the diagonal ( $y = x$ , red solid line), suggesting the existence of critical frequencies controlling keyhole stability and pore-shedding behaviors.

Guided by Eqs. 1 and 2, we investigate within 800 $\mu s$ snippets the correlation between the time histories of $\dot{v}_e$ and $\ddot{L}$, independently extracted and numerically calculated from experiment data of airborne acoustic emissions and side-view X-ray images of vapor cavities. Figure 2A showcases two example snippets with close temporal agreement between the two modalities, revealing that the acoustic signal captures the rate of vapor-jet through $\partial\Omega_e$ and, consequently, the evolution history and dominant frequencies of cavity depth oscillations. Results clearly show that the acoustic amplitude parallels its corresponding cavity depth evolution with the same oscillation frequency and a phase lag. Reconciling with our mechanistic interpretation, the discovered correlation and its associated interpretation enable noninvasive characterization of vapor-jet dynamics and cavity oscillations on sub-millisecond scales, demonstrating the potentials of quantifying vaporization-related properties and parameters by simply using airborne acoustic emission data.

**Analysis of the vapor-jet-driven cavity dynamics**

We further extend the above analysis to more general pore-free cavity oscillations. Starting from the integral mass and momentum balances over the domain enclosing the vapor cavity and introducing a few simplifying assumptions (see Supplementary Information), we arrive at the following new nonlinear ordinary differential equation for $L$, which we call the *Vapor-jet–Cavity (keyhole) Oscillation Dynamics Equation* (VCODE):

$$\rho_V \phi_e \left( L\ddot{L} + \dot{L}^2 \right) + \left( 2\varphi\rho_V\Omega - \dot{m} \right)\dot{L} + \left( 2\pi C\gamma\cos\theta - \ddot{m} \right)L = \left( p_e + \rho_V v_e^2 \right)\phi_e \qquad \text{(Eq. 3)}$$

where:

- $p_e$—local pressure at $\partial\Omega_e$.

- $\gamma$, $\theta$—coefficient of surface tension at $\partial\Omega_V$ and the cavity characteristic angle (*e.g.*, can be interpreted as the front wall inclination angle).
- $C$, $\varphi$—phenomenological parameters for interfacial forces at $\partial\Omega_V$ and the vertical vapor-velocity gradient, respectively.

Full derivation details are provided in the Supplementary Information (see Extended Data Fig. S1). Fundamentally, VCODE models the vapor-jet cavity as a vertical closed-end vapor depression column whose depth is under nonlinear oscillations sustained by continuous laser energy input. It is worth noting that $\dot{v}_e = -\ddot{L}$, previously used for perturbative analysis, does not hold at this point, since we don't have $\ddot{m} = 0$ anymore. From Eq. 3, we point out the following observations:

1. VCODE captures the interplay between vaporization-induced recoil effects ($\left(p_e + \rho_V v_e^2\right)\phi_e$) and depth-dependent restoring actions at $\partial\Omega_V$ ($2\pi C\gamma\cos\theta \cdot L$; more details in the Supplementary Information), the driving forces governing the vapor-cavity oscillation.
2. Additional theoretical analysis (Supplementary Information) further shows that the cavity depth co-oscillates with the acoustic pressure (*i.e.*, same frequency as the driving force), with a damping-dependent phase lag.
3. It introduces a new dynamical-systems perspective: $\dot{m}$, $\Omega$, $\dot{L}$, and $\varphi$ in Eq. 3 regulate whether cavity oscillations decay or grow, providing a mechanistic description that spans energy-dissipating responses and self-amplifying, exponentially growing depth dynamics. As their balance shifts, the damping can vary, cross zero, and change sign, thereby moving the vapor-jet cavity oscillator between stable and unstable regimes and motivating mechanistic descriptions and further exploration of self-oscillation, nonlinear saturation, and limit-cycle behavior.

4. Mathematically, it not only parallels but also further extends the well-known *Rayleigh–Plesset equation*[34,35], which assumes spherical cavity symmetry and is widely used to model bubble oscillations and cavitation in liquids. By instead treating an axisymmetric cavity and explicitly accounting for vapor-velocity gradient and vapor-jet flux along cavity depth, VCODE underscores its novelty and distinction from established physical descriptions of cavity dynamics.

In essence, VCODE describes cavity-depth fluctuations as the motion of a sustained nonlinear oscillator whose properties and transient evolution are governed by geometric factors, material parameters, and boundary variables (*i.e.*, $L$ ($\dot{L}$), $\dot{m}$, $v_e$, $\theta$, $\gamma$).

Next, under small perturbations, we substitute $\delta\dot{v}_e = -\delta\ddot{L}$ into Eq. 3 and obtain the following perturbative form of VCODE:

$$\rho_V \phi_e \left[ L_0 \, \delta\ddot{L} + \left( 2\varphi L_0 + v_{e0} \right) \delta\dot{L} \right] + 2\pi C\gamma \cos\theta_0 \cdot \delta L = 0 \qquad \text{(Eq. 4)}$$

More details are provided in the Supplementary Information. At a specific moment of perturbative oscillation, Eq. 4 takes the canonical form of a damped oscillator for $\delta L$, where the first term corresponds to effective mass inertia ($M = \rho_V \phi_e L_0$), the second term to effective damping ($c = \rho_V \phi_e \left( 2\varphi L_0 + v_{e0} \right)$, where the sign of $\varphi$ depends on the direction of vapor-velocity gradient), and the third term to the effective restoring stiffness arising from interfacial actions including capillary surface tensions ( $k = 2\pi C\gamma \cos\theta_0$ ) (shown in Fig. 1C). Under perturbative pore-free oscillation of the cavity depth, an immediate first-order estimate of its frequency is the natural frequency ( $f_N$ ) derived from Eq. 4, writing as:

$$f_N = \frac{\omega_L}{2\pi} \approx \frac{1}{2\pi}\sqrt{\frac{2\pi C\gamma\cos\theta_0}{\rho_V \Omega_0}} = \sqrt{\frac{C\gamma\cos\theta_0}{2\pi\rho_V \phi_e L_0}} \qquad \text{(Eq. 5)}$$

Under perturbative cavity oscillations, Eq. 5 serves as an effective linear approximation for the actual cavity depth oscillation frequency, which is related to the material properties and cavity's geometric factors. When damping becomes non-negligible, the actual frequency should be closer to the system's resonant frequency that is considerably smaller than $f_N$, making it an effective upper bound of the actual vapor-cavity oscillation frequency.

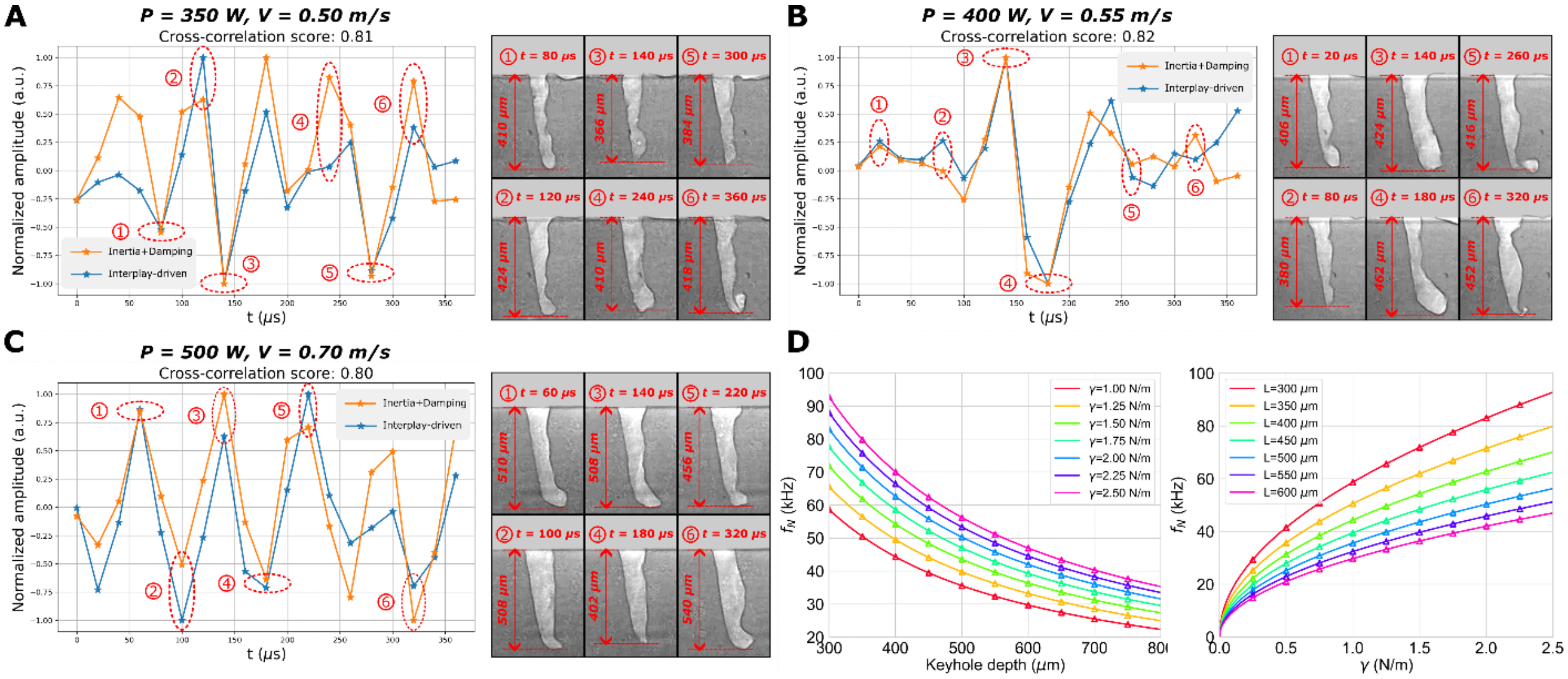

**Fig. 3. VCODE-enabled cross-modal agreement over sub-millisecond windows and sensitivity analysis.** (**A**)(**B**)(**C**) All measurements in the showcased three snippet subplots are temporally aligned and segmented into 400 *μs* windows. For each snippet, we compute the normalized *Inertia+Damping* and *Interplay-driven* terms of the VCODE (Eq. 3) using acquired experimental data after synchronization. In the left panel, despite a few mismatching outliers, the geometry-driven and acoustic-driven time traces show consistent trend alignment across different operating conditions. The corresponding X-ray frames annotated with time stamps (the right panel) visualize the cavity

shapes at the labeled points. Similar results have been obtained over a total of fifty snippets across different conditions. (**D**) Sensitivity of estimated natural frequency with respect to average cavity depth (left) and surface tension coefficient (right) according to the proposed theory.

Combining Eqs. 1, 4, and 5, our theory further implies that we can use acoustic frequencies to estimate the actual frequency of the steady perturbative cavity depth oscillation and, in turn, infer the geometric parameters of the cavity itself within sub-millisecond time windows. We demonstrate this potential predictive power of Eq. 5 on experimental data across different process parameters. For each 400 $\mu s$ snippet, we either predict the oscillation frequency by substituting cavity geometric features or predict the mean depth by substituting the dominant airborne acoustic frequency obtained from CWT scalograms, shown in Fig. 2B. In Fig. 2C, we demonstrate the inference results of frequency given depth (left) and depth given frequency (right) and summarize the following three major takeaways:

1. For N-KHs (solid markers), theoretical predictions agree with both the dominant acoustic frequencies and X-ray measured mean cavity depths (frequency prediction $R^2 \geq 0.87$, depths prediction $R^2 \geq 0.85$).
2. For P-KHs (hollow markers), predictions degrade as expected, consistent with the imposed limits of validity (see Supplementary Information) that our theory collapses when non-negligible pores are generated.
3. Vapor cavities tend to break down and shed pores within low-frequency resonance regimes. In the left plot, specifically, we observe two critical threshold frequencies ( $f_c$ ) along each axis (orange dashed lines—42 kHz and 53 kHz for the observed frequency; and brown dashed lines—42 kHz and 60 kHz for the prediction frequency) that roughly partition the frequency spectrum into three parts based on the cavity types—predominantly P-KHs, predominantly N-KHs, and a mixed region—with a few outliers. Another frequency ($f_c = 48$ kHz, cyan dashed lines) emerges as a critical intersection between N-KHs and P-KHs along the diagonal ($y = x$, red solid line). These frequencies fall within the range previously reported for Ti-6Al-4V vapor-cavity oscillations[19], suggesting a critical vapor-cavity stability threshold that may be governed by resonance of the coupled cavity-liquid system.

We further combine Eq. 3 with experimental measurements to qualitatively validate the proposed VCODE theory. Because converting each sensor modality to absolute physical units is extremely challenging, we instead compare the normalized temporal trends of two aggregated terms: (i) the sum of the nonlinear inertial and damping contributions (*Inertia+Damping*, the first two terms on the left of Eq. 3), and (ii) the net "competitive interplay" defined as the difference between the right-hand side of Eq. 3 and the restoring term (*Interplay-driven*, the right-hand side of Eq. 3 subtracts the last term on the left). This decomposition is motivated by the observation that the *Inertia+Damping* contribution is typically an order of magnitude smaller than the *Interplay-driven* term, so it can be interpreted as a dynamical residual arising from the near-balance between recoil forcing and capillary restoration. We computed the two summation terms purely from experimentally collected data across a series of out-of-sample snippets with different process conditions, evaluating their cross-correlation scores to demonstrate the effectiveness of VCODE. It is worth noting that since we don't have $\ddot{m} = 0$, we use the auxiliary photodiode modality to capture laser absorption that can be used to infer $\dot{m}$. As shown in Figs. 3A–3C, we discover and demonstrate that after calibrating phenomenological parameters in Eq. 3 using only one 400 $\mu s$ snippet, *Inertia+Damping* and *Interplay-driven* terms show consistent trend alignment across three snippets encompassing different *P–V* settings. The results of more snippets are available in the Supplementary Information (Extended Data Fig. S2).

Finally, we perform sensitivity analysis using Eqs. 4 and 5 regarding how $f_N$ changes with varying vaporization parameters. Figure 3D clearly shows that $f_N$ decreases with increasing cavity depth and decreasing $\gamma$; from Eq. 5, we also realize that $f_N$ should increase with decreasing $\theta$. Theoretically, $\cos\theta$ depends on the aspect ratio of the keyhole $\left(L/2R,\ \phi_e = \pi R^2\right)$ (see Supplementary Information); from a physical perspective, however, we can interpret $\theta$ as the vapor-cavity front-wall inclination angle, which sets the direction of capillary forces along the front wall and does not materially affect the interpretation of VCODE. Therefore, these observations link the frequency of cavity-depth fluctuations to geometric parameters and liquid-vapor interfacial properties, providing new perspectives for a physics-based understanding of cavity oscillations that can, in turn, govern defect formation[19]. The implications and impacts of this observation will be further discussed in the next section.

**Implications and impacts in metallic additive manufacturing**

Additional analyses—including a dimensionless stability number for the vapor-jet-driven open cavity and the correspondence between acoustic and thermal signatures—are provided in the Supplementary Information (Extended Data Figs. S3–S6). Collectively, they support a mechanistic picture of the vapor-jet–cavity as a damped, driven dynamical system: cavity depth behaves as a lossy resonator continuously forced by vapor-jet/recoil-pressure fluctuations. In this regime, the cavity response is frequency-locked to dominant forcing components, while broadband excitation is preferentially amplified near the cavity's resonant band. Because the airborne acoustic field is generated by (and radiated approximately linearly from) the same unsteady jet and pressure field that both drives and responds to cavity motion, the emitted tone inherits the system's dominant response frequency. Accordingly, under conditions where cavity depth is governed by nonlinear coupled oscillations, the dominant acoustic tone near the cavity exit tracks the transient cavity-depth oscillation frequency, with a damping-induced phase lag between forcing and response. This coupling generalizes across materials, as shown by AlSi10Mg bare-plate measurements (Extended Data Fig. S3; Extended Data Video S1) and by cross-correlation analyses between cavity-depth and acoustic-amplitude evolutions (mean absolute correlation > 0.6; Extended Data Fig. S4; Extended Data Videos S1–S3). Together, these results clarify the physical origins of the measured signals and establish sub-millisecond, deterministic links among vapor-jet dynamics, cavity oscillations, and airborne acoustics—enabling accurate, rapid (few-hundred-microsecond) and low-cost, noninvasive inference of cavity depth and oscillation frequency from airborne measurements.

To extend the impact of this work and further illustrate the broader implications of our theory, we contextualize the newly developed framework in laser-PBF, a widely adopted LMP and additive manufacturing technique in which intense vaporization is common, to explore further implications and practical applications. We utilize the newly identified thresholding keyhole oscillation frequency in Fig. 2C and VCODE formulations to construct a new interpretation of the established KH-bound state. As shown in Fig. 4A, inspired by Eq. 5 and based on the data measured and extracted from X-ray synchrotron images, we discovered that experimentally derived quantities, namely $\frac{\cos\theta}{L}$ *vs.* $\frac{P-P_0}{V^n}$ (in Fig. 4A, we have $n=4$ and $P_0=50$ W for Ti-6Al-4V from experiments, where $P_0$ denotes the maximum zero-scan-speed power for pore-free keyhole oscillation and is introduced based on the fact that there always exists such a non-zero keyhole depth when $V$ vanishes), exhibit a strong reciprocal relationship. Motivated by the observation in Fig. 2C, we hypothesize a single critical keyhole oscillation frequency, $f_c$, whose iso-frequency contour in the *P*–*V* space delineates the keyhole-defect regime and defines the corresponding process window. Assuming that all N-KHs remain in the perturbative-oscillation regime until they become unstable and transition to P-KHs, we use $f_N$ from Eq. 5 to estimate $f_c$ at the critical boundary. By keeping $n$ as an undetermined parameter and substituting $\frac{\cos\theta}{L}=B\frac{V^n}{P-P_0}+b$ (where $B$ and $b$ are fitting constants with specific units) into Eq. 5, we obtain:

$$P=\frac{BC\gamma}{2\pi\rho_V\phi_e f_c^2-bC\gamma}V^n+P_0 \qquad \text{(Eq. 6)}$$

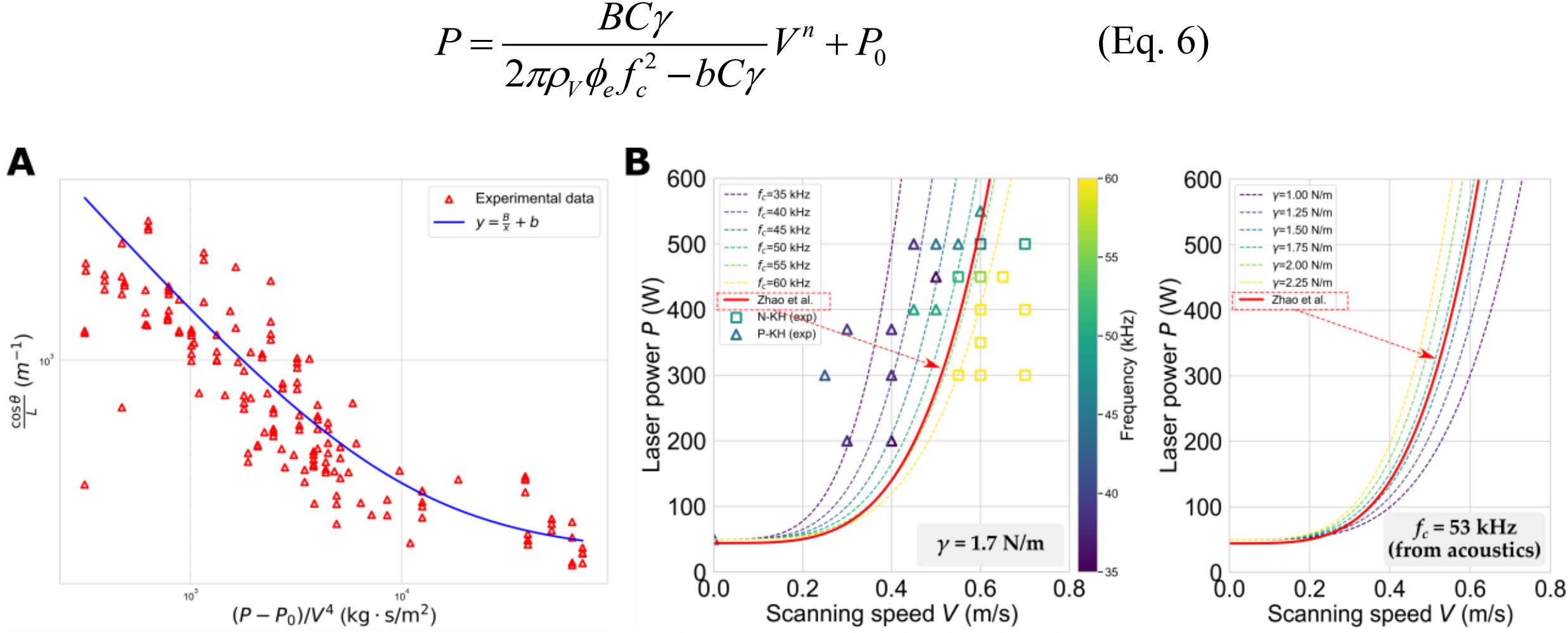

**Fig. 4. Experimentally discovered interpretation, scaling, and modeling of the KH-bound and vapor keyhole frequency threshold.** (**A**) Experimental measurements show the inversely proportional empirical relationship (with constants $B$ and $b$) between two ratios derived from (with $n=4$) collected data as a log-log plot. The corresponding fitting parameters as well as state variables are listed in the Supplementary Information and will be used for the visualization of Eq. 6. Red triangular markers denote the data points, and the blue curve is a best-fit inverse function. (**B**) Interpretation of the KH-bound as an iso-frequency contour in the *P*–*V* space. We plot the KH-bound reported by Zhao *et al.*[19] (red), meanwhile superimposing iso-frequency contours with $n=4$ and examining their sensitivities with respect to varying $f_c$ and $\gamma$. Specifically, these iso-frequency contours are obtained: (1) for $f_c$ ranging from 35–60

kHz (left, with $\gamma$ fixed near that of liquid Ti-6Al-4V at its boiling point, *i.e.*, $\gamma \approx 1.7$ N/m), and the result shows that the reported boundary aligns closely with the $f_c = 53$ kHz contour, matching what we observed from the experiments; (2) for $\gamma$ ranging from 1–2.25 N/m (right, with $f_c = 53$ kHz). Moreover, we scatter in the left plot datapoints of acoustic frequency (annotated with colors) measured for both P-KHs (triangular markers) and N-KHs (square markers) from multimodal experiments of different *P*–*V* conditions; we observe a general frequency gradient perpendicular to the visualized iso-frequency contours. These findings provide potential explanations and interpretations for the experimentally characterized KH-bound in LMP and specifically laser-PBF community.

We observe that Eq. 6 reveals a polynomial relationship between *P* and *V* on the KH-bound, a general mathematical expression of the iso-frequency contour in the *P*–*V* space that corresponds to the hypothesized $f_c$. As shown in Fig. 4B, when using fitted parameters from Fig. 4A and Figs. 3A–3C, $n = 4$, and specific material parameters of Ti-6Al-4V ($\gamma = 1.7$ N/m), the superimposed KH-bound reported by Zhao *et al.*[19] closely agrees with the iso-frequency contour corresponding to $f_c = 53$ kHz, matching one of the thresholding frequencies independently found in Fig. 2C. In the left plot of Fig. 4B, we scatter datapoints of acoustic frequency (annotated with colors) measured for both P-KHs (triangular markers) and N-KHs (square markers) from multimodal experiments of different *P*–*V* conditions, demonstrating a general frequency gradient perpendicular to the visualized iso-frequency contours. Furthermore, Fig. 4B also showcases how the KH-bound may vary in the *P*–*V* space when we change $\gamma$ or $f_c$. Aside from the above, this result also links to a recent study[37], which reported that the KH-bound depends on keyhole aspect ratio and inclination angle, consistent with our prediction of a critical iso-frequency contour related to the same quantities. Accordingly, Eq. 6 offers a theoretical basis for the KH-bound and yields an interpretable process map that can guide parameter selection and optimization.

From a mechanistic perspective, VCODE offers an alternative physical picture of keyhole oscillations. As shown in Eqs. 3 and 4, VCODE essentially describes the vapor-jet–cavity as a nonlinear oscillatory dynamical system; the effective damping coefficient of this system, namely $\left(2\varphi\rho_V\Omega-\dot{m}\right)$ or $\rho_V\phi_e\left(2\varphi L_0+v_{e0}\right)$, is fundamentally controlled by (1) the direction of vapor-velocity gradient and (2) the relation between $\dot{m}$ and $L$ (see Supplementary Information). In other words, this effective damping coefficient can take either positive (energy-dissipating) or negative signs (energy-injecting) depending on the transient status of the vapor-jet–cavity system, thereby capturing different keyhole dynamics regimes. When the damping coefficient crosses zero, system trajectories may transition from self-amplifying exponential growth to nonlinear saturation, potentially restabilizing into bounded oscillations or abruptly collapsing, producing pores and resetting the keyhole state.

Notably, during collapse, uncontrolled pore evacuation can violate the strict mass-conservation assumptions, stating the theory's domain of validity: quasi-axisymmetric keyholes with negligible mass loss to pores satisfy the assumptions; whereas extremely shallow cavities, strongly asymmetric geometries, or conditions with significant pore ejection violate mass conservation and lie outside its scope. What's more, we showed that $\dot{m}$ can be inferred indirectly from photodiode-based laser-reflection measurements, because it depends strongly on the total laser absorption within the cavity, which in turn is governed by the evolving cavity geometry and absorptivity. We identify the above as key limitations and avenues for future work, where additional modeling concepts and sensing modalities, such as self-excited oscillations, chaotic dynamics, or sophisticated photodiode sensing, may be required to capture unstable vapor cavity behaviors.

More broadly, we believe this work lays a general physical foundation for cost-efficient physical characterization of vapor-jet–cavity dynamics during intense evaporation, which arises across a wide range of scientific and engineering settings[2–8]. With the support of the new theoretical framework and synchronized observations from our vaporization prototype, airborne emissions can be elevated from qualitative indicators to quantitative, physics-informed observables that recover key dynamical variables and stability-relevant thresholds. Moreover, the coupled modeling strategy developed here—linking first-principles cavity dynamics with aeroacoustic radiation and validating the mapping against time-resolved geometry—offers a transferable route for interrogating other non-equilibrium vaporization systems in which driven vapor jets interact with compliant liquid-vapor interfaces. This perspective opens opportunities for scalable monitoring, interpretable process conditionings, and feedback-ready control strategies in vaporization-mediated manufacturing and beyond.

## Conclusions

We recast ubiquitous aeroacoustic signatures from intense laser-induced vaporizations into cost-effective microsecond-level probes of transient vapor-cavity depth acceleration and cavity oscillation frequency, enabling a noninvasive and scalable route to resolve extremely fast vapor-jet-driven cavity dynamics. Building on this observability, we develop the *Vapor-jet–Cavity (keyhole) Oscillation Dynamics Equation* (VCODE), which models the excessive-vaporization cavity as a nonlinear dynamical system whose depth fluctuations regulate vapor ejection and radiated sound. VCODE linearization yields predictive expressions for oscillation frequency, amplitude, and stability. Coupling VCODE with aeroacoustic formulations leads to the *Vapor-jet–Cavity (keyhole) Aeroacoustic Equation* (VCAE), which connects vapor-ejection kinematics to measurable acoustic pressure near the cavity exit. Synchronized operando experiments validate predictions from VCODE and VCAE, showing that airborne acoustics enable accurate inference of keyhole oscillation frequency and mean depth in pore-free regimes within sub-millisecond windows (frequency: $R^2 \geq 0.87$ ; depth: $R^2 \geq 0.85$ ). We further identify stability-relevant thresholding frequencies that delineate pore-shedding from pore-free keyhole behavior. Together, these results establish a physics-based foundation for real-time, control-relevant inference of vapor-jet-driven cavity dynamics in metal additive manufacturing.

Our framework supports a deterministic dynamical-systems view in which vapor cavities behave as nonlinear, self-sustained oscillators, and their acoustic emissions encode internal state variables that can be recovered from airborne measurements within a well-defined validity regime. Within this regime, the KH-bound results establish quantitative stability delineations—via critical thresholding frequencies that separate pore-free perturbative oscillations from pore-shedding or collapse-prone behaviors—thereby providing an interpretable map of operating conditions and actionable margins for process design. Finally, the coupled strategy advanced, *i.e.*, deriving first-principles cavity dynamics, linking them to aeroacoustic radiation, and validating that mapping against time-resolved geometry, offers a transferable route for quantitative, cost-efficient characterization of other non-equilibrium vaporization systems in which driven vapor jets interact with compliant liquid–vapor interfaces, enabling interpretable monitoring and feedback-ready control beyond laser powder bed fusion.

# Methods

## Vapor-jet-cavity prototype based on laser-induced evaporation

Controllable laser-induced evaporation were performed on metallic alloy using a modulated, steerable fiber laser (maximum power 700 W, wavelength 1064 nm) with a Gaussian beam profile and an 82 $\mu m$ focal-spot diameter. Experiments were conducted using a focused laser across a range of power–scan speed ($P$–$V$) combinations to systematically probe vapor-cavity behavior. The scan length was 4 mm when the laser traveling speed was non-zero. The experimental environment was a closed airtight chamber, filled with argon at atmospheric pressure and oxygen level <100 ppm. The substrate material, dimensions, and surface preparation were: Ti-6Al-4V alloy and AlSi10Mg alloy, 50 mm × 5 mm × 0.5 mm, with all surfaces well-polished up to P800.

## Synchronized multimodal *in situ* diagnostics

A synchronized sensing suite captured complementary process responses across timescales and spatial resolutions, including: (i) an ultrasonic microphone sampled at 1 MHz to record airborne ultrasonic emissions; (ii) a photodiode sampled at 1 MHz to monitor laser reflection; (iii) a near-infrared (NIR) thermal camera operated at 250 kfps with 1 μs exposure to resolve melt-pool and cavity-opening dynamics; and (iv) high-speed synchrotron X-ray radiography operated at 50 kfps with 2 μs exposure and 2 μm/pixel to directly visualize cavity geometry evolution.

The ultrasonic microphone (PCB Piezotronics, model 378C01) was positioned near the top of the fusion zone to capture near-source acoustic signals from vapor jets exiting the cavity while minimizing turbulence contamination. The microphone was placed approximately 5 mm from the fusion spot. The microphone was mounted at the top of the plate, oriented nearly perpendicular relative to the vapor exit surface. Raw acoustic time series were digitized at 1 MHz using the NIST-DAQ system.

A photodiode was used to monitor laser reflection at a roughly 50° reception angle. The spectral filtering was a 1064 nm bandpass filter for the collection of laser reflections. The photodiode signal was digitized at 1 MHz on the same acquisition system as the acoustic channel. The photodiode voltage was used as a relative measure of reflected intensity.

Synchrotron X-ray radiography. Side-view high-speed X-ray radiography was conducted at the Advanced Photon Source (beamline 32-ID) to visualize transient cavity geometry at 50 kfps (2 $\mu s$ exposure, 2 $\mu m$/pixel). The imaging plane was aligned to capture the cavity; the angle between scan direction and the imaging plane was 90°. NIR thermal imaging was performed using a high-speed camera operated at 250 kfps with 1 µs exposure. The NIR signal was interpreted as relative thermal intensity.

Timing, triggering, and synchronization. All sensor channels were registered onto a common global timeline to enable direct correlation between acoustic emissions and X-ray-resolved cavity geometry. Triggering hardware was achieved via a manual switch. To account for acoustic time-of-flight from the effective vibration source near the cavity's opening to the microphone, an approximately 15 µs temporal correction was applied to the acoustic channel. After synchronization, the registered measurements were segmented into short snippets of duration 400 µs (and, when noted, 800 $\mu s$) for subsequent correlation analysis. Snippet start times were selected along the scanning path. Snippets used for calibration were defined as the initial snippets in each dataset, and independent validation used separate snippets not included in calibration.

## Multimodal data processing analysis

Acoustic time–frequency analysis. Airborne acoustic signals were analyzed in the time domain and, when needed, using continuous wavelet transforms (CWT) to obtain time–frequency scalograms and extract dominant frequencies and spectro-temporal features. The preprocessing applied prior to CWT (detrending, filtering, normalization) was deseasonalization, which removes obvious global trends/DC components. The wavelet family selected was Morlet, and the analyzed frequency range and frequency sampling were 5–100 kHz. All signal processing was implemented using ImageJ, Python, and MATLAB Wavelet Toolbox.

X-ray segmentation and geometric feature extraction. Cavity geometries were segmented from synchrotron X-ray images using a U-Net-based convolutional neural network trained on manually annotated frames. The inference pipeline produced binary masks of the vapor cavity for each frame. From segmented masks, we quantified instantaneous characteristic length (depth) of the cavity, opening area, and inclination. Cavity depth was defined as the distance between the vapor cavity exit and the bottom pixel belonging to the detected cavity geometry. Opening area was defined as the area of a circle with a diameter equal to the opening width, and inclination was defined as the slope of the front wall. Mask postprocessing (morphological operations, contour smoothing, removal of spurious regions) was implemented using Python. Measurement uncertainty was assessed by the intersection-over-union and the Dice score, and the resulting geometric time series were co-registered with the acoustic and photodiode signals.

Cross-modality correlation and model validation. Co-registered time series were used for correlation analysis and model validation. Numerical derivatives of the depth signal (*e.g.*, velocity and acceleration) were computed using the Python SciPy package. Cross-correlation analysis between acoustic pressure and depth-derived metrics used cross-correlations, with lag search bounds of 10 μs. Calibration was performed on one reference snippet, and predictive performance was reported on independent validation snippets. According to VCAE, we have for low-Mach jet flow $p_{acc} \propto \dot{v}_e$. In both perturbative and finite nonlinear vapor-jet-driven cavity oscillations, the fluctuation magnitude of $v_e$ is no bigger than its long-time-average, *i.e.*, $v_{e0}$. Therefore, we substitute phase-shifted airborne acoustic signals to $v_e$-dependent terms—such as $v_e$ or $v_e^2$— in JCODE for corresponding qualitative analysis since $\dot{v}_e$ mathematically shares the same frequency and variational trend with $v_e = v_{e0} + A_{Ve}e^{i\omega_V t}$ $(v_{e0} \geq A_{Ve})$ as well as $v_e^2$. The inverse of photodiode-collected laser-reflection signal is substituted qualitatively as $\dot{m}$. Fluctuations in vapor-jet-driven cavity features can be measured either by X-ray imaging of the cavity–compliance profile or by thermal/NIR imaging of the emerging jet at the cavity exit; the resulting oscillation frequency tracks the cavity's compliance ("breathing") evolutions and agrees with the acoustic dominant frequency, consistent with our new VCODE/VCAE framework and useful for the subsequent physics-guided quantitative inference with sub-millisecond time window.

## Data and materials availability

All data needed to evaluate the conclusions are available in the main text and Supplementary Information (Supplementary Figures 1-6 and Supplementary Table 1). Supplementary Data 1 (data_S1.zip) contains the processed datasets used for figures and model validation. Supplementary Videos 1-3 (movie_S1.mp4-movie_S3.mp4) provide representative synchronized recordings. Raw X-ray radiography and raw acoustic time series are available from the corresponding authors upon reasonable request.

## Code availability

Custom code used for acoustic time-frequency analysis, X-ray segmentation, parameter identification, and acoustic-to-state inference is available from the corresponding authors upon reasonable request.

## Acknowledgements

We thank **K. Mumm** and **L. Balderson** for assistance with the Advanced Photon Source beamline experiments. We thank **C. Li, R. Jiang, K. Ferguson, X. Li, J. Malen, S. Feng, S. Khairallah, N. Calta, J. Cao, Y. J. Zhang, A. Myer, Y. Sun, I. Matthews, D. Wu, N. Lamprinakos, J. Beuth, T. Reddy,** and **Z. Taylor** for helpful discussions. We thank **V. Rajan** for assistance with dataset preparation and **S. Liu** for support. This research was performed on APS beam time award(s) (**DOI: 10.46936/APS-184522/60012099**) from the **Advanced Photon Source**, a U.S. Department of Energy (DOE) Office of Science user facility operated for the DOE Office of Science by **Argonne National Laboratory** under Contract No. **DE-AC02-06CH11357**. **Z. R.** acknowledges partial efforts at the **18-ID (FXI) beamline** of the **National Synchrotron Light Source II (NSLS-II)**, a U.S. DOE Office of Science User Facility operated for the DOE Office of Science by **Brookhaven National Laboratory** under Contract No. **DE-SC0012704**.

## Fundings

Eaton Corporation Award Number **001145471** (HL, ADR, LBK).

MIT Postdoctoral Fellowship Program for Engineering Excellence (**PFPFEE**) (SKN).

A*STAR MTC Individual Research Project **M24N7c0082** (WY, YZ).

National Science Foundation grants **DMS 2108784** and **DMREF 1921857** (KD).

Army Research Office MURI **W911NF-24-2-0184** (KD).

## Author contributions

Conceptualization: HL, ADR, LBK;

Methodology: HL, SKN, ZR, YZ, JS, SJC, KF, XZ, LG, WY, NW, KD, TS, ADR, LBK;

Investigation: HL, SKN, ZR, YZ, JS, SJC, KF, XZ, LG, NW, KD, TS, ADR, LBK;

Resources and experiments: HL, ZR, SJC, KF, LG, TS, ADR;

Visualization: HL, SKN, ZR, YZ, XZ;

Funding acquisition: KD, WY, TS, ADR, LBK;

Project administration: HL, TS, ADR, LBK;

Supervision: HL, NW, KD, WY, TS, ADR, LBK;

Writing – original draft: HL, SKN, ZR, ADR, LBK;

Writing – review & editing: HL, SKN, ZR, YZ, JS, SJC, KF, XZ, LG, WY, NW, KD, TS, ADR, LBK.

## Competing interests

The authors declare that they have no competing interests.

# Supplementary Information

# Aeroacoustic signatures reveal fast transient dynamics of vapor-jet-driven cavity oscillations in metallic additive manufacturing

Haolin Liu[1,2*,†], S. Kiana Naghibzadeh[3†], Zhongshu Ren[4,5,6], Yanming Zhang[7], Jiayun Shao[4], Samuel J. Clark[8], Kamel Fezzaa[8], Xuzhe Zeng[1], Lin Gao[9], Wentao Yan[7], Noel Walkington[10], Kaushik Dayal[1,10,11], Tao Sun[4,8], Anthony D. Rollett[2*], Levent Burak Kara[1*]

[1]Department of Mechanical Engineering, Carnegie Mellon University; Pittsburgh, PA, USA.

[2]Department of Materials Science and Engineering, Carnegie Mellon University; Pittsburgh, PA, USA.

[3]Department of Civil and Environmental Engineering, Massachusetts Institute of Technology; Cambridge, MA, USA.

[4]Department of Mechanical Engineering, Northwestern University; Evanston, IL, USA.

[5]Department of Materials Science and Engineering, University of Virginia; Charlottesville, VA, USA.

[6]Physical Sciences and Research Operations Division, National Synchrotron Light Source II, Brookhaven National Laboratory; Upton, NY, USA.

[7]Department of Mechanical Engineering, National University of Singapore; Singapore.

[8]X-ray Science Division, Advanced Photon Source, Argonne National Laboratory; Lemont, IL, USA.

[9]Department of Mechanical Engineering, The University of Alabama; Tuscaloosa, AL, USA.

[10]Center for Nonlinear Analysis, Department of Mathematical Sciences, Carnegie Mellon University; Pittsburgh, PA, USA.

[11]Department of Civil and Environmental Engineering, Carnegie Mellon University; Pittsburgh, PA, USA.

*Corresponding authors. Email: haolinl@andrew.cmu.edu; rollett@andrew.cmu.edu; lkara@andrew.cmu.edu.

†These authors contribute equally to this work.

## This PDF file includes:

Supplementary Text

Supplementary Figures S1 to S6

Supplementary Table S1

## Other Supplementary Information for this manuscript:

Supplementary Videos S1 to S3

Supplementary Data S1

# 1 Supplementary Text

## 1.1 First-principles modeling of vapor-jet cavity oscillation

### 1.1.1 Modeling schematic of the vapor-jet cavity

The derivation is conducted using symbolic and tensorial notations within the framework of continuum mechanics, specifically fluid mechanics, which focuses on describing the conservation of mass, momentum, and energy in a mechanical system, along with the corresponding boundary conditions. It is important to note that this investigation does not prioritize numerical simulations or the solution of differential equations over a discretized time and space domain. Instead, our primary objective is to establish a simplified, idealized, yet phenomenologically effective theoretical framework to develop a heuristic understanding of the fundamental physics underlying laser-metal interactions.

To initiate this analysis, we introduce key notations and critical assumptions to construct an idealized and conceptually clear physical model representing a vapor-jet compliant cavity in laser-induced vaporization. All relevant variables (assumed to be tensors by default) and their corresponding notations are outlined in Fig. S1 and a bullet-point list in Sec. 1.1.2.

### 1.1.2 Nomenclature

We define the following concepts and variables, including those shown in Fig. S1:

- $\Omega(t)$ & $\partial\Omega(t)$—the vapor-jet cavity region & its boundaries mainly containing only the vapor phase of the alloy.
    - Specifically, one can further decouple $\partial\Omega(t)$ into two partial boundaries—$\partial\Omega_e(t)$ and $\partial\Omega_V(t)$ (namely, $\partial\Omega(t) = \partial\Omega_e(t) \cup \partial\Omega_V(t)$), which are the vapor-jet cavity opening and the vapor-jet cavity wall (*i.e.*, the liquid–vapor interface), respectively.
    - It is easily seen that $\partial\Omega_e(t)$ and $\partial\Omega_V(t)$ are two types of boundary — they are a fixed (control) boundary and a moving (with the liquid–vapor interface) boundary, respectively.

- $\Omega(t)$ has an outward normal vector $\vec{n}$ defined as shown above.

- $\Pi(t)$ & $\partial\Pi(t)$ — a selected small material volume region & its boundaries at a local interface $\partial\Sigma(t)$ (Note that $\partial\Sigma(t)$ is only an infinitesimal portion of $\partial\Omega_V(t)$).

  - $\Pi(t)$ & $\partial\Pi(t)$ should contain both the liquid (also denoted as "+" phase) and the vapor phase (also denoted as "-" phase) of the alloy, necessary for characterizing mass transportation across $\partial\Sigma(t)$ and characterize the boundary conditions.
  - It is easily seen that $\partial\Pi(t) = \partial\Pi^+(t) \cup \partial\Pi^-(t)$ is a material (convective) boundary, implying zero mass transportation across it.
  - Similar to $\partial\Omega(t)$, one can further decouple $\Pi(t)$ and $\partial\Pi(t)$ into $\Pi^+(t) \cup \Pi^-(t)$ and $\partial\Pi^+(t) \cup \partial\Pi^-(t)$, respectively, with superscripts "+" and "-" indicating different sides of $\partial\Sigma(t)$.
  - Two phases of the alloy in $\Pi(t)$ is divided by $\partial\Sigma(t)$.
    - Therefore, superscripts ("+" and "-") of $\Pi(t)$ and $\partial\Pi(t)$, as well as other corresponding variables, should be used for different sides of $\partial\Sigma(t)$.
  - $\Pi(t)$ has an outward normal vector $\vec{m}$ (for different sides/phases, it should be either $\vec{m}^+$ or $\vec{m}^-$ depending on which portion of $\Pi(t)$ we are working on) defined as shown above.

- $\rho_V$, $p_V$, $T_V$ and $\vec{v}_V$—state variables of the vapor phase of the alloy, namely density, pressure, temperature, and velocity in the vapor-jet cavity.

- Similarly, we have $\rho_e$, $p_e$, $T_e$ and $\vec{v}_e$—state variables of the vapor phase of the alloy at $\partial\Omega_e(t)$, namely density, pressure, temperature, and velocity at the vapor-jet cavity opening. These quantities determine the vibration source that emits airborne acoustic emissions, which are subsequently collected by the ultrasonic microphone.

- $\rho_L$, $p_L$, $T_L$ and $\vec{v}_L$—state variables of the liquid phase of the alloy, namely density, pressure, temperature, and velocity in the vapor-jet cavity.

- $\vec{v}_b$—the velocity of the vapor-jet cavity wall, a.k.a. the liquid–vapor interface $\partial\Omega_V(t)$.

- $p_{\text{recoil}}$ and $p_{\text{sat}}$—the total effective recoil pressure yielded at the liquid–vapor interface and the saturation vapor pressure at a point in $\Omega(t)$, respectively.

- $\sigma_{\mathbf{V}}$ and $\sigma_{\mathbf{L}}$—the Cauchy stress tensors of the vapor and liquid phases of the alloy, respectively.

  – When necessary (*e.g.*, momentum/energy conservation equations), we use $\mu_L$ to denote the viscosity of the liquid alloy.

- $\Omega$, $\phi_e$ and $L$—vapor-jet cavity volume, cross-sectional area (or vapor-jet cavity opening area), and vapor-jet cavity depth, respectively.

  – If assuming the vapor-jet cavity has an ideal cylindrical shape, we have $\Omega = \phi_e L$.

- $\dot{m}$—the mass rate (unit: kg/s) of vaporization leaving the liquid phase of the alloy (entering the vapor phase of the alloy) at $\partial\Omega_V(t)$.

- $\xi$—the total laser reflectance.

- $P$ and $w$—the power and scanning speed of the fusion laser (unit: W and m/s, respectively).

Table S1 summarizes the above nomenclature for easy reference.

### 1.1.3 Initial assumptions

In essence, the vapor-jet cavity region, represented by the control volume $\Omega(t) \cup \partial\Omega(t)$, serves as the central focus of our modeling. We assume that $\Omega(t)$ undergoes continuous fluctuations, driven either by vapor-jet cavity instability or the complex interplays of two-phase fluid dynamics. These dynamic variations play a crucial role in linking vapor-jet cavity geometric features with physically meaningful quantities that are emitted, generated, or altered throughout the printing process, thereby influencing the overall stability and efficiency of laser-based manufacturing.

Based on the above interpretation, we make the following essential assumptions (some of them may be re-emphasized in the later derivations):

- For simplicity, we do not formulate the problem in a moving-frame reference, despite acknowledging that the vapor-jet cavity continuously travels along the laser scanning direction at a velocity nearly identical to the laser's travel speed.
  - One rationale supporting this assumption is that we disregard the laser's travel speed, as we believe it is nearly negligible compared to the velocity of vapor ejection within the vapor-jet cavity.
- In analyzing the vapor-jet cavity formation and fluctuation process discussed in this document, we assume that only two material phases are involved: the liquid and vapor phases of the alloy.
  - Therefore, we neglect complex phenomena such as plasma generation, metal powder or substrate melting, and vapor-plasma expansion resulting from laser decoupling.
  - This assumption may intuitively imply: the newly generated vapor enters $\Omega(t)$ from the liquid phase of the alloy via the liquid–vapor interface $\partial\Omega_V(t)$, *i.e.*, the evaporation of the alloy happens only at $\partial\Omega_V(t)$ as a form of mass transportation across this singular (Knudsen) boundary.
- The absorbed laser energy at $\partial\Omega_V(t)$ is assumed to be predominantly utilized in overcoming the energy barrier associated with the vaporization process, specifically the enthalpy of vaporization of the alloy.
- We assume that evaporation occurs primarily at the liquid–vapor interface $\partial\Omega_V(t)$, which induces a recoil pressure arising from the rapid ejection of vaporized material. Physically, this recoil manifests as an inertial force, resulting from the net momentum flux normal to the interface boundary $\partial\Omega_V(t)$.
- It is assumed that the liquid phase of the alloy is a homogeneous, incompressible Newtonian fluid, *i.e.*,
  - $\nabla\rho_L = \vec{0}$, and

- $\frac{D\rho_L}{Dt} = \frac{\partial \rho_L}{\partial t} + \vec{v}_L \cdot \nabla \rho_L = 0$

- It is assumed that the density of the vapor phase, $\rho_V$, within $\Omega(t)$ is homogeneous, *i.e.*, $\nabla \rho_V|_{\Omega(t)} = \vec{0}$.
  - Consequently, we have: $\rho_V(x,t) = \rho_V(t) = \rho_e(t)$.
  - Although this may not seem realistic at first glance, it is actually a reasonable assumption. One of the reasons is that we treat the vapor-jet cavity as a vibrational source; thus, the spatial distribution of the vapor density within $\Omega(t)$ should not significantly affect the eventual emission outside the vapor-jet cavity opening, such as airborne acoustic emissions. Assuming vapor density homogeneity within $\Omega(t)$ also helps to significantly reduce the complexity of the modeling process (*e.g.*, in the subsequent mass conservation equation). Nevertheless, we must ensure that this assumption remains consistent with other components of the physical model.
  - This assumption, *per se*, does not imply the incompressibility of the vapor phase. In fact, we acknowledge and require that $\rho_V$ is a function of time, *i.e.*, $\rho_V = \rho_V(t)$, and that $\dot{\rho_V} = \frac{d\rho_V}{dt} \neq 0$.
    * We will discuss later, however, some specific modeling cases with the assumption of vapor incompressibility for the sake of simplification.

- It is assumed that all physical quantities of the vapor phase are homogeneous only if they are located at the vapor-jet cavity opening $\partial\Omega_e(t)$, *i.e.*,
  - Consequently, we have: $(\cdot)|_{\partial\Omega_e(t)}(x,t) = (\cdot)|_{\partial\Omega_e(t)}(t)$.
  - $\nabla(\cdot)|_{\partial\Omega_e(t)} = \vec{0}$ or $\mathbf{0}$, a.k.a., all intensive variables defined at the vapor-jet cavity opening are uniformly distributed.
  - Note: This does not imply that the quantities (except for $\rho_V$) of the vapor phase are homogeneous in the remainder of $\Omega(t)$. In fact, we explicitly allow that $\nabla(\cdot)|_{\Omega(t)\setminus\Omega_e(t)} \neq \vec{0}$ or $\mathbf{0}$, meaning that all intensive variables defined in $\Omega(t) \setminus \Omega_e(t)$ are still location-dependent.

- It is assumed that the vapor at $\partial\Omega_e(t)$ obeys the ideal gas law.
  - Therefore, $\rho_V(t)$ can be fully determined by $p_e$ and $T_e(t)$.
  - Since $\rho_V$ is homogeneous, we then have: $p_V(x,t) = \eta\rho_V(t)T_V(x,t)$, where $x$ denotes the spatial coordinate within $\Omega(t)$.
- It is assumed that $\vec{v}_e$ at $\partial\Omega_e(t)$ follows:

$$\vec{v}_e \cdot \vec{n}|_{\partial\Omega_e(t)} = \underbrace{\vec{v}_e \cdot \vec{e}_k = \|\vec{v}_e\|}_{\text{vertical ejection}} = \underbrace{v_e \geq 0}_{\text{upward}} \tag{1}$$

  where $\vec{e}_k = \vec{n}|_{\partial\Omega_e(t)}$.

- We assume that the mass rate of vaporization at $\partial\Omega_V(t)$, denoted by $\dot{m}$, depends on the input laser power $P$ and the laser absorptance $\varepsilon = 1 - \xi$, where $\varepsilon$ is further influenced by the geometric characteristics of the vapor-jet cavity $\Omega(t)$. This dependence can be expressed as

$$\dot{m} = \frac{bP}{L_V} \cdot \varepsilon = \frac{bP}{L_V}(1-\xi) \tag{2}$$

  where $b$ is a tunable proportionality coefficient and $L_V$ represents the latent heat of vaporization. This formulation captures the intuitive coupling between energy input, absorption efficiency, and mass removal rate during vapor-jet cavity evaporation, providing a physically interpretable basis for subsequent modeling.

Moreover, beyond the key assumptions discussed above, we introduce several additional assumptions that emphasize the negligibility of certain physical quantities under specific conditions. These assumptions serve to simplify the subsequent derivations without compromising generality, particularly in regimes where the neglected terms contribute minimally to the system dynamics:

- Negligible magnitude of $\vec{v}_L$, *i.e.*, $|\vec{v}_L|$, compared with that of $\vec{v}_V$, *i.e.*, $|\vec{v}_V|$.
- Negligible body-force acceleration $\vec{b}$ within the control volume $\Omega(t)$.
- Negligible heat conduction $\vec{q}(t)$ across $\partial\Omega_V(t)$, rendering the vapor flow within the vapor-jet cavity approximately isentropic.

In summary, the above physical picture can be interpreted as follows: assuming the vapor-jet cavity oscillates with a recoil force (due to the vaporization of the alloy) continuously driving it downward (*i.e.*, deeper) and compete against the surface tension, we apply conservation of mass, conservation of momentum, and state equations if necessary, to relate $p_e$, $\rho_e$, $\vec{v}_e$, and $T_e$ to the dynamic vapor-jet cavity geometric features. The above assumptions, however, are *not* all the assumptions we will use in the subsequent derivations—some additional assumptions will be introduced as we progress through the derivation.

### 1.1.4 Mechanics of vapor-jet cavity oscillation system

We formulate the dynamics of the vapor-jet cavity within the evolving control volume $\Omega(t)$ (illustrated in Fig. S1) under the framework of continuum mechanics. To establish the governing equations that describe vapor-jet cavity oscillation, we begin by invoking the fundamental physical conservation laws.

**The conservation of mass.** First, we consider the conservation of mass in its integral form, derived via the Reynolds Transport Theorem (RTT), which provides a rigorous basis for analyzing the change of mass and mass fluxes across the deforming boundaries of $\Omega(t)$:

$$\frac{d}{dt}\int_{\Omega(t)} \rho_V d\Omega = \underbrace{\int_{\Omega(t)} \frac{\partial \rho_V}{\partial t} d\Omega + \int_{\partial\Omega(t)} \rho_V \vec{v}_b \cdot \vec{n} dS = -\int_{\partial\Omega(t)} \rho_V \left(\vec{v}_V - \vec{v}_b\right) \cdot \vec{n} dS}_{\text{conservation of mass}} \tag{3}$$

It is important to emphasize that an explicit mass source term is not required in this formulation. The generation of metal vapor at $\partial\Omega_V(t)$ is inherently captured by the mass flux term on the right-hand side (R.H.S.) of Eq. (3), namely, $-\int_{\partial\Omega(t)} \rho_V \left(\vec{v}_V - \vec{v}_b\right) \cdot \vec{n} dS$. This term accounts for the relative motion of the vapor phase with respect to the control volume boundary and fully encompasses the effect of vaporization-driven mass outflow from the vapor-jet cavity.

By transferring the mass flux term from the R.H.S. to the left-hand side (L.H.S.) of Eq. (3), we isolate the material derivative of the total mass within the control volume $\Omega(t)$. This yields a compact expression that directly quantifies the rate of change of mass enclosed by $\Omega(t)$:

$$\begin{aligned}\int_{\Omega(t)} \frac{\partial \rho_V}{\partial t} d\Omega + \int_{\partial\Omega(t)} \rho_V \vec{v}_b \cdot \vec{n} dS + \int_{\partial\Omega(t)} \rho_V \left(\vec{v}_V - \vec{v}_b\right) \cdot \vec{n} dS \\ = \underbrace{\int_{\Omega(t)} \frac{\partial \rho_V}{\partial t} d\Omega + \int_{\partial\Omega(t)} \rho_V \vec{v}_V \cdot \vec{n} dS}_{=\int_{\Omega(t)}\left[\frac{\partial \rho_V}{\partial t} + \nabla\cdot(\rho_V \vec{v}_V)\right] d\Omega} = \underbrace{0}_{\text{no source}}\end{aligned} \tag{4}$$

Evidently, Eq. (4) represents the classical statement of mass conservation within $\Omega(t)$ in the absence of any internal mass sources or sinks:

$$\underbrace{\int_{\Omega(t)} \left[\frac{\partial \rho_V}{\partial t} + \nabla \cdot (\rho_V \vec{v}_V)\right] d\Omega}_{\text{material derivative of the mass}} = \frac{D}{Dt} \int_{\Omega(t)} \rho_V d\Omega = \underbrace{0}_{\text{no source}} \tag{5}$$

By decomposing the boundary of the vapor-jet cavity domain as $\partial\Omega(t) = \partial\Omega_e(t) \cup \partial\Omega_V(t)$ and appropriately reorganizing the terms in Eq. (5), we arrive at the following form, which separates the contributions from $\partial\Omega_V(t)$ and $\partial\Omega_e(t)$:

$$\int_{\Omega(t)} \frac{\partial \rho_V}{\partial t} d\Omega + \int_{\partial\Omega_V(t)} \rho_V \vec{v}_V \cdot \vec{n} dS + \int_{\partial\Omega_e(t)} \rho_V \vec{v}_V \cdot \vec{n} dS = 0 \tag{6}$$

Invoking the assumption of spatial homogeneity in vapor density $\rho_V(t)$ within the vapor-jet cavity domain $\Omega(t)$, along with the uniformity of relevant vapor state variables at the vapor-jet cavity opening $\partial\Omega_e(t)$, Eq. (6) admits a simplified form, expressed as:

$$\begin{aligned}\text{L.H.S. (6)} &= \frac{\partial \rho_V}{\partial t} \int_{\Omega(t)} d\Omega + \rho_V \int_{\partial\Omega_V(t)} \vec{v}_V \cdot \vec{n} dS + \rho_e \underbrace{\vec{v}_e \cdot \vec{e}_k}_{\text{Eq. (1)}} \int_{\partial\Omega_e(t)} dS \\ &= \frac{\partial \rho_V}{\partial t} \Omega + \rho_V \int_{\partial\Omega_V(t)} \vec{v}_V \cdot \vec{n} dS + \rho_e v_e \phi_e = 0 \\ &= \text{R.H.S. (6)}\end{aligned} \tag{7}$$

Under the assumption of vapor density homogeneity, the vapor density at $\partial\Omega_e(t)$ satisfies $\rho_e(t) = \rho_V(t)$. With this simplification, only one non-trivial term remains in Eq. (7)—namely, the surface integral $\rho_V \int_{\partial\Omega_V(t)} \vec{v}_V \cdot \vec{n} dS$, which encapsulates the net vapor outflow across $\partial\Omega_V(t)$ and requires further interpretation.

To further elucidate the detailed behavior of vapor transport across $\partial\Omega_V(t)$, we consider a localized material subdomain $\Pi(t)$ that straddles $\partial\Sigma(t)$, an infinitesimal surface element of $\partial\Omega_V(t)$. By applying RTT to the mass contained within this differential control volume, we obtain the localized integral form of the mass conservation law in $\Pi(t)$, which captures the instantaneous exchange of mass across $\partial\Sigma(t)$:

$$
\begin{aligned}
\frac{d}{dt}\int_{\Pi(t)} \rho d\Pi &= \underbrace{\frac{d}{dt}\int_{\Pi^+(t)} \rho^+ d\Pi + \frac{d}{dt}\int_{\Pi^-(t)} \rho^- d\Pi}_{\Pi(t)=\Pi^+(t)\cup\Pi^-(t)} \\
&= \underbrace{\left(\int_{\Pi^+(t)} \frac{\partial\rho^+}{\partial t} d\Pi + \int_{\partial\Pi^+(t)} \rho^+ \vec{v}^+ \cdot \vec{m}^+ dS + \int_{\partial\Sigma^+(t)} \rho^+ \vec{v}_b \cdot (-\vec{n})\, dS\right)}_{=\frac{d}{dt}\int_{\Pi^+(t)} \rho^+ d\Pi} + \\
&\quad \underbrace{\left(\int_{\Pi^-(t)} \frac{\partial\rho^-}{\partial t} d\Pi + \int_{\partial\Pi^-(t)} \rho^- \vec{v}^- \cdot \vec{m}^- dS + \int_{\partial\Sigma^-(t)} \rho^- \vec{v}_b \cdot \vec{n} dS\right)}_{=\frac{d}{dt}\int_{\Pi^-(t)} \rho^- d\Pi}
\end{aligned} \tag{8}
$$

Taking the limit of the L.H.S. of Eq. (8) as the material volume $\Pi(t) \to 0$, we further derive the mass conservation law at $\partial\Sigma(t)$ as:

$$
\begin{aligned}
\lim_{\Pi(t)\to 0}\left(\frac{d}{dt}\int_{\Pi(t)} \rho d\Pi\right) &= \lim_{\Pi(t)\to 0}\left(\frac{d}{dt}\int_{\Pi^+(t)} \rho^+ d\Pi + \frac{d}{dt}\int_{\Pi^-(t)} \rho^- d\Pi\right) \\
&= \lim_{\Pi^+(t)\to 0}\left[\underbrace{\int_{\Pi^+(t)} \frac{\partial\rho^+}{\partial t} d\Pi}_{\to 0} + \underbrace{\int_{\partial\Pi^+(t)} \rho^+ \vec{v}^+ \cdot \vec{m}^+ dS}_{\partial\Pi^+(t)\to\partial\Sigma(t)} - \int_{\partial\Sigma(t)} \rho^+ \vec{v}_b \cdot \vec{n} dS\right] + \\
&\quad \lim_{\Pi^-(t)\to 0}\left[\underbrace{\int_{\Pi^-(t)} \frac{\partial\rho^-}{\partial t} d\Pi}_{\to 0} + \underbrace{\int_{\partial\Pi^-(t)} \rho^- \vec{v}^- \cdot \vec{m}^- dS}_{\partial\Pi^-(t)\to\partial\Sigma(t)} + \int_{\partial\Sigma(t)} \rho^- \vec{v}_b \cdot \vec{n} dS\right] \\
&= \underbrace{0}_{\text{no source}}
\end{aligned} \tag{9}
$$

As $\Pi(t) \to 0$, we have $\vec{m}^+ \to \vec{n}$, $\vec{m}^- \to -\vec{n}$. We then merge all integral terms in Eq. (9) into a unified expression, yielding:

$$\int_{\partial\Sigma(t)} \left(\rho^+\vec{v}^+ - \rho^+\vec{v}_b - \rho^-\vec{v}^- + \rho^-\vec{v}_b\right) \cdot \vec{n} dS = 0 \tag{10}$$

Given that $\Pi(t)$, and consequently $\partial\Sigma(t)$, was chosen arbitrarily, the integrand in Eq. (10) must vanish at any point across the entire liquid–vapor interface $\partial\Omega_V(t)$. By replacing the "+" and "–" superscripts with the corresponding phase-specific labels for the alloy system, we then obtain:

$$(\rho_L\vec{v}_L - \rho_V\vec{v}_V - \rho_L\vec{v}_b + \rho_V\vec{v}_b) \cdot \vec{n} = [\![\rho\vec{v}]\!] \cdot \vec{n} - [\![\rho]\!]\, \vec{v}_b \cdot \vec{n} = 0 \tag{11}$$

$$[\![\rho\vec{v}]\!] \cdot \vec{n} = [\![\rho]\!]\, \vec{v}_b \cdot \vec{n} \tag{12}$$

Here, the operator $[\![\cdot]\!]$ in Eq. (12) denotes the jump across the interface at $\partial\Sigma(t)$, formally defined as:

$$[\![\cdot]\!] = \left[(\cdot)^+ - (\cdot)^-\right]\Big|_{\partial\Sigma(t)} = \left[(\cdot)_L - (\cdot)_V\right]\big|_{\partial\Sigma(t)} \tag{13}$$

The jump operator arises naturally from the presence of the Knudsen layer at the liquid–vapor interface $\partial\Omega_V(t)$. Within the framework of the continuum hypothesis—applied consistently throughout this derivation—this layer manifests as a singular boundary across which various physical quantities experience discontinuities, effectively represented as step functions. By further manipulating Eq. (11), we obtain the following refined expression that holds at $\partial\Sigma(t)$:

$$\rho_L\left(\vec{v}_L - \vec{v}_b\right) \cdot \vec{n} = \rho_V\left(\vec{v}_V - \vec{v}_b\right) \cdot \vec{n} \tag{14}$$

Equation (14) represents the precise boundary condition for mass conservation in the presence of discontinuities at the infinitesimal singular interface $\partial\Sigma(t)$. Intuitively, this condition embodies the physical principle that the mass flux entering or leaving the liquid phase within $\Pi^+(t)$ is exactly balanced by the mass flux leaving or entering the vapor phase within $\Pi^-(t)$ across the interface, thereby ensuring the continuity of mass.

By integrating both sides of Eq. (14) over the entire liquid–vapor interface $\partial\Omega_V(t)$, and recognizing that the net mass flux entering or leaving the liquid phase at time $t$ corresponds directly to the negative of the vaporization rate of the alloy's liquid phase plus porosity removal rate from the alloy's vapor phase. Denoting the total mass flux as $\dot{m}$, we obtain:

$$\rho_L \int_{\partial\Omega_V(t)} \left(\vec{v}_L - \vec{v}_b\right) \cdot \vec{n} dS = \rho_V \int_{\partial\Omega_V(t)} \left(\vec{v}_V - \vec{v}_b\right) \cdot \vec{n} dS = -\dot{m} \tag{15}$$

Introducing $\dot{m}_s$—the local mass flux (*i.e.*, the vaporization and porosity-generation rate per unit area) on $\partial\Omega_V(t)$, we can express Eq. (15) in the following refined form:

$$\begin{aligned}\dot{m} = \int_{\partial\Omega_V(t)} \dot{m}_s dS = \overbrace{-\rho_L \int_{\partial\Omega_V(t)} \left(\vec{v}_L - \vec{v}_b\right)\cdot\vec{n}dS = -\rho_V \int_{\partial\Omega_V(t)} \left(\vec{v}_V - \vec{v}_b\right)\cdot\vec{n}dS}^{\text{Eq. (15)}} \\ = \rho_V \left(\int_{\partial\Omega_V(t)} \vec{v}_b \cdot \vec{n}dS - \int_{\partial\Omega_V(t)} \vec{v}_V \cdot \vec{n}dS\right)\end{aligned} \tag{16}$$

$$\rho_V \int_{\partial\Omega_V(t)} \vec{v}_V \cdot \vec{n}dS = \rho_V \int_{\partial\Omega_V(t)} \vec{v}_b \cdot \vec{n}dS - \int_{\partial\Omega_V(t)} \dot{m}_s dS = \rho_V \int_{\partial\Omega_V(t)} \vec{v}_b \cdot \vec{n}dS - \dot{m} \tag{17}$$

**Corollary 1.1** (from Eq. (16) and Eq. (17))**.** *We can re-write Eq.* (14) *as:*

$$\dot{m}_s = -\rho_L \left(\vec{v}_L - \vec{v}_b\right)\cdot\vec{n} = -\rho_V \left(\vec{v}_V - \vec{v}_b\right)\cdot\vec{n} \tag{18}$$

By substituting Eq. (17) into the general mass balance expression given in Eq. (6), we obtain the following relation, which incorporates the localized vaporization flux into the global mass conservation framework:

$$\frac{\partial\rho_V}{\partial t}\Omega + \rho_V \int_{\partial\Omega_V(t)} \vec{v}_b \cdot \vec{n}dS - \dot{m} + \rho_e v_e \phi_e = 0 \tag{19}$$

$$\dot{m} = \frac{\partial\rho_V}{\partial t}\Omega + \rho_V \int_{\partial\Omega_V(t)} \vec{v}_b \cdot \vec{n}dS + \rho_e v_e \phi_e \tag{20}$$

Notice that since $\vec{v}_b|_{\partial\Omega_e(t)} = \vec{0}$, we have the following:

$$\begin{aligned}\int_{\partial\Omega_V(t)} \vec{v}_b \cdot \vec{n}dS &= \int_{\partial\Omega_V(t)} \vec{v}_b \cdot \vec{n}dS + \underbrace{\int_{\partial\Omega_e(t)} \vec{v}_b \cdot \vec{n}dS}_{=0} \\ &= \underbrace{\int_{\partial\Omega(t)} \vec{v}_b \cdot \vec{n}dS = \overbrace{\int_{\Omega(t)} \nabla\cdot\vec{v}_b d\Omega}^{\text{Re-applying RTT}}}_{\text{Gauss's divergence theorem}} \overbrace{= \frac{d}{dt}\int_{\Omega(t)} d\Omega}^{} \\ &= \frac{\partial\Omega}{\partial t}\end{aligned} \tag{21}$$

By combining Eq. (20) with Eq. (21), we arrive at a compact and analytically simplified expression that encapsulates the essential form of mass conservation for the system:

$$\boxed{\dot{m} = \int_{\partial\Omega_V(t)} \dot{m}_s dS = \frac{\partial\rho_V}{\partial t}\Omega + \rho_V\frac{\partial\Omega}{\partial t} + \rho_e v_e \phi_e = \frac{d}{dt}\left(\rho_V\Omega\right) + \rho_V v_e \phi_e} \tag{22}$$

Equation (22) formalizes the implications of mass conservation within the evolving vapor-jet cavity domain $\Omega(t)$, under the set of assumptions and boundary conditions previously established. If the vapor-jet cavity is totally porosity-free, we can subsequently extend this formulation by substituting Eq. (2) into Eq. (22) to derive a new governing equation that jointly enforces the conservation of mass and energy within $\Omega(t)$ under pure vaporization:

$$\boxed{\frac{d}{dt}\left(\rho_V \Omega\right) + \rho_V v_e \phi_e = \frac{bP}{L_V}\left(1-\xi\right)} \tag{23}$$

where $\xi$ denotes the total laser reflection ratio. In addition, we once again invoke the assumption of spatial homogeneity in vapor density throughout the domain $\Omega(t)$, allowing us to replace $\rho_e$ with $\rho_V$ in Eq. (23) for analytical consistency.

**The conservation of momentum.** Departing from the mass-conservation framework, we now turn to the formulation of momentum conservation within $\Omega(t)$. Before presenting the formal derivation, it is helpful to outline the governing high-level picture: $\Omega(t)$ is a finite, time-evolving vapor cavity whose momentum responds to a balance between driving influences and restoring/resistive influences, with the outcome expressed through the cavity's evolving geometry and its oscillatory motion. Because the control volume is deformable, changes in key geometric descriptors (*e.g.*, depth, effective cross-section, and interface shape) do not merely result from the dynamics—they also enter the momentum balance by reshaping the system's effective inertia and the way external interactions act on the cavity. Establishing this qualitative context is essential for interpreting the momentum equation that follows, and for understanding how the derived momentum balance connects to vapor-jet cavity geometry evolution and oscillation dynamics.

We believe that the dominant driving force responsible for the downward penetration of the vapor-jet cavity into the powder bed or build plate is the recoil pressure force, $\vec{F}_{\text{recoil}}$, which arises from the intense metal vaporization occurring at $\partial\Omega_V(t)$. This localized vapor expulsion generates a force acting normal to the liquid–vapor interface, effectively pushing the vapor-jet cavity deeper into the substrate. In response, a restoring force emerges, oriented generally opposite to $\vec{F}_{\text{recoil}}$. This restorative action is primarily provided by two components: the hydrostatic pressure force $\left(\int_{\partial\Omega_V(t)} \sigma_{\mathbf{L}} \cdot \vec{n} dS\right)$ and the resultant surface tension force, $\vec{F}_{\text{surface}}$. Notably, owing to the microscale geometry of the vapor-jet cavity, the hydrostatic contribution is orders of magnitude smaller than that of surface tension, making it negligible in comparison. This force imbalance suggests that the dynamic interplay between $\vec{F}_{\text{recoil}}$ and $\vec{F}_{\text{surface}}$ should govern the oscillatory behavior of the vapor-jet cavity, highlighting a fundamental mechanism underlying its unsteady evolution.

From an alternative mechanistic standpoint, the entire vapor-jet cavity—together with the enclosed vapor phase bounded by $\partial\Omega_e(t)$ and $\partial\Omega_V^-(t)$—can be effectively conceptualized as a rocket jet engine. This analogy stems from the recognition that the vapor cavity acts as a transient propulsion chamber, expelling mass through a narrow, nozzle-like exit and thereby generating reactive thrust. In this context, the vapor-jet cavity functions as a self-driven downward advancing 'rocket', initiated from the top surface of the melt pool and penetrating into the underlying substrate, with the reactive thrust transmitted through $\partial\Omega_e(t)$.

To sustain this thrust, a continuous internal source of force is required—physically provided by $\vec{F}_{\text{recoil}}$, the recoil force induced by vigorous metal vaporization along $\partial\Omega_V(t)$. $\vec{F}_{\text{recoil}}$ in general propels the vapor upward at high velocity ($\vec{v}_e$) through $\partial\Omega_e(t)$, establishing a momentum flux that drives the thrust in a manner directly analogous to classical rocket propulsion. Since it is assumed that $\vec{v}_e$ is oriented along the vertical direction, *i.e.*, $\left(\vec{v}_e \cdot \vec{n}|_{\partial\Omega_e(t)} = \|\vec{v}_e\| = v_e > 0\right)$, the corresponding thrust force, $\vec{F}_{\text{jet}}$, must act vertically downward, with its magnitude $\left\|\vec{F}_{\text{jet}}\right\|$ governed by the following relation:

$$\left\|\vec{F}_{\text{jet}}\right\| = (p_e - p_\infty)\,\phi_e + v_e \cdot \rho_e v_e \phi_e = \left[(p_e - p_\infty) + \rho_e v_e^2\right]\phi_e \tag{24}$$

where $p_e$ denotes the static pressure of the ejected vapor at $\partial\Omega_e(t)$. Denoting the central vertical dimension of the vapor-jet cavity as the $z$-axis, we now propose the following hypothesis based on the mechanistic framework and physical interpretations described above:

**Proposition 1.1.** *For an oscillating vapor-jet cavity with negligible (but still existing) vapor momentum change inside the cavity,* $\left\|\vec{F}_{\text{jet}}\right\|$ *should approximately equal to the magnitude of the total upward effective recoil force,* $F_{\text{recoil}}^{(z)}$*, namely:*

$$\left\|\vec{F}_{\text{jet}}\right\| \approx F_{\text{recoil}}^{(z)} = \left[(p_e - p_\infty) + \rho_e v_e^2\right]\phi_e \tag{25}$$

We call the above analysis and Prop. 1.1 the *rocket thrust analog*. Hypothesizing Prop. 1.1, we acknowledge that the recoil force imposed on the vapor inside the vapor-jet cavity is dominant compared to other forces. It is also important to emphasize that the total magnitude of the recoil force, $\left\|\vec{F}_{\text{recoil}}\right\|$, is not in general equal to $F^{(z)}_{\text{recoil}}$, since the vapor-jet cavity system is not assumed to be axisymmetric about the $z$-axis at this point. To further simplify Eq. (25), we introduce an effective vertical recoil pressure, $p^{(z)}_{\text{recoil}}$, which acts on $\partial\Omega_e(t)$ and encapsulates the net vertical momentum flux. With this definition, the governing equation becomes the following:

$$F^{(z)}_{\text{recoil}} = p^{(z)}_{\text{recoil}}\phi_e = \left[(p_e - p_\infty) + \rho_e v_e^2\right]\phi_e \tag{26}$$

$$p^{(z)}_{\text{recoil}} = (p_e - p_\infty) + \rho_V v_e^2 \tag{27}$$

Note that we replace $\rho_e$ with $\rho_V$ in Eq. (27). Equations (26) and (27) link $F^{(z)}_{\text{recoil}}$ and $p^{(z)}_{\text{recoil}}$ to specific state variables and $v_e$ at $\partial\Omega_e(t)$. The interpretation implied in Eq. (27) will be discussed later in the modeling of the mechanisms of airborne acoustic emissions.

Next, in direct analogy to the procedure previously employed to derive the mass conservation equation within $\Omega(t)$, we now formulate the governing equation for momentum conservation over the same control volume:

$$\begin{aligned}\frac{d}{dt}\int_{\Omega(t)}\rho_V\vec{v}_V d\Omega &= \overbrace{\int_{\Omega(t)}\frac{\partial\rho_V\vec{v}_V}{\partial t}d\Omega + \int_{\partial\Omega(t)}\rho_V\vec{v}_V\otimes\vec{v}_b\cdot\vec{n}dS}^{\text{RTT for }\rho_V\vec{v}_V} \\ &= \underbrace{-\int_{\partial\Omega(t)}\rho_V\vec{v}_V\otimes(\vec{v}_V-\vec{v}_b)\cdot\vec{n}dS + \int_{\partial\Omega(t)}\sigma_{\mathbf{V}}\cdot\vec{n}dS}_{\text{conservation of momentum}}\end{aligned} \tag{28}$$

It's important to emphasize that we neglect the influence of body forces (*e.g.*, gravity or buoyancy) acting on the vapor within the vapor-jet cavity. Thus, we have the following equation for the vapor stress tensor:

$$\int_{\partial\Omega(t)}\sigma_{\mathbf{V}}\cdot\vec{n}dS = \int_{\partial\Omega(t)}(-p_V\mathbf{I}+\tau_{\mathbf{V}})\cdot\vec{n}dS \tag{29}$$

Combining Eq. (28) with Eq. (29), and recognizing vapor density homogeneity and the uniformity of the variables at $\partial\Omega_e(t)$, we get:

$$\begin{aligned}\frac{d}{dt}\left(\rho_V\int_{\Omega(t)}\vec{v}_V d\Omega\right) &= -\int_{\partial\Omega(t)}\rho_V\vec{v}_V\otimes(\vec{v}_V-\vec{v}_b)\cdot\vec{n}dS + \int_{\partial\Omega(t)}\sigma_{\mathbf{V}}\cdot\vec{n}dS \\ &= -\rho_V\int_{\partial\Omega_V(t)}\vec{v}_V\otimes(\vec{v}_V-\vec{v}_b)\cdot\vec{n}dS + \int_{\partial\Omega(t)}(-p_V\mathbf{I}+\tau_{\mathbf{V}})\cdot\vec{n}dS - \left(p_e+\rho_V v_e^2\right)\phi_e\vec{e}_k\end{aligned} \tag{30}$$

Analogously, by adopting a strategy similar to that used in deriving the mass conservation boundary condition at $\partial\Omega_V(t)$, we now formulate the corresponding momentum balance across this discontinuous interface. Beginning with the application of RTT to the momentum of an infinitesimal material volume $\Pi(t)$ straddling the boundary, we arrive at the following expression:

$$
\begin{aligned}
\frac{d}{dt}\int_{\Pi(t)} \rho\vec{v}\,d\Pi &= \underbrace{\frac{d}{dt}\int_{\Pi^+(t)} \rho^+\vec{v}^+ d\Pi + \frac{d}{dt}\int_{\Pi^-(t)} \rho^-\vec{v}^- d\Pi}_{\Pi(t)=\Pi^+(t)\cup\Pi^-(t)} \\
&= \underbrace{\left(\int_{\Pi^+(t)} \frac{\partial\rho^+\vec{v}^+}{\partial t} d\Pi + \int_{\partial\Pi^+(t)} \rho^+\vec{v}^+\otimes\vec{v}^+\cdot\vec{m}^+ dS + \int_{\partial\Sigma^+(t)} \rho^+\vec{v}^+\otimes\vec{v}_b\cdot(-\vec{n})\,dS\right)}_{=\frac{d}{dt}\int_{\Pi^+(t)}\rho^+\vec{v}^+ d\Pi} + \\
&\quad \underbrace{\left(\int_{\Pi^-(t)} \frac{\partial\rho^-\vec{v}^-}{\partial t} d\Pi + \int_{\partial\Pi^-(t)} \rho^-\vec{v}^-\otimes\vec{v}^-\cdot\vec{m}^- dS + \int_{\partial\Sigma^-(t)} \rho^-\vec{v}^-\otimes\vec{v}_b\cdot\vec{n}dS\right)}_{=\frac{d}{dt}\int_{\Pi^-(t)}\rho^-\vec{v}^- d\Pi}
\end{aligned}
\tag{31}
$$

By taking the limit of the L.H.S. of Eq. (31), we get:

$$
\begin{aligned}
\lim_{\Pi(t)\to 0}\left(\frac{d}{dt}\int_{\Pi(t)} \rho\vec{v}\,d\Pi\right) &= \lim_{\Pi(t)\to 0}\left(\frac{d}{dt}\int_{\Pi^+(t)} \rho^+\vec{v}^+ d\Pi + \frac{d}{dt}\int_{\Pi^-(t)} \rho^-\vec{v}^- d\Pi\right) \\
&= \lim_{\Pi^+(t)\to 0}\left[\underbrace{\int_{\Pi^+(t)} \frac{\partial\rho^+\vec{v}^+}{\partial t} d\Pi}_{\to 0} + \underbrace{\int_{\partial\Pi^+(t)} \rho^+\vec{v}^+\otimes\vec{v}^+\cdot\vec{m}^+ dS}_{\partial\Pi^+(t)\to\partial\Sigma(t)} - \int_{\partial\Sigma(t)} \rho^+\vec{v}^+\otimes\vec{v}_b\cdot\vec{n}dS\right] + \\
&\quad \lim_{\Pi^-(t)\to 0}\left[\underbrace{\int_{\Pi^-(t)} \frac{\partial\rho^-\vec{v}^-}{\partial t} d\Pi}_{\to 0} + \underbrace{\int_{\partial\Pi^-(t)} \rho^-\vec{v}^-\otimes\vec{v}^-\cdot\vec{m}^- dS}_{\partial\Pi^-(t)\to\partial\Sigma(t)} + \int_{\partial\Sigma(t)} \rho^-\vec{v}^-\otimes\vec{v}_b\cdot\vec{n}dS\right] \\
&= \underbrace{\int_{\partial\Pi^+(t)} \sigma^+\cdot\vec{m}^+ dS}_{\partial\Pi^+(t)\to\partial\Sigma(t)} + \underbrace{\int_{\partial\Pi^-(t)} \sigma^-\cdot(-\vec{n})\,dS}_{\partial\Pi^-(t)\to\partial\Sigma(t)} + \int_{\partial\Sigma(t)} \vec{f}dS
\end{aligned}
\tag{32}
$$

where $\vec{f}$ represents the external surface force density acting on the interface $\partial\Sigma$. As $\Pi(t)$ collapses onto the singular surface (*i.e.*, $\Pi(t)\to 0$), the unit outward normals converge such that $\vec{m}^+\to\vec{n}$ and $\vec{m}^-\to -\vec{n}$. By consolidating all integral contributions in Eq. (32), we arrive at the following

expression:

$$\int_{\partial\Sigma(t)} \left(\rho^{+}\vec{v}^{+} \otimes \vec{v}^{+} - \rho^{+}\vec{v}^{+} \otimes \vec{v}_b\right) \cdot \vec{n} dS - \int_{\partial\Sigma(t)} \left(\rho^{-}\vec{v}^{-} \otimes \vec{v}^{-} - \rho^{-}\vec{v}^{-} \otimes \vec{v}_b\right) \cdot \vec{n} dS = \int_{\partial\Sigma(t)} \left[\left(\sigma^{+} - \sigma^{-}\right) \cdot \vec{n} + \vec{f}\right] dS \tag{33}$$

Given that the choice of $\Pi(t)$—and consequently the location of the discontinuity surface $\partial\Sigma(t)$—is arbitrary, the resulting condition must hold at every point across $\partial\Omega_V(t)$. Accordingly, we replace the generic "+" and "−" superscripts with the specific phase labels corresponding to the vapor and liquid states of the alloy and get:

$$\rho_L \vec{v}_L \otimes (\vec{v}_L - \vec{v}_b) \cdot \vec{n} - \rho_V \vec{v}_V \otimes (\vec{v}_V - \vec{v}_b) \cdot \vec{n} = (\sigma_{\mathbf{L}} - \sigma_{\mathbf{V}}) \cdot \vec{n} + \vec{f} \tag{34}$$

An alternative form of Eq. (34) is:

$$[\![\rho\vec{v} \otimes \vec{v}]\!] \cdot \vec{n} - [\![\rho\vec{v}]\!] \otimes \vec{v}_b \cdot \vec{n} = [\![\sigma]\!] \cdot \vec{n} + \vec{f} \tag{35}$$

Equation (34) (or equivalently, Eq. (35)) serves as the jump condition governing momentum conservation across a discontinuous interface. By integrating both sides of Eq. (34) over the entire liquid–vapor interface $\partial\Omega_V(t)$, we obtain:

$$\int_{\partial\Omega_V(t)} \rho_L \vec{v}_L \otimes (\vec{v}_L - \vec{v}_b) \cdot \vec{n} dS - \int_{\partial\Omega_V(t)} \rho_V \vec{v}_V \otimes (\vec{v}_V - \vec{v}_b) \cdot \vec{n} dS = \int_{\partial\Omega_V(t)} (\sigma_{\mathbf{L}} - \sigma_{\mathbf{V}}) \cdot \vec{n} dS + \int_{\partial\Omega_V(t)} \vec{f} dS \tag{36}$$

From Eq. (18), we have the interfacial mass flux continuity condition on $\partial\Omega_V(t)$, namely $\dot{m}_s = -\rho_L (\vec{v}_L - \vec{v}_b) \cdot \vec{n} = -\rho_V (\vec{v}_V - \vec{v}_b) \cdot \vec{n}$. Therefore, we can recast Eq. (36) into the following equivalent form:

$$\begin{aligned} \text{L.H.S. (36)} &= \int_{\partial\Omega_V(t)} \left[\rho_L (\vec{v}_L - \vec{v}_b) \cdot \vec{n}\right] \vec{v}_L dS - \int_{\partial\Omega_V(t)} \left[\rho_V (\vec{v}_V - \vec{v}_b) \cdot \vec{n}\right] \vec{v}_V dS \\ &= \int_{\partial\Omega_V(t)} \dot{m}_s \vec{v}_V dS - \int_{\partial\Omega_V(t)} \dot{m}_s \vec{v}_L dS \\ &= \int_{\partial\Omega_V(t)} \dot{m}_s (\vec{v}_V - \vec{v}_L) dS = \text{R.H.S. (36)} \\ &= \int_{\partial\Omega_V(t)} \sigma_{\mathbf{L}} \cdot \vec{n} dS - \int_{\partial\Omega_V(t)} \sigma_{\mathbf{V}} \cdot \vec{n} dS + \int_{\partial\Omega_V(t)} \vec{f} dS \end{aligned} \tag{37}$$

Given the earlier assumption that $\|\vec{v}_L\| \ll \|\vec{v}_V\|$, Eq. (37) can be further simplified, yielding the following:

$$\begin{aligned}\int_{\partial\Omega_V(t)} \dot{m}_s \left(\vec{v}_V - \vec{v}_L\right) dS \approx \int_{\partial\Omega_V(t)} \dot{m}_s \vec{v}_V dS &= -\int_{\partial\Omega_V(t)} \left[\rho_V \left(\vec{v}_V - \vec{v}_b\right) \cdot \vec{n}\right] \vec{v}_V dS \\ &= -\rho_V \int_{\partial\Omega_V(t)} \vec{v}_V \otimes \left(\vec{v}_V - \vec{v}_b\right) \cdot \vec{n} dS \\ &= \int_{\partial\Omega_V(t)} \sigma_{\mathbf{L}} \cdot \vec{n} dS - \int_{\partial\Omega_V(t)} \sigma_{\mathbf{V}} \cdot \vec{n} dS + \int_{\partial\Omega_V(t)} \vec{f} dS\end{aligned} \tag{38}$$

We now proceed to analyze the two surface integrals $\int_{\partial\Omega_V(t)} \vec{f} dS$ and $\int_{\partial\Omega_V(t)} \sigma_{\mathbf{L}} \cdot \vec{n} dS$ in Eq. (38). Guided by our physical interpretation of the vapor-jet cavity oscillation dynamics, we postulate the following hypothesis:

**Assumption 1.1.** *At $\partial\Omega_V(t)$, the dominant portion of $\int_{\partial\Omega_V(t)} \vec{f} dS$ comes from surface tension on $\partial\Omega_V(t)$, while the rest of the forces are trivial.*

It is worth mentioning that $\vec{F}_{\text{recoil}}$ should be understood as an effective inertial force arising from the net momentum flux across the $\partial\Omega_V(t)$. As such, its contribution is inherently embedded within the momentum conservation framework already established and does not require separate treatment.

Moreover, owing to the relatively small characteristic length scale of the vapor-jet cavity—on the order of a few hundred micrometers—we further hypothesize the following:

**Assumption 1.2.** *At $\partial\Omega_V(t)$, we have $\left\|\int_{\partial\Omega_V(t)} \sigma_{\mathbf{L}} \cdot \vec{n} dS\right\| \ll \left\|\int_{\partial\Omega_V(t)} \vec{f} dS\right\|$.*

As established in Assum. 1.2, the contribution of the hydrostatic pressure from the surrounding melt pool is considered negligible and can thus be omitted from the momentum balance. Incorporating this simplification and Eq. (29), along with the preceding approximations, we arrive at the following expression:

$$-\int_{\partial\Omega_V(t)} \rho_V \vec{v}_V \otimes \left(\vec{v}_V - \vec{v}_b\right) \cdot \vec{n} dS = -\int_{\partial\Omega_V(t)} \sigma_{\mathbf{V}} \cdot \vec{n} dS + \int_{\partial\Omega_V(t)} \vec{f} dS \tag{39}$$

Substituting Eq. (39) into Eq. (30), we obtain:

$$\begin{aligned}
\left(p_e+\rho_V v_e^2\right)\phi_e\vec{e}_k+\frac{d}{dt}\left(\rho_V\int_{\Omega(t)}\vec{v}_V d\Omega\right) &= -\int_{\partial\Omega_V(t)}\rho_V\vec{v}_V\otimes(\vec{v}_V-\vec{v}_b)\cdot\vec{n}dS+\int_{\partial\Omega_V(t)}\sigma_{\mathbf{V}}\cdot\vec{n}dS \\
&= -\int_{\partial\Omega_V(t)}\sigma_{\mathbf{V}}\cdot\vec{n}dS+\int_{\partial\Omega_V(t)}\vec{f}dS+\int_{\partial\Omega_V(t)}\sigma_{\mathbf{V}}\cdot\vec{n}dS \\
&= \int_{\partial\Omega_V(t)}\vec{f}dS
\end{aligned} \tag{40}$$

$$\boxed{\left(p_e+\rho_V v_e^2\right)\phi_e\vec{e}_k+\frac{d}{dt}\left(\rho_V\int_{\Omega(t)}\vec{v}_V d\Omega\right)=\vec{F}_{\text{surface}}} \tag{41}$$

It is worth noting that the term $\int_{\partial\Omega_V(t)}\sigma_{\mathbf{V}}\cdot\vec{n}dS$, including both pressure and viscous stress tensors, is canceled during substitution. To sum up the above derivation, we get two fundamental equations, Eq. (22) and Eq. (41), for mass and momentum conservation of a vapor-jet cavity $\Omega(t)$, respectively, based on our prescribed assumptions and hypotheses:

$$\boxed{\begin{cases}\dot{m}=\dfrac{\partial\rho_V}{\partial t}\Omega+\rho_V\dfrac{\partial\Omega}{\partial t}+\rho_e v_e\phi_e=\dfrac{d}{dt}\left(\rho_V\Omega\right)+\rho_V v_e\phi_e \\ \left(p_e+\rho_V v_e^2\right)\phi_e\vec{e}_k+\dfrac{d}{dt}\left(\rho_V\displaystyle\int_{\Omega(t)}\vec{v}_V d\Omega\right)=\vec{F}_{\text{surface}}=\displaystyle\int_{\partial\Omega_V(t)}\vec{f}dS\end{cases}} \tag{42}$$

## 1.2 Vapor-jet–cavity oscillation dynamics equation (JCODE)

To further distill Eq. (42) into a more tractable and physically interpretable form, we introduce an additional simplifying assumption. While this comes at the cost of some quantitative accuracy, it enables a substantial reduction in model complexity, allowing us to retain the dominant physics governing the essential dynamics of the system.

The first additional hypothesis is shown as follows:

**Assumption 1.3.** *The vapor in $\Omega(t)$ is assumed to be incompressible, i.e.:*

$$\frac{D\rho_V}{Dt}=\frac{\partial\rho_V}{\partial t}+\vec{v}\cdot\nabla\rho_V=0 \tag{43}$$

Assumption 1.3, although physically invalid due to the inherently compressible nature of the vapor phase, has been widely adopted in vaporization simulations for computational simplification. In the context of our theoretical framework, this assumption proves similarly advantageous: it significantly simplifies the subsequent derivation by eliminating terms associated with minor vapor density fluctuations, which are presumed to have negligible influence on the vapor-jet cavity oscillation dynamics.

Second, we hypothesize the following:

**Assumption 1.4.** *Both the geometry and the field quantities of the vapor-jet cavity exhibit axisymmetry.*

This assumption enables a focused analysis of the dominant physics along the $z$-axis. Experimental observations consistently reveal that vapor-jet cavitys, particularly those that are deep and prone to defect formation, tend to adopt a nearly cylindrical shape oriented vertically.

---

**Discussion 1.1.** *With the introduction of Assum. 1.4, we are ready to discuss the physical interpretation of the recoil force, $\vec{F}_{\text{recoil}}$, in greater detail. It remains essential to elucidate the physical origin of $\vec{F}_{\text{recoil}}$ and its relationship to other boundary conditions at $\partial\Omega_V(t)$ and $\partial\Omega_e(t)$. Moreover, such clarification is critical for interpreting the previously hypothesized Prop. 1.1 and for informing the subsequent modeling of acoustic emissions.*

*It is widely known that there exists a special layer called the Knudsen layer, where the mass and momentum exchange takes place, between the continuum liquid and vapor phases during the vaporization process. Physically, the Knudsen layer spans only a few mean free paths and constitutes a non-equilibrium region, where $\vec{F}_{\text{recoil}}$ originates. While the continuum hypothesis is assumed throughout the derivation, the analysis of and within the Knudsen layer may break it down. As such, we fundamentally treat the entire Knudsen layer as a singular discontinuous boundary located at $\partial\Omega_V(t)$, where externally applied forces—such as surface tension and hydrostatic pressure—are accordingly exerted. Using this model, $\vec{F}_{\text{recoil}}$ must be interpreted as an inertial reaction force arising from the drastic mass and momentum exchange across $\partial\Omega_V(t)$, balanced by the external forces acting on the surrounding continua. We refer to this understanding as the conventional view of $\partial\Omega_V(t)$ and $\vec{F}_{\text{recoil}}$.*

*There exists, however, an alternative interpretation that introduces a new discontinuous boundary, $\partial\Omega_V^-(t)$, which lies within the vapor phase just a Knudsen-layer-thickness away from $\partial\Omega_V(t)$*

*and perfectly delineates the vapor-Knudsen interface. In this framework, the system comprises three distinct phases—namely liquid, Knudsen, and vapor—forming a sandwich-like structure. Unlike the conventional view, this interpretation permits us* to convert fictitious $\vec{F}_{\text{recoil}}$ into two separate applied forces/pressures acting on either $\partial\Omega_V(t)$ side or $\partial\Omega_V^-(t)$ side of the non-equilibrium Knudsen layer. *Consequently,* the necessity to model cross-boundary mass and momentum transfer is eliminated*, allowing the application of Cauchy's fundamental postulate and Newton's third law to formulate two distinct force-balance equations at the two discontinuities.*

**Proposition 1.2.** *Denoting $p_r$ and $p_r^-$ as the local recoil pressures at $\partial\Omega_V(t)$ and $\partial\Omega_V^-(t)$, respectively. Then we hypothesize:*

$$\vec{F}_{\text{recoil}} = \int_{\partial\Omega_V(t)} \left(p_r^- - p_r\right) \mathbf{I} \cdot \vec{n} dS \tag{44}$$

*On the liquid side of the Knudsen layer, i.e., $\partial\Omega_V(t)$, we hypothesize the following equation for the balance of forces:*

$$\int_{\partial\Omega_V(t)} (p_r \mathbf{I} + \sigma_{\mathbf{L}}) \cdot \vec{n} dS + \int_{\partial\Omega_V(t)} \vec{f} dS = \vec{0} \tag{45}$$

*On the vapor side of the Knudsen layer, i.e., $\partial\Omega_V^-(t)$, we hypothesize:*

$$\int_{\partial\Omega_V^-(t)} \left(p_r^- - p_V\right) \mathbf{I} \cdot \vec{n} dS = \vec{0} \tag{46}$$

*where we neglect the viscous stress in the vapor phase for the convenience of discussion.*

*Proposition 1.2, specifically Eq.* (45)*, states that the recoil pressure at $\partial\Omega_V(t)$ is balanced by the surface tension and hydrostatic pressure from the melt pool, while Eq.* (46) *indicates that the recoil pressure $p_r^-$ at $\partial\Omega_V^-(t)$ is equal to the corresponding local vapor pressure $p_V$.*

*Finally, combining the above hypothesized relations with Assum. 1.4, we have the following corollary:*

**Corollary 1.2** (from Prop. 1.1 and Assum. 1.4)**.** *Given Eq.* (29) *and Eq.* (46)*, we have the following at $\partial\Omega_V^-(t)$:*

$$F_{\text{recoil}}^{(z)} = \int_{\partial\Omega_V^-(t)} p_r^- \mathbf{I} \cdot \vec{n} dS = \left\| \int_{\partial\Omega_V(t)} \sigma_{\mathbf{V}} \cdot \vec{n} dS \right\| \approx \left\| - \int_{\partial\Omega_V(t)} p_V \mathbf{I} \cdot \vec{n} dS \right\| \tag{47}$$

*which, combined with Eq.* (25) *and Eq.* (26)*, becomes:*

$$F_{\text{recoil}}^{(z)} = p_{\text{recoil}}\phi_e = \left\| -\int_{\partial\Omega_V(t)} p_V \mathbf{I}\cdot\vec{n}dS \right\| = \left\| \vec{F}_{\text{jet}} \right\| = \left[ (p_e - p_\infty) + \rho_V v_e^2 \right] \phi_e \tag{48}$$

*Equation* (48) *provides a clear physical interpretation of the vapor pressure at* $\partial\Omega_V(t)$ *as the source of a vertical pressure gradient force plus the net ejection momentum flux across* $\partial\Omega_e(t)$*, a.k.a. the thrust force. This interpretation is consistent with our earlier hypothesis in Prop. 1.1 and aligns with the conventional understanding of recoil pressure in jer-driven cavitation dynamics. It also clarifies that the recoil pressure is not an independent external force but rather a manifestation of the internal pressure dynamics within the vapor-jet cavity, balanced by surface tension and hydrostatic pressures at the liquid–vapor interface.*

---

Given that the depth $L$ typically far exceeds its lateral cross-sectional diameter $R$ for a deep vapor-jet cavity, we speculate that variations along the radial direction are negligible in comparison. Based on this geometric and physical reasoning, we further introduce the following hypothesis:

**Assumption 1.5.** *For estimating the volume of the vapor-jet cavity, we use:*

$$\Omega = \int_{\Omega(t)} d\Omega = L\phi_e = L\pi R^2 = L\int_{\partial\Omega_e(t)} dS \tag{49}$$

It should be noted that Eq. (51) employs a cylindrical formulation solely to approximate the volume of the vapor-jet cavity $\Omega(t)$. While the actual vapor-jet cavity may indeed exhibit an axisymmetric geometry that closely resembles a vertical cylinder, it is not required to be a perfect one. In fact, we deliberately avoid imposing a perfectly cylindrical shape, particularly near the vapor-jet cavity opening $\partial\Omega_e(t)$, as doing so affords greater flexibility in modeling $F_{\text{surface}}^{(z)}$, which is sensitive to local curvature and interface topology.

**Corollary 1.3** (from Assum. 1.4)**.** *We have the following at* $\partial\Omega_V(t)$*:*

$$\vec{F}_{\text{surface}} = \int_{\partial\Omega_V(t)} \vec{f} dS = F_{\text{surface}}^{(z)} \vec{e}_k \tag{50}$$

Third, we turn our attention to simplifying the temporal derivative term $\frac{d}{dt}\left(\int_{\Omega(t)} \vec{v}_V d\Omega\right)$, which represents the rate of change of linear momentum within the vapor domain. To facilitate analytical tractability while retaining physical relevance, we introduce the following hypothesis:

**Assumption 1.6.** *For all the vapor-jet cavitys we study, the following approximation holds:*

$$\int_{\Omega(t)} \vec{v}_V d\Omega \approx \frac{\Omega}{\beta} v_e \vec{e}_k \tag{51}$$

*where $\beta$ is a dimensionless phenomenological constant.*

In other words, using $\langle\cdot\rangle$ to denote the volume-averaged quantity, Assum. 1.6 posits that the $z$-component of the average vapor velocity, $\left|\left\langle \vec{v}_V^{(z)} \right\rangle\right|$, can be effectively approximated by a scalar multiple of $|\vec{v}_e|$. This hypothesis is supported by numerical simulations, which demonstrate that the average vertical vapor velocity remains close to the vapor ejection velocity at $\partial\Omega_e(t)$. Consequently, the proportionality factor $\beta$ may be reasonably approximated as unity.

Fourth, motivated by empirical observations from prototypical experiments, we advance the following hypothesis:

**Assumption 1.7.** *With a laser with a fixed beam diameter, stable beam profile, and a moderately slow scanning speed, it is reasonable to assume that the temporal variation of the lateral vapor-jet cavity cross-sectional area is negligible, i.e., we set* $\dot{\phi}_e = \frac{d\phi_e}{dt} = 0$.

Although Assum. 1.7 represents a first-order approximation, it is supported by our experimental observations through a wide range of X-ray vapor-jet cavity imaging. Furthermore, several independent studies have reported that, under consistent conditions, the lateral cross-sectional area of the vapor-jet cavity remains approximately constant across both spatial and temporal domains. This quasi-stationary behavior lends further credence to the validity of our hypothesis.

Finally, we develop an approximate model for the total surface tension force $\left(\vec{F}_{\text{surface}}\right)$ acting on the vapor-jet cavity. A consistent experimental observation across deep and shallow vapor-jet cavitys is the presence of an inclination angle between the front vapor-jet cavity wall and the top surface of the melt pool. We denote this inclination as $\theta$, the characteristic angle between the vapor-jet cavity front wall and the horizontal surface, and postulate the following hypothesis to capture its role in shaping the resultant surface tension force:

**Assumption 1.8.** *By taking a cross-sectional rim near the vapor-jet cavity opening,* $F_{\text{surface}}^{(z)}$ *can be*

*formulated as:*

$$F^{(z)}_{\text{surface}} = \underbrace{C \cdot 2\pi R \cdot \gamma \cdot \frac{L}{\sqrt{L^2+R^2}} \approx 2\pi C \gamma \cos\theta \cdot L}_{\text{Assuming } \cos\theta \approx \frac{R}{\sqrt{L^2+R^2}}} \tag{52}$$

*where $\gamma$ is the surface tension coefficient for the liquid metal, $\theta$ is the vapor-jet cavity front wall characteristic angle, and C is a phenomenological constant.*

Summing up all the above new hypotheses and assumptions, we now further simplify Eq. (42) and obtain two new scalar equations:

$$\begin{cases} \dot{m} = \rho_V \dfrac{d\Omega}{dt} + \rho_e v_e \phi_e \\ \left(p_e + \rho_V v_e^2\right)\phi_e + \dfrac{\rho_V}{\beta}\left(\dfrac{dv_e}{dt}\Omega + \dfrac{d\Omega}{dt} v_e\right) = \dfrac{2\pi C \gamma R L}{\sqrt{L^2+R^2}} \end{cases} \tag{53}$$

Substituting Assum. 1.3 and Assum. 1.7 into Eq. (53), we have:

$$v_e = \frac{\dot{m} - \rho_V \frac{d\Omega}{dt}}{\rho_V \phi_e} \tag{54}$$

$$\frac{dv_e}{dt} = \frac{\frac{d\dot{m}}{dt} - \rho_V \frac{d^2\Omega}{dt^2}}{\rho_V \phi_e} \tag{55}$$

Substituting Eq. (54) and Eq. (55) into Eq. (53) and rearranging terms, we finally obtain:

$$\boxed{\rho_V \Omega \frac{\partial^2 L}{\partial t^2} + \rho_V \phi_e \left(\frac{\partial L}{\partial t}\right)^2 - \dot{m}\frac{\partial L}{\partial t} + \left(\frac{2\pi\beta C \gamma R}{\sqrt{L^2+R^2}} - \frac{d\dot{m}}{dt}\right) L = \beta\left(p_e + \rho_V v_e^2\right)\phi_e} \tag{56}$$

Or alternatively, using the dot notation,

$$\boxed{\rho_V \phi_e \left(L\ddot{L} + \dot{L}^2\right) - \dot{m}\dot{L} + \left(\alpha\gamma\cos\theta - \ddot{m}\right) L = \beta\left(p_e + \rho_V v_e^2\right)\phi_e = \beta\left(F^{(z)}_{\text{recoil}} + p_\infty \phi_e\right)} \tag{57}$$

$$\boxed{L\ddot{L} - v_e \dot{L} + \frac{\alpha\gamma\cos\theta - \ddot{m}}{\rho_V \phi_e} L = \beta\left(\frac{p_e}{\rho_V} + v_e^2\right)} \tag{58}$$

where $\alpha = 2\pi\beta C$. Equations (56) to (58) characterize the temporal evolution of vapor-jet cavity depth $L$ as a nonlinear damped oscillator.

Alternatively, a comparable equation governing vapor-jet cavity depth oscillations can be derived without invoking Assum. 1.6. This approach, however, requires introducing additional assumptions.

***Importantly, the resulting derivation also yields novel insights into the dynamical-system properties of an oscillating vapor-jet cavity.***

First, we assume that $v_V^{(z)}$—the $z$-component of $\vec{v}_V$—is a spatially smooth function of $z$ under the assumption of vapor-jet cavity axisymmetry. Assuming $z = 0$ and $z = L$ represent the vapor-jet cavity opening ($\partial\Omega_e(t)$) and the vapor-jet cavity bottom, respectively, we therefore have:

$$v_V^{(z)} = v_V^{(z)}(z), \; v_V^{(z)}\Big|_{z=0} = v_e \tag{59}$$

Denoting the average velocity as $\left\langle v_V^{(z)} \right\rangle$, we have the following:

$$\left\langle v_V^{(z)} \right\rangle = \frac{1}{L}\int_0^L v_V^{(z)} dz \tag{60}$$

Reformulating $v_V^{(z)}$ using Taylor expansion around $z = 0$, we have:

$$v_V^{(z)} = v_V^{(z)}\Big|_{z=0} + \left.\frac{\partial v_V^{(z)}}{\partial z}\right|_{z=0} z + \left.\frac{\partial^2 v_V^{(z)}}{\partial z^2}\right|_{z=0} \frac{z^2}{2} + O\left(z^3\right) \tag{61}$$

Substituting Eq. (61) into Eq. (60) and integrating, we obtain:

$$\begin{aligned}
\left\langle v_V^{(z)} \right\rangle &= \frac{1}{L}\int_0^L \left( v_V^{(z)}\Big|_{z=0} + \left.\frac{\partial v_V^{(z)}}{\partial z}\right|_{z=0} z + \left.\frac{\partial^2 v_V^{(z)}}{\partial z^2}\right|_{z=0} \frac{z^2}{2} + O\left(z^3\right)\right) dz \\
&= \frac{1}{L}\left[ v_e z + \frac{1}{2}\left.\frac{\partial v_V^{(z)}}{\partial z}\right|_{z=0} z^2 + \frac{1}{6}\left.\frac{\partial^2 v_V^{(z)}}{\partial z^2}\right|_{z=0} z^3 + \cdots \right]\Bigg|_0^L \\
&= v_e + \frac{\varphi_1}{2}L + \frac{\varphi_2}{6}L^2 + \cdots
\end{aligned} \tag{62}$$

where $\varphi_1 = \left.\frac{\partial v_V^{(z)}}{\partial z}\right|_{z=0}$ and $\varphi_2 = \left.\frac{\partial^2 v_V^{(z)}}{\partial z^2}\right|_{z=0}$. **Assuming** $\dot{\varphi}_1 = \dot{\varphi}_2 = 0$, we take the time derivative of both sides of Eq. (62) and obtain:

$$\frac{d\left\langle v_V^{(z)} \right\rangle}{dt} = \frac{dv_e}{dt} + \frac{\varphi_1}{2}\frac{dL}{dt} + \frac{\varphi_2}{3}L\frac{dL}{dt} + \cdots \Rightarrow \dot{\overline{\left\langle v_V^{(z)} \right\rangle}} = \dot{v}_e + \frac{\varphi_1}{2}\dot{L} + \frac{\varphi_2}{3}L\dot{L} + \cdots \tag{63}$$

Truncating the series in Eq. (63) after the first-order term, we can subsequently obtain the linear approximation of $\dot{\overline{\left\langle v_V^{(z)} \right\rangle}}$. Higher order terms can be included if necessary, at a cost of increased number of parameters and model complexity.

*The following derivation is inspired by and based on the discussion above. Assuming that all the remaining aforementioned assumptions (**except for** Assum. 1.6), propositions, and corollaries continue to be valid in the upcoming derivations*, we now propose the following:

**Assumption 1.9** (replacing Assum. 1.6)**.** *Recalling the axisymmetry of the vapor-jet cavity system assumed in Assum. 1.4, and denoting:*

$$\langle \vec{v}_V \rangle = \frac{1}{\Omega} \int_{\Omega(t)} \vec{v}_V d\Omega = \langle v_V \rangle \vec{e}_k \tag{64}$$

$$\langle v_V \rangle = \frac{1}{\Omega} \left( \int_{\Omega(t)} \vec{v}_V d\Omega \right) \cdot \vec{e}_k = \frac{1}{\Omega} \int_{\Omega(t)} \vec{v}_V \cdot \vec{e}_k d\Omega = \frac{1}{\Omega} \int_{\Omega(t)} v_V^{(z)} d\Omega \tag{65}$$

*where $\langle \cdot \rangle$ represents the averaging operator. Since the direction of $\langle \vec{v}_V \rangle$ aligns with z-axis (same as that of $\vec{v}_e$), we hypothesize the below relation between their magnitudes $v_e$ and $\langle v_V \rangle$:*

$$\frac{dv_e}{dt} = \frac{d \langle v_V \rangle}{dt} + \varphi \frac{\partial L}{\partial t} = \frac{d}{dt} \left( \frac{1}{\Omega} \int_{\Omega(t)} v_V^{(z)} d\Omega \right) + \varphi \frac{\partial L}{\partial t} \tag{66}$$

*where $\varphi$ is a real constant and possesses a dimension of $s^{-1}$.*

Assumption 1.9 can also be interpreted as the time derivative of the first-order Taylor approximation of a *linear* velocity profile with a *constant* velocity gradient inside a nozzle-like closed-end tube. In essence, $\varphi$ is a phenomenological constant that captures the influence of vapor velocity gradient ($\nabla \mathbf{v_V}$) along $z$-axis. Using the above hypothesis, we proceed to develop Eq. (42) further to extract information about the vapor-jet cavity depth oscillations. Specifically, by differentiating both sides of Eq. (22) with respect to time, we obtain a scalar equation for $\frac{dv_e}{dt}$:

$$\frac{d\dot{m}}{dt} = \rho_V \frac{d^2\Omega}{dt^2} + \rho_V \phi_e \frac{dv_e}{dt} \tag{67}$$

Replacing $\frac{dv_e}{dt}$ with Eq. (66) and rearranging the equation a little bit, we have:

$$
\begin{aligned}
\frac{d\dot{m}}{dt} &= \rho_V \frac{d^2\Omega}{dt^2} + \rho_V \phi_e \left[ \frac{d\langle v_V \rangle}{dt} + \varphi \frac{\partial L}{\partial t} \right] \\
&= \rho_V \frac{d^2\Omega}{dt^2} + \frac{\rho_V \phi_e}{\Omega} \left[ \Omega \frac{d\langle v_V \rangle}{dt} \right] + \varphi \rho_V \phi_e \frac{\partial L}{\partial t} \\
&= \rho_V \frac{d^2\Omega}{dt^2} + \frac{\rho_V \phi_e}{\Omega} \left[ \frac{d}{dt} \left( \Omega \langle v_V \rangle \right) - \frac{d\Omega}{dt} \langle v_V \rangle \right] + \varphi \rho_V \phi_e \frac{\partial L}{\partial t} \\
&= \rho_V \frac{d^2\Omega}{dt^2} + \frac{\rho_V \phi_e}{\Omega} \left[ \frac{d}{dt} \left( \Omega \cdot \frac{1}{\Omega} \int_{\Omega(t)} v_V^{(z)} d\Omega \right) - \frac{d\Omega}{dt} \cdot \frac{1}{\Omega} \int_{\Omega(t)} v_V^{(z)} d\Omega \right] + \varphi \rho_V \phi_e \frac{\partial L}{\partial t} \\
&= \rho_V \frac{d^2\Omega}{dt^2} + \frac{\rho_V \phi_e}{\Omega} \left[ \frac{d}{dt} \left( \int_{\Omega(t)} v_V^{(z)} d\Omega \right) - \frac{1}{\Omega} \frac{d\Omega}{dt} \int_{\Omega(t)} v_V^{(z)} d\Omega \right] + \varphi \rho_V \phi_e \frac{\partial L}{\partial t}
\end{aligned}
\tag{68}
$$

Denoting $A = \int_{\Omega(t)} v_V^{(z)} d\Omega = \Omega \langle v_V \rangle$, we then have:

$$
\frac{d\dot{m}}{dt} = \rho_V \frac{d^2\Omega}{dt^2} + \frac{\rho_V \phi_e}{\Omega} \left( \frac{dA}{dt} - \frac{A}{\Omega} \frac{d\Omega}{dt} \right) + \varphi \rho_V \phi_e \frac{\partial L}{\partial t} \tag{69}
$$

Combining Eq. (40), Eq. (64) and Assum. 1.8, we get another scalar equation:

$$
\left( p_e + \rho_V v_e^2 \right) \phi_e + \rho_V \frac{d}{dt} \left( \int_{\Omega(t)} v_V^{(z)} d\Omega \right) = \left( p_e + \rho_V v_e^2 \right) \phi_e + \rho_V \frac{dA}{dt} = F_{\text{surface}}^{(z)} \tag{70}
$$

$$
\frac{dA}{dt} = \frac{F_{\text{surface}}^{(z)} - \left( p_e + \rho_V v_e^2 \right) \phi_e}{\rho_V} \tag{71}
$$

Substituting Eq. (71) into Eq. (69), we obtain:

$$
\frac{d\dot{m}}{dt} = \rho_V \frac{d^2\Omega}{dt^2} + \frac{\rho_V \phi_e}{\Omega} \left[ \frac{F_{\text{surface}}^{(z)} - \left( p_e + \rho_V v_e^2 \right) \phi_e}{\rho_V} - \frac{A}{\Omega} \frac{d\Omega}{dt} \right] + \varphi \rho_V \phi_e \frac{\partial L}{\partial t} \tag{72}
$$

$$
\rho_V \Omega^2 \frac{d^2\Omega}{dt^2} - \rho_V \phi_e A \frac{d\Omega}{dt} - \frac{d\dot{m}}{dt} \Omega^2 + \varphi \rho_V \phi_e \Omega^2 \frac{\partial L}{\partial t} + \phi_e \left[ F_{\text{surface}}^{(z)} - \left( p_e + \rho_V v_e^2 \right) \phi_e \right] \Omega = 0 \tag{73}
$$

Using Assum. 1.5 ($\Omega = L\phi_e$) and Eq. (52), and dividing both sides of Eq. (73) by $L\phi_e^2$, we get:

$$
\rho_V \Omega \frac{\partial^2 L}{\partial t^2} + \underbrace{\rho_V \left( \varphi \Omega - \frac{A}{L} \right)}_{\text{damping}} \frac{\partial L}{\partial t} + \left[ \underbrace{2\pi C \gamma \cos\theta}_{\text{surface tension}} - \frac{d\dot{m}}{dt} \right] L = \left( \underbrace{p_e - p_\infty + \rho_V v_e^2}_{\text{recoil pressure}} + p_\infty \right) \phi_e \tag{74}
$$

Recognizing that $A = \Omega \langle v_V \rangle = \phi_e L \langle v_V \rangle$ and using the dot notation, we obtain the following:

$$\boxed{\begin{aligned}&\rho_V \Omega \ddot{L} + \rho_V \left(\varphi \Omega - \phi_e \langle v_V \rangle\right) \dot{L} + \left(2\pi C \gamma \cos\theta - \dot{m}\right) L \\ &\qquad = \left(p_e + \rho_V v_e^2\right) \phi_e = F_{\text{recoil}}^{(z)} + p_\infty \phi_e\end{aligned}} \tag{75}$$

Finally, dividing both sides of Eq. (75) by $\rho_V \phi_e$, we have:

$$\boxed{L \ddot{L} + \left(\varphi L - \langle v_V \rangle\right) \dot{L} + \frac{2\pi C \gamma \cos\theta - \dot{m}}{\rho_V \phi_e} L = \frac{p_e}{\rho_V} + v_e^2} \tag{76}$$

It is worth noting that we only use two phenomenological constants, namely $\varphi$ for velocity gradient and $C$ for surface tension, in Eq. (76). Compared with Eq. (58), Eq. (76) does not contain the additional parameter $\beta$ and has the damping coefficient reformulated as $\varphi L - \langle v_V \rangle$. Equations (75) and (76), though nonlinear, do not directly take the form of the *Rayleigh-Plesset equation*, yet the nature that $\varphi L - \langle v_V \rangle$ can be either positive or negative allows Eq. (76) to describe both ordinary damped and self-exciting oscillations, where a limit cycle can be reached before the vapor-jet cavity collapse due to its inherent instability. Finally, despite containing unsolvable $\langle v_V \rangle$, this equation still provides a useful framework for understanding the vapor-jet cavity depth dynamics, especially the natural frequency of the vapor-jet cavity depth oscillation.

If we further assume:

$$v_e = \langle v_V \rangle + \varphi L \tag{77}$$

which is a stronger assumption that can replace Assum. 1.9. Then, by substituting Eq. (77) and Eq. (54) into Eq. (75), we have:

$$\boxed{\rho_V \phi_e \left(L \ddot{L} + \dot{L}^2\right) + \left(2\varphi \rho_V \Omega - \dot{m}\right) \dot{L} + \left(2\pi C \gamma \cos\theta - \dot{m}\right) L = \left(p_e + \rho_V v_e^2\right) \phi_e} \tag{78}$$

Or alternatively, by neglecting $\dot{m}$ for stable vapor-jet cavity oscillations, we have:

$$\boxed{\rho_V \phi_e \left(L \ddot{L} + \dot{L}^2\right) + \left(2\varphi \rho_V \Omega - \dot{m}\right) \dot{L} + F_{\text{stiffness}}\left(L\right) = p_\infty \phi_e + F_{\text{jet}}\left(\dot{m}\right)} \tag{79}$$

$$\boxed{\rho_V \phi_e \left[L \ddot{L} + \left(2\varphi L - v_e\right) \dot{L}\right] + 2\pi C \gamma \cos\theta \cdot L = \left(p_e + \rho_V v_e^2\right) \phi_e = \left(p_\infty + p_{\text{recoil}}^{(z)}\right) \phi_e} \tag{80}$$

where the general denotations of $F_{\text{stiffness}}$ and $F_{\text{jet}}$ are $F_{\text{surface}}^{(z)}$ and $F_{\text{recoil}}^{(z)}$ in this problem setting, respectively. Equations (78) to (80) parallels the *Rayleigh-Plesset equation*, with an additional damping term $2\varphi\rho_V\Omega\ (2\varphi L)$. This additional term, together with $-\dot{m}$, effectively switches the system between positive and negative damping regimes, thereby enabling the modeling of both stable and unstable vapor-jet cavity oscillations as a self-exciting oscillator. From this point, further theoretical analysis can be conducted to study the limit cycle (attractor) and the stability of the system under certain conditions, the results of which should provide insights into the onset of vapor-jet cavity instability and the transition to pore formation in laser-induced vaporization processes.

If, when necessary, we need to include the second-order term for $v_e$, the above assumptions and equations will undergo certain adjustments. To discuss this, we will alternatively assume the following relation based on Eq. (63):

$$v_e = \langle v_V \rangle + \varphi L + \frac{\varphi'}{2} L^2 \ \Rightarrow\ \frac{dv_e}{dt} = \frac{d\langle v_V \rangle}{dt} + \varphi \frac{\partial L}{\partial t} + \varphi' L \frac{\partial L}{\partial t} \tag{81}$$

where we assume that $\dot{\varphi} = \dot{\varphi}' = 0$. $\varphi'$ needs to possess a unit of $\mathrm{m}^{-1} \cdot \mathrm{s}^{-1}$ in order to keep consistent dimensions throughout equations. Substituting Eq. (81) into Eq. (67) and Eq. (70) and deriving following a similar procedure deduced above, we eventually obtain new versions of the equations with additional nonlinear coupling effects:

$$\boxed{\begin{aligned}\rho_V \phi_e \left(L\ddot{L} + \dot{L}^2\right) + \left[\left(2\varphi + \frac{3}{2}\varphi' L\right)\rho_V \Omega - \dot{m}\right]\dot{L} \\ + \left(2\pi C\gamma\cos\theta - \ddot{m}\right) L = \left(p_e + \rho_V v_e^2\right)\phi_e\end{aligned}} \tag{82}$$

$$\boxed{L\ddot{L} + \dot{L}^2 + \left[\left(2\varphi + \frac{3}{2}\varphi' L\right) L - \frac{\dot{m}}{\rho_V \phi_e}\right]\dot{L} + \frac{2\pi C\gamma\cos\theta - \ddot{m}}{\rho_V \phi_e} L = \frac{p_e}{\rho_V} + v_e^2} \tag{83}$$

$$\boxed{\rho_V \phi_e \left\{L\ddot{L} + \left[\left(2\varphi + \frac{3}{2}\varphi' L\right) L - v_e\right]\dot{L}\right\} + \left(2\pi C\gamma\cos\theta - \ddot{m}\right) L = \left(p_e + \rho_V v_e^2\right)\phi_e} \tag{84}$$

It can be easily seen that the introduction of the nonlinear assumption in Eq. (81) essentially replaces $2\varphi$ in Eq. (78) to Eq. (80) to $2\varphi + \frac{3}{2}\varphi' L$ when developing Eq. (82) to Eq. (84). For the convenience of further discussions in the document, though, we will stick to *linear* versions of the equations, *i.e.*, everything before Eq. (81).

---

**Discussion 1.2.** *It is also worth commenting that assuming* $\dot{\varphi} = 0$ *should be considered reasonable under stable vapor-jet cavity oscillation conditions: since* $\varphi$ *essentially represents the velocity gradient at* $\partial\Omega_e(t)$ *, if we heuristicly (not rigorously correct, though) exchange the order of spatial and time derivatives and re-write* $\dot{\varphi} = 0$ *as:*

$$\dot{\varphi} = \frac{\partial}{\partial t}\left(\left.\frac{\partial v_V^{(z)}}{\partial z}\right|_{z=0}\right) = \frac{\partial}{\partial z}\left.\left(\frac{\partial v_V^{(z)}}{\partial t}\right)\right|_{z=0} \tag{85}$$

*Based on Eq.* (85), $\dot{\varphi} = 0$ *can accordingly be interpreted as the following: the rate of change of the vapor ejection velocity at and around* $\partial\Omega_e(t)$ *is roughly spatially uniform. This agrees with the intuition for the non-equilibrium vapor dynamics, as vapor will typically expand freely around the vapor-jet cavity opening with an invariant adiabatic constant.*

---

For the convenience of reference, we call all different forms of the vapor-jet cavity depth oscillation equations (Eq. (56) to Eq. (84)) as the *Vapor-jet–Cavity Oscillation Dynamics Equations*, abbreviated as *JCODEs*.

---

**Discussion 1.3.** *It turns out that an alternative form of JCODE can also be derived without assuming constant* $\varphi$*—but instead assuming a* constant *bottom vapor velocity* $\vec{v}_B$ $(v_B = \|\vec{v}_B\| = \vec{v}_B \cdot \vec{e}_k)$ *and defining* $\varphi$ *as a velocity-gradient-dependent parameter. In the meantime, we still keep the essential assumption of linear vapor velocity profile along the vapor-jet cavity depth—i.e., Eq.* (77) *still holds, but now we have a non-zero constant vapor velocity at the bottom of the vapor-jet cavity so* $\dot{\varphi}$ *can be obtained accordingly. This leads to the following expression for* $\dot{v}_e$ *and* $\varphi$*:*

$$\dot{v_e} = \dot{\overline{\langle v_V \rangle}} + \varphi\dot{L} + \dot{\varphi}L \tag{86}$$

$$\varphi = \frac{v_e - v_B}{2L} \tag{87}$$

$$\dot{\varphi} = \frac{\dot{v_e}L - (v_e - v_B)\,\dot{L}}{2L^2} \tag{88}$$

*We further revise Assum. 1.5 to* $\Omega = \lambda L\phi_e$*, where* $\lambda$ *is a geometry-specific constant, to make the cavity geometry more general. Following similar derivation steps, we can obtain the following*

*forms of JCODE:*

$$\boxed{\begin{aligned}&\lambda^2\rho_V\phi_e\left(L\ddot{L}-\dot{L}^2\right)+\lambda\left(3\dot{m}-\rho_V\phi_e v_B\right)\dot{L}\\&\qquad\qquad+\left(4\pi C\gamma\cos\theta-\lambda\ddot{m}\right)L=2\left(p_e\phi_e+\frac{\dot{m}^2}{\rho_V\phi_e}\right)\end{aligned}} \tag{89}$$

$$\boxed{\lambda\rho_V\phi_e\left[\lambda\left(L\ddot{L}+2\dot{L}^2\right)+\left(3v_e-v_B\right)\dot{L}\right]+\left(4\pi C\gamma\cos\theta-\lambda\ddot{m}\right)L=2\left(p_e\phi_e+\frac{\dot{m}^2}{\rho_V\phi_e}\right)} \tag{90}$$

*when* $\lambda = 1$, *we then have:*

$$\boxed{\rho_V\phi_e\left[\left(L\ddot{L}+2\dot{L}^2\right)+\left(3v_e-v_B\right)\dot{L}\right]+\left(4\pi C\gamma\cos\theta-\ddot{m}\right)L=2\left(p_e\phi_e+\frac{\dot{m}^2}{\rho_V\phi_e}\right)} \tag{91}$$

---

Moreover, it is worth noting that the motivation of writing JCODE as the form in Eq. (79) and Eq. (80) is to apply the rocket thrust analogy to the R.H.S., *i.e.*, replacing the R.H.S. of JCODE with terms related to $F^{(z)}_{\text{recoil}}$. If treating $F^{(z)}_{\text{recoil}}$ as merely a function of $\dot{m}$, we can not only clearly see the interplay between it and the surface tension at $\partial\Omega_V(t)$ in governing the cavity depth oscillation dynamics, but also perform linearization without expanding the $v_e^2$ terms, potentially bringing new dynamical system structures into the vapor-jet cavity depth oscillation.

Fundamentally, JCODE models the vapor-jet cavity as a vertical, closed-end vapor-depression column whose depth undergoes nonlinear oscillations driven by continuous laser energy input. In essence, JCODE describes cavity-depth fluctuations as the motion of a sustained nonlinear oscillator whose properties and transient evolution are governed by geometric factors, material parameters, and boundary variables. To summarize, we point out the following three takeaways:

- JCODE captures the interplay between vaporization-induced recoil effects ($\left(p_e+\rho_V v_e^2\right)\phi_e$) and depth-dependent restoring actions at $\partial\Omega_e\,(t)$ ($2\pi C\gamma\cos\theta\cdot L$), the driving forces governing the vapor-cavity oscillation.

- From a dynamical-systems perspective, $\dot{m}$, $\Omega$, $\dot{L}$, and $\varphi$ (i.e., general vapor-velocity gradient—the relative magnitude between $v_e$ and $v_B$) govern the damping behavior of the cavity oscillations, enabling mechanistic descriptions spanning both energy-dissipating responses and self-amplifying exponential growth of the cavity depth.

- Mathematically, it not only parallels but also further extends the well-known *Rayleigh–Plesset equation*, which assumes spherical cavity symmetry and is widely used to model bubble oscillations and cavitation in liquids. By treating an axisymmetric cavity and explicitly accounting for the vapor-velocity gradient and vapor-jet flux along the cavity depth, JCODE underscores its novelty and distinguishes it from established physical descriptions of cavity dynamics.

## 1.3 Perturbative linearization of JCODE and its utility values for stable porosity-free deep vapor-jet cavities

We move on to demonstrate the utility value of JCODEs. To investigate the oscillation of a stable porosity-free deep vapor-jet cavity, we can linearize JCODEs around their steady-state values. The mathematical tool for performing this task is inevitably the *perturbation theory*. For the convenience of analysis, we start from Eq. (80) (with which $v_e = \langle v_V \rangle + \varphi L$ is assumed). We now assume that both the vapor-jet cavity depth $L$ and the vapor ejection speed $v_e$ oscillate around its steady-state value (denoted as $L_0$ and $v_{e0}$, respectively) with a very small linear perturbation (denoted as $\delta L$ and $\delta v_e$, respectively), written in mathematics as:

$$L = L_0 + \delta L, \ \delta L \ll L_0 \tag{92}$$

$$v_e = v_{e0} + \delta v_e, \ \delta v_e \ll v_{e0} \tag{93}$$

Next, we assume that under specified perturbation, $\dot{m}$ remains approximately constant, *i.e.*, $\dot{m} \approx$ const., thus $\ddot{m} = \frac{d\dot{m}}{dt} \approx 0$. This assumption is considered reasonable since under stable oscillation with no porosities, $\dot{m}$ is primarily determined by the rate of vaporization across the liquid–vapor interface ($\partial\Omega_V(t)$), which is typically stable over short time scales without changing laser power ($P$) and absorptance ($\varepsilon$). To further reduce the complexity of derivation, we also choose to neglect the variation of $p_e$ and $\theta$, *i.e.*, we have $p_e = p_{e0}$ and $\theta = \theta_0$, respectively, throughout the perturbation analysis.

Combining all the above relationships, we immediately realize that at steady state of the vapor-jet cavity system—where we have $\dot{L} = \ddot{L} = 0$—an equilibrium state can be formulated. Specifically,

Eq. (22) reduces to:

$$\dot{m} = \rho_v \phi_e v_{e0} \;\Rightarrow\; \ddot{m} = 0 \tag{94}$$

and Eq. (80) reduces to:

$$2\pi C\gamma \cos\theta_0 L_0 = \left(p_{e0} + \rho_V v_{e0}^2\right)\phi_e \;\Rightarrow\; L_0 = \frac{\left(p_{e0} + \rho_V v_{e0}^2\right)\phi_e}{2\pi C\gamma\cos\theta_0} \tag{95}$$

Departing from the steady state and introducing linear deviations defined in Eq. (92) and Eq. (93), we can derive the perturbation form of Eq. (22) as follows:

$$\begin{aligned} \dot{m} = \rho_v \phi_e \left(\dot{L} + v_e\right) &= \rho_v \phi_e \underbrace{\dot{\overline{(L_0 + \delta L)}}}_{\dot{L}_0 = 0} + \rho_V \phi_e \left(v_{e0} + \delta v_e\right) \\ &= \rho_v \phi_e \left(\dot{\delta L} + \delta v_e\right) + \underbrace{\rho_v \phi_e v_{e0}}_{\text{Eq. (94)}} \end{aligned} \tag{96}$$

$$\rho_v \phi_e \left(\dot{\delta L} + \delta v_e\right) = 0 \;\Rightarrow\; \delta v_e = -\dot{\delta L} \tag{97}$$

Equation (97) clearly shows that $\delta L$ and $\delta v_e$ are interdependent for linearly perturbed porosity-free deep vapor-jet cavitys, as the system must always satisfy the law of the conservation of mass to maintain its stability. If further taking time derivative for both sides of Eq. (97), we have:

$$\dot{\delta v_e} = -\ddot{\delta L} \tag{98}$$

Equation (98) indicates that the perturbation in vapor ejection acceleration ($\dot{\delta v_e}$) is directly proportional to the negative of the perturbation in vapor-jet cavity depth acceleration ($\ddot{\delta L}$) for stable deep vapor-jet cavitys. This relationship underscores the dynamic coupling between the vapor-jet cavity geometry and the vapor flow characteristics under stable oscillation, and plays a crucial role in the subsequent analysis of the acoustic emission mechanisms.

We move on to substitute Eq. (92) and Eq. (93) into Eq. (80) and obtain:

$$\begin{aligned} \rho_V \phi_e \left(L_0 + \delta L\right) \ddot{\delta L} + \rho_V \phi_e \left[2\varphi \left(L_0 + \delta L\right) - \left(v_{e0} + \delta v_e\right)\right] \dot{\delta L} \\ + 2\pi C\gamma \cos\theta_0 \left(L_0 + \delta L\right) = \left[p_{e0} + \rho_V \left(v_{e0} + \delta v_e\right)^2\right] \phi_e \end{aligned} \tag{99}$$

Neglecting all the higher-order terms of $\delta L$ and $\delta v_e$ (as well as their time derivatives), we arrive at the following linearized equation for $\delta L$:

$$
\begin{aligned}
&\rho_V \phi_e L_0 \delta\ddot{L} + \rho_V \phi_e \left(2\varphi L_0 - v_{e0}\right) \delta\dot{L} + 2\pi C\gamma \cos\theta_0 \delta L \\
&\qquad + \underbrace{2\pi C\gamma \cos\theta_0 L_0 = \left(p_{e0} + \rho_V v_{e0}^2\right) \phi_e}_{\text{Eq. (95)}} + 2\rho_V \phi_e v_{e0} \delta v_e
\end{aligned}
\tag{100}
$$

$$
\rho_V \phi_e \left[L_0 \delta\ddot{L} + \left(2\varphi L_0 - v_{e0}\right) \delta\dot{L}\right] + 2\pi C\gamma \cos\theta_0 \delta L = 2\rho_V \phi_e v_{e0} \delta v_e \tag{101}
$$

It is worth noting that the underbraced terms in Eq. (100) cancel each other because they are exactly equivalent according to the steady-state relationship as stated in Eq. (95). Substituting Eq. (97) into Eq. (101), we finally have:

$$
\boxed{\rho_V \phi_e \left[L_0 \delta\ddot{L} + \left(2\varphi L_0 + v_{e0}\right) \delta\dot{L}\right] + 2\pi C\gamma \cos\theta_0 \delta L = 0} \tag{102}
$$

If we replace $v_{e0}$ with $\dot{m}$ using the steady-state relationship in Eq. (94), we can write Eq. (102) alternatively as:

$$
\boxed{\rho_V \phi_e L_0 \delta\ddot{L} + \left(2\varphi \rho_V \phi_e L_0 + \dot{m}\right) \delta\dot{L} + 2\pi C\gamma \cos\theta_0 \delta L = 0} \tag{103}
$$

We notice that both Eq. (102) and Eq. (103) take the canonical form of a damped harmonic oscillator for $\delta L$, where the first term corresponds to inertia, the second to damping, and the third to the restoring force arising from surface tension and Marangoni effects, establishing a *mass-spring analog* for the oscillation of the vapor-jet cavity depth. It's also worth emphasizing that Eq. (102) and Eq. (103) do not describe self-oscillation anymore, since the perturbed form of the governing equation essentially formulates stable oscillation processes. For a stable vapor-jet cavity for which damping contribution to the oscillation frequency can be considered negligible, the system oscillates at its natural frequency, expressed as:

$$
\boxed{f_N = \frac{\omega_L}{2\pi} \approx \frac{1}{2\pi}\sqrt{\frac{2\pi C\gamma \cos\theta_0}{\rho_V \Omega_0}} = \sqrt{\frac{C\gamma \cos\theta_0}{2\pi \rho_V \phi_e L_0}}} \tag{104}
$$

where $\omega_L$ represents the angular frequency of the oscillation of $\delta L$, and $\Omega_0 = L_0 \phi_e$. It is evident that Eq. (104) is the same as the natural frequency derived from Eq. (80). With actual observations we obtained through multimodal X-ray synchrotron experiments, we can effectively replace $\Omega_0$, $L_0$, and $\cos\theta$ with $\langle\Omega\rangle$, $\langle L\rangle$, and $\langle\cos\theta\rangle$, respectively. Therefore we have:

$$\boxed{\langle f_N\rangle = \frac{\langle\omega_L\rangle}{2\pi} = \sqrt{\frac{C\gamma\,\langle\cos\theta\rangle}{2\pi\rho_V\,\langle\Omega\rangle}} = \sqrt{\frac{C\gamma}{2\pi\rho_V}\cdot\left\langle\frac{\cos\theta}{\Omega}\right\rangle} = \sqrt{\frac{C\gamma}{2\pi\rho_V\phi_e}\cdot\left\langle\frac{\cos\theta}{L}\right\rangle}} \tag{105}$$

Equation (105) offers an effective linear estimate of the natural frequency of vapor-jet cavity depth oscillations. When combined with Eq. (97) and Eq. (98), it establishes a direct link to the frequency of airborne acoustic emissions detected by a microphone positioned near the center of $\partial\Omega_e(t)$. It is important to note that Eq. (105) is derived under a set of simplifying assumptions—such as constant vapor-jet cavity radius, linearized dynamics, stable oscillation, and negligible damping—which may limit its applicability across all conditions and vapor-jet cavity regimes. Nevertheless, the expression provides a valuable first-order approximation that captures the essential coupling between vapor-jet cavity geometry, material properties, and the resulting acoustic response.

It is important to emphasize that the angle $\theta$ appearing in the derivations represents a characteristic angle, which can sometimes be approximated by the inclination of the vapor-jet cavity front wall relative to the horizontal plane. However, this approximation may break down when the vapor-jet cavity geometry deviates significantly from idealized assumptions, such as in shallow vapor-jet cavitys, where the front-wall inclination no longer accurately reflects $\theta$. In such cases, $\cos\theta$ must be redefined in accordance with Assum. 1.8. Under this correction, the complete expression for the natural frequency can be written as:

$$\langle f_N\rangle = \sqrt{\frac{C\gamma}{2\pi\rho_V\,\langle\Omega\rangle}\cdot\left\langle\frac{R}{\sqrt{L^2+R^2}}\right\rangle} \Rightarrow \langle f_N\rangle^2 = \frac{C\gamma}{2\pi\rho_V}\cdot\left\langle\frac{R}{\Omega\sqrt{L^2+R^2}}\right\rangle \tag{106}$$

Alternatively, we can replace $\cos\theta$ with $\tan\theta = \frac{L}{R}$, which exactly represents the half of the aspect ratio of the vapor-jet cavity. Denoting the aspect ratio as $\upsilon$, *i.e.*, $\upsilon = \frac{L}{2R} = \frac{1}{2}\tan\theta$, we have:

$$\langle f_N\rangle = \sqrt{\frac{C\gamma}{2\pi\rho_V\,\langle\Omega\rangle}\cdot\left\langle\frac{1}{\sqrt{4\upsilon^2+1}}\right\rangle} \Rightarrow \langle f_N\rangle^2 = \frac{C\gamma}{2\pi\rho_V}\cdot\left\langle\frac{1}{\Omega\sqrt{4\upsilon^2+1}}\right\rangle \tag{107}$$

which indicates that the larger the stable volume and the aspect ratio $\upsilon$ of the vapor-jet cavity, the lower the linear approximation of the perturbed natural frequency. It can be easily seen that Eq. (106) and Eq. (107) are identical, as Eq. (107) can be obtained by dividing $R$ for both the numerator and denominator of Eq. (106).

Aside from characterizing vapor-jet cavity depth oscillation frequency, we can also obtain physical insights directly from Eq. (98) by rewriting it using Eq. (92) and Eq. (93) as:

$$\underbrace{\delta\dot{v}_e = -\delta\ddot{L}}_{\text{Eq. (98)}} \Rightarrow \overline{(\delta v_e + v_{e0})}^{\cdot} = -\overline{(\delta L + L_0)}^{\cdot\cdot} \Rightarrow \dot{v_e} = -\ddot{L} \tag{108}$$

which indicates that the amplitude of acoustic emission (proportional to $\dot{v_e}$) is directly related to the vapor-jet cavity depth acceleration ($\ddot{L}$) for linearly perturbed porosity-free deep vapor-jet cavitys. From Eq. (108), two conclusions can be drawn:

- Under linear perturbation around the equilibrium, the frequency of acoustic emission is the same as that of the vapor-jet cavity depth oscillation, namely $\langle f_N \rangle$ as given in Eq. (105).
  - It can be seen by substituting $\delta L = A_L e^{i\omega_L t}$ into Eq. (108), from which we obtain:

$$\begin{aligned} \delta\dot{v}_e = -\delta\ddot{L} = -A_L\,(i\omega_L)^2\,e^{i\omega_L t} &= A_L\omega_L^2 e^{i\omega_L t} \\ = -A_L\omega_L^2 e^{i\pi} e^{i\omega_L t} &= -A_L\omega_L^2 e^{i(\omega_L t + \pi)} \end{aligned} \tag{109}$$

    where $A_L$ denotes the amplitude of $\delta L$ oscillation. Equation (109) sufficiently demonstrates that $\delta\dot{v}_e$, representing the emitted airborne acoustic wave, should oscillate at the same angular frequency as $\delta\ddot{L}$, *i.e.*, $\omega_L = 2\pi\,\langle f_N \rangle$ according to Eq. (105). A phase shift of $\pi$ is also observed, indicating that the acoustic wave is out of phase with the vapor-jet cavity depth oscillation. This phase difference is expected, as the acoustic wave is generated by the acceleration of the vapor-jet cavity depth, leading to a temporal lag between the two phenomena.

- The amplitude of acoustic emission is proportional to the amplitude of vapor-jet cavity depth acceleration. In other words, a more violently oscillating vapor-jet cavity (with a higher $\ddot{L}$) would generate a stronger acoustic signal (with a higher $\dot{v_e}$).

Equations (108) and (109) provide a theoretical basis for using acoustic measurements to infer vapor-jet cavity dynamics during laser-induced vaporization processes.

---

**Discussion 1.4.** *Following Discussion 1.3, we derive the corresponding perturbative linearization formulations that includes* $v_B$.

*Assuming* $\ddot{m} = 0$ *under stable perturbative oscillations, the corresponding linearized forms of Eq.* (89) *and Eq.* (90) *would be:*

$$\boxed{\lambda^2 \rho_V \phi_e L_0 \delta\ddot{L} + \lambda\left(3\dot{m} - \rho_V \phi_e v_B\right)\delta\dot{L} + 4\pi C\gamma\cos\theta_0 \cdot \delta L = 0} \tag{110}$$

$$\boxed{\lambda \rho_V \phi_e \left[\lambda L_0 \delta\ddot{L} + \left(3v_{e0} - v_B\right)\delta\dot{L}\right] + 4\pi C\gamma\cos\theta_0 \cdot \delta L = 0} \tag{111}$$

*From Eq.* (110) *and Eq.* (111)*, we get the corresponding natural frequencies with negligible effects from damping:*

$$\boxed{f_N = \frac{1}{2\pi}\sqrt{\frac{4\pi C\gamma\cos\theta_0}{\lambda^2\rho_V\phi_e L_0}} = \sqrt{\frac{C\gamma\cos\theta_0}{\pi\lambda^2\rho_V\phi_e L_0}}} \tag{112}$$

$$\boxed{\langle f_N\rangle = \sqrt{\frac{C\gamma}{\pi\lambda\rho_V\langle\Omega\rangle}\cdot\left\langle\frac{1}{\sqrt{4\upsilon^2+1}}\right\rangle} \Rightarrow \langle f_N\rangle^2 = \frac{C\gamma}{\pi\lambda\rho_V}\cdot\left\langle\frac{1}{\Omega\sqrt{4\upsilon^2+1}}\right\rangle} \tag{113}$$

*Meanwhile, Eq.* (97) *also needs to be revised accordingly:*

$$\boxed{\rho_v\phi_e\left(\lambda\delta\dot{L} + \delta v_e\right) = 0 \Rightarrow \delta v_e = -\lambda\delta\dot{L} \Rightarrow \dot{v_e} = -\lambda\delta\ddot{L}} \tag{114}$$

*All the new formulations presented here and in Discussion 1.3 do not change the essential physics and implications of the JCODEs and the associated theoretical framework. It can be easily seen that when* $\lambda = 1/3$*, the vapor-jet cavity geometry is an axisymmetric cone; when* $\lambda = 1$*, the vapor-jet cavity geometry is an axisymmetric cylinder. The introduction of* $v_B$ *or* $\varphi$ *does not fundamentally change the JCODEs, but rather provides an alternative perspective on the vapor-jet cavity vapor velocity profile.*

---

From a mechanistic perspective, JCODE as well as its perturbative linearization offers an alternative physical picture of vapor-jet cavity oscillations. As shown in Eqs (78) and (102), JCODE essentially describes the vapor-jet-cavity as a nonlinear oscillatory dynamical system; the effective damping coefficient of this system, namely $(2\varphi\rho_V\Omega - \dot{m})$ or $\rho_V\phi_e\left(2\varphi L_0 + v_{e0}\right)$, is fundamentally controlled by (1) the direction of vapor-velocity gradient and (2) the relation between $\dot{m}$ and $L$. In other words, the depth of a vapor-jet cavity can be interpreted analogously as a nonlinear mass-spring oscillator, whose effective damping coefficient can take either positive (energy-dissipating) or negative signs (energy-injecting) depending on the transient status of the vapor-jet-cavity system, thereby capturing different vapor-jet cavity dynamics regimes. When the damping coefficient crosses zero, system trajectories may transition from self-amplifying exponential growth to nonlinear saturation, potentially restabilizing into bounded oscillations or abruptly collapsing, producing pores and resetting the vapor-jet cavity state.

---

**Discussion 1.5.** *According to Discussion 1.4, it seems interesting that the damping coefficient in these alternative forms of JCODEs now explicitly depends on the relationship between $v_e$ and $v_B$, or $\dot{m}$ and $\rho_V \phi_e v_B$, which may provide additional insights into the vapor-jet cavity oscillation mode. Roughly speaking, we realize that when the damping is positive, the JCODE describes a normal damped harmonic oscillation; when the damping switches to negative, it may experience temporally self-amplifying exponential growth, which leads to either divergence or a pseudo-stable limit cycle. We leave further explorations of these alternative JCODEs and associated nonlinear oscillations to future work.*

---

## 1.4 Mechanisms of airborne acoustics emitted from an oscillating vapor-jet cavity

To begin, we draw on the well-established field of aeroacoustics, using Lighthill's equation and acoustic analogies to clarify what an acoustic sensor actually measures. For clarity and ease of reference, we refer to Tab. S1 for the nomenclature of the following analysis.

### 1.4.1 Aeroacoustic equations and Lighthill's analogies

According to *Lighthill's acoustic analogy*, the sound generated by general fluids under motions can be described by an inhomogeneous wave equation, where the source terms are related to the flow properties. The fundamental equations governing fluid dynamics are the Navier-Stokes equations, which consist of the continuity equation (mass conservation) and the momentum equation (Newton's second law for fluids) for the medium at rest and without the presence of any external mass and momentum sources:

$$\begin{cases} \dfrac{\partial \rho}{\partial t} + \nabla \cdot (\rho \vec{v}) = 0 \\ \dfrac{\partial \rho \vec{v}}{\partial t} + \nabla \cdot (\rho \vec{v} \otimes \vec{v}) = \nabla \cdot (-p\mathbf{I} + \tau) = -\nabla p + \nabla \cdot \tau \end{cases} \tag{115}$$

By taking the time derivative of the continuity equation and the divergence of the momentum equation, and then combining them, we can merge and derive Eq. (115) into one single wave equation for the density fluctuations:

$$\frac{\partial^2 \rho}{\partial t^2} - c_0^2 \nabla^2 \rho = \underbrace{\nabla\nabla : \left[\rho \vec{v} \otimes \vec{v} + \left(p - c_0^2 \rho\right) \mathbf{I} - \tau\right]}_{\text{Acoustical source}} \tag{116}$$

Equation (116) is the so-called *Lighthill equation for aeroacoustics*, with R.H.S. being the Hessian of the *Lighthill turbulence stress tensor*. Neglecting the viscous stress tensor $\tau$ and re-arranging terms, Eq. (116) reduces to:

$$\frac{\partial^2 \rho}{\partial t^2} \approx \nabla\nabla : \left[\rho \vec{v} \otimes \vec{v} + \left(p - c_0^2 \rho\right) \mathbf{I}\right] + c_0^2 \nabla^2 \rho = \nabla\nabla : \left[\rho \vec{v} \otimes \vec{v} + p\mathbf{I}\right] \tag{117}$$

Equations (116) and (117) indicate that the sound wave generated by the vapor flow at $\partial\Omega_e(t)$ can be attributed to an integral acoustical source as an effective stress tensor, encompassing the Reynolds stress tensor $\rho \vec{v} \otimes \vec{v}$ and the static pressure fluctuations $p\mathbf{I}$. The Reynolds stress tensor represents the momentum flux due to turbulent velocity fluctuations, while the pressure term accounts for compressibility effects outside $\partial\Omega_e(t)$. The resultant term $\mathbf{T} = \rho \vec{v} \otimes \vec{v} + p\mathbf{I}$ is a second-order tensor (simplified linear acoustical stress tensor), and the Hessian operator $\nabla\nabla : \mathbf{T}$ effectively captures how variations in these quantities lead to the generation of density waves.

If, when necessary, there exist sources consistently adding mass or momentum at a point into the medium, the above equations may need to be revised. Denoting $S_m$ and $\vec{F}$ as the volumetric mass and momentum sources, respectively, we can revise Eq. (115) as the following:

$$\begin{cases} \dfrac{\partial \rho}{\partial t} + \nabla \cdot (\rho \vec{v}) = S_m \\ \dfrac{\partial \rho \vec{v}}{\partial t} + \nabla \cdot (\rho \vec{v} \otimes \vec{v}) = \nabla \cdot (-p\mathbf{I} + \tau) + \vec{F} = -\nabla p + \nabla \cdot \tau + \vec{F} \end{cases} \tag{118}$$

It can be easily seen that we have $S_m \propto \rho_V \phi_e v_e$ and $\vec{F} \propto \left[ (p_e - p_\infty) + \rho_V v_e^2 \right] \phi_e \vec{e}_k$ at $\partial \Omega_e(t)$, but this does not hold for the mass and momentum sources in the rest of the medium. Since the sources are designed to be volumetric, it is evident that both $S_m$ and $\vec{F}$ are intensive properties with units of $\mathrm{kg} \cdot \mathrm{m}^{-3} \cdot \mathrm{s}^{-1}$ and $\mathrm{N/m}^3$, respectively. By including these additional terms into the derivation, the Lighthill equation for aeroacoustics are updated to the following:

$$\frac{\partial^2 \rho}{\partial t^2} - c_0^2 \nabla^2 \rho = \underbrace{\frac{\partial S_m}{\partial t} - \nabla \cdot \vec{F} + \nabla\nabla : \left[ \rho \vec{v} \otimes \vec{v} + \left( p - c_0^2 \rho \right) \mathbf{I} - \tau \right]}_{\text{Multipole acoustic source}} \tag{119}$$

$$\begin{aligned} \frac{\partial^2 \rho}{\partial t^2} &\approx \frac{\partial S_m}{\partial t} - \nabla \cdot \vec{F} + \nabla\nabla : \left[ \rho \vec{v} \otimes \vec{v} + \left( p - c_0^2 \rho \right) \mathbf{I} \right] + c_0^2 \nabla^2 \rho \\ &= \frac{\partial S_m}{\partial t} - \nabla \cdot \vec{F} + \nabla\nabla : \left[ \rho \vec{v} \otimes \vec{v} + p\mathbf{I} \right] \end{aligned} \tag{120}$$

Equation (119) is also known as the *Ffowcs Williams–Hawkings (FW–H) equation*, which is widely used in aeroacoustics to predict noise generated by moving rigid surfaces. Comparing Eq. (119) with Eq. (120), the additional volumetric mass and momentum sources enter the R.H.S. through temporal and spatial derivatives, and, together with the original terms, form a mixed multipole acoustic source according to *Lighthill's aeroacoustic analogies*: specifically, the mass source term corresponds to a monopole, the momentum source to a dipole, and the stress tensor term to a quadrupole. In our problem setting, except at $\partial \Omega_e(t)$, both $S_m$ and $\vec{F}$ vanish throughout the medium, since no explicit mass or momentum sources exist elsewhere. Consequently, the canonical Lighthill equation for aeroacoustics (*i.e.*, Eq. (116)) remains valid over the vast majority of the domain surrounding the vapor-jet cavity opening.

### 1.4.2 Acoustic emission under perturbative vapor-jet from a vapor-jet cavity

Next, following the procedures of perturbative linearization used in Sec. 1.3 and applying separation of variables (only for the convenience of derivations), we perturb $p$, $\rho$, and $\vec{v}$ around their equilibrium points as the following:

$$p\left(\vec{x}, t\right) = p_0\left(\vec{x}\right) + \delta p\left(t\right),\ \delta p \ll p_0 \tag{121}$$

$$\rho\left(\vec{x}, t\right) = \rho_0\left(\vec{x}\right) + \delta \rho\left(t\right),\ \delta \rho \ll \rho_0 \tag{122}$$

$$\vec{v}\left(\vec{x}, t\right) = \vec{v}_0\left(\vec{x}\right) + \delta \vec{v}\left(t\right),\ \|\delta \vec{v}\| \ll \|\vec{v}_0\| \tag{123}$$

At equilibrium, we substitute $p_0$, $\rho_0$, and $\vec{v}_0$ into Eq. (117) and reduce it to:

$$\nabla\nabla : \left[\rho_0 \vec{v}_0 \otimes \vec{v}_0 + p_0 \mathbf{I}\right] = 0 \tag{124}$$

where the second-order time derivative of $\rho_0$ vanishes. Substituting Eq. (121) to Eq. (123) into Eq. (117), we get:

$$\frac{\partial^2\left(\rho_0 + \delta\rho\right)}{\partial t^2} = \nabla\nabla : \left[\left(\rho_0 + \delta\rho\right) \cdot \left(\vec{v}_0 + \delta\vec{v}\right) \otimes \left(\vec{v}_0 + \delta\vec{v}\right) + \left(p_0 + \delta p\right)\mathbf{I}\right] \tag{125}$$

Re-arranging terms, substituting in Eq. (124), and neglecting higher-order perturbation terms, Eq. (125) can be simplified as:

$$\frac{\partial^2 \delta\rho}{\partial t^2} = \nabla\nabla : \left[\delta\rho \vec{v}_0 \otimes \vec{v}_0 + \rho_0 \vec{v}_0 \otimes \delta\vec{v} + \rho_0 \delta\vec{v} \otimes \vec{v}_0 + \delta p \mathbf{I}\right] \tag{126}$$

Under linear perturbations, we may use the following relation:

$$\delta p = c_0^2 \delta\rho \ \Rightarrow\ \nabla^2 \delta p = \nabla\nabla : \delta p \mathbf{I} = \nabla\nabla : c_0^2 \delta\rho \mathbf{I} = c_0^2 \nabla^2 \delta\rho \tag{127}$$

Substituting Eq. (127) into Eq. (126) and re-arranging terms, we have:

$$\left(\frac{\partial^2}{\partial t^2} - c_0^2 \nabla^2\right)\delta\rho = \underbrace{\nabla\nabla : \left[\delta\rho \vec{v}_0 \otimes \vec{v}_0 + \rho_0 \vec{v}_0 \otimes \delta\vec{v} + \rho_0 \delta\vec{v} \otimes \vec{v}_0\right]}_{\text{Perturbative acoustic source}} = \nabla\nabla : \delta\mathbf{T} \tag{128}$$

According to Lighthill's acoustic analogy, $\nabla\nabla : \delta\mathbf{T}$ essentially results in an acoustic quadrupole, as it mathematically represents curvatures of the perturbed components of $\delta\mathbf{T}$. As such, Eq. (128)

demonstrates that under linear perturbation, the emitted airborne acoustic waves are generated from local quadrupole acoustic sources at each point in the medium.

Similarly, we can express the full acoustic-emission model that includes contributions generated at $\partial\Omega_e(t)$ through perturbations of $v_e$ (recalling Eq. (93)) as well as those arising from the surrounding medium:

$$\left(\frac{\partial^2}{\partial t^2} - c_0^2\nabla^2\right)\delta\rho' = \left.\left(\frac{\partial\delta S_m}{\partial t} - \nabla\cdot\delta\vec{F} + \nabla\nabla : \delta\mathbf{T}\right)\right|_{\partial\Omega_e(t)} + \nabla\nabla : \delta\mathbf{T}\left(\vec{x}\right) \tag{129}$$

where $\delta\rho'$ represents the collected airorne acoustic wave at $\vec{x}$. Meanwhile, we have the following:

$$\delta S_m|_{\partial\Omega_e(t)} = \rho_V\phi_e\left(v_{e0} + \delta v_e - v_{e0}\right) = \rho_V\phi_e\delta v_e \tag{130}$$

$$\left.\delta\vec{F}\right|_{\partial\Omega_e(t)} = \rho_V\left[\left(v_{e0} + \delta v_e\right)^2 - v_{e0}^2\right]\phi_e\vec{e}_k \approx 2\rho_V\phi_e v_{e0}\delta v_e\vec{e}_k \tag{131}$$

where we ignore all components along $x$-axis and $y$-axis and keep only the terms along $z$-axis for the convenience of analysis. It is worth noting that here $\delta\mathbf{T}|_{\partial\Omega_e(t)}$ does not equal to $\delta\rho_V\left(\vec{v}_{e0} + \delta\vec{v}_e\right)\otimes\left(\vec{v}_{e0} + \delta\vec{v}_e\right)$ anymore, since this term has been effectively modeled as separate mass and momentum sources, *i.e.*, $\delta S_m|_{\partial\Omega_e(t)}$ and $\left.\delta\vec{F}\right|_{\partial\Omega_e(t)}$, respectively. Therefore, we have $\delta\mathbf{T}|_{\partial\Omega_e(t)} \approx \mathbf{0}$ with negligible vapor viscosity and local turbulence. Likewise, the last component of the R.H.S. can typically be omitted in both theoretical analyses and practical applications unless intense turbulence is present locally.

### 1.4.3 Vapor-jet–Cavity Acoustic Equation (VCAE)

Based on the above analysis, the leftover R.H.S. terms of Eq. (129) show there are two types of acoustic sources: (a) a monopole with its strength being $\left.\frac{\partial\delta S_m}{\partial t}\right|_{\partial\Omega_e(t)}$ (Eq. (130)), and (b) dipoles (first-order spatial derivatives) with their strength being $\delta\vec{F}$ (Eq. (131)). Employing the delayed *Green's function* to solve Eq. (129) near $\partial\Omega_e(t)$ and *retaining only the leading terms that scale inversely with distance from the sources*, we obtain:

$$\begin{aligned}\delta\rho'\left(\vec{x},t\right) =& \rho'\left(\vec{x},t\right) - \rho_0' \\ \approx& \frac{1}{4\pi c_0^2}\int_{\Xi(\partial\Omega_e(t))}\left\{\frac{1}{\|\vec{x}-\vec{y}\|}\frac{\partial}{\partial t}\left[\delta S_m\left(\vec{y}, t - \frac{\|\vec{x}-\vec{y}\|}{c_0}\right)\right]\right\}d^3y \\ &+ \frac{1}{4\pi c_0^2}\int_{\Xi(\partial\Omega_e(t))}\left\{\frac{\vec{x}-\vec{y}}{c_0\|\vec{x}-\vec{y}\|^2}\cdot\frac{\partial}{\partial t}\left[\delta\vec{F}\left(\vec{y}, t - \frac{\|\vec{x}-\vec{y}\|}{c_0}\right)\right]\right\}d^3y\end{aligned} \tag{132}$$

where $\vec{x}$ denotes the location of the sources, and $\vec{y}$ denotes the inner-source location relative to $\vec{x}$. It is worth noting that a retardance (*i.e.*, $t - \frac{\|\vec{x}-\vec{y}\|}{c_0}$) has been specified for each acoustical source. Oftentimes, we can assume that the volumetric sources are concentrated at $\partial\Omega_e(t)$ and therefore $\|\vec{x}-\vec{y}\| \approx \|\vec{x}\|$. Denoting $r = \|\vec{x}\| = x_i\vec{e}_i$ (using *Einstein's notation*), exchanging the order of integration and differentiation, and substituting $x_i \approx r$, we can then simplify Eq. (132) as:

$$\begin{aligned}\delta\rho'(\vec{x},t) \approx\ & \frac{1}{4\pi c_0^2}\int_{\Xi(\partial\Omega_e(t))}\left\{\frac{1}{r}\frac{\partial}{\partial t}\left[\delta S_m\left(\vec{y},t-\frac{r}{c_0}\right)\right]\right\}d^3y \\ & +\frac{1}{4\pi c_0^2}\int_{\Xi(\partial\Omega_e(t))}\left\{\frac{x_i}{c_0 r^2}\cdot\frac{\partial}{\partial t}\left[\delta F_i\left(\vec{y},t-\frac{r}{c_0}\right)\right]\right\}d^3y\end{aligned} \tag{133}$$

$$\begin{aligned}\delta\rho'(\vec{x},t) \propto\ & \frac{1}{4\pi c_0^2}\cdot\frac{1}{r}\frac{\partial}{\partial t}\left[\delta\int_{\Xi(\partial\Omega_e(t))}S_m\left(\vec{y},t-\frac{r}{c_0}\right)d^3y\right] \\ & +\frac{1}{4\pi c_0^2}\cdot\frac{1}{c_0 r}\cdot\frac{\partial}{\partial t}\left\{\sum_i\left[\delta\int_{\Xi(\partial\Omega_e(t))}F_i\left(\vec{y},t-\frac{r}{c_0}\right)d^3y\right]\right\}\end{aligned} \tag{134}$$

From Eq. (128) to Eq. (134), we see that for a microphone positioned at a finite distance (well beyond several wavelengths) on top of $\partial\Omega_e(t)$, the measured acoustic emission is essentially the linearized wave radiated by sources at $\partial\Omega_e(t)$ (*i.e.*, $\delta\rho'(\vec{x},t)$). It can be easily seen that the amplitude of the wave decays inversely with the distance between the source and the microphone, as expected. By substituting the boundary conditions at $\partial\Omega_e(t)$ into Eq. (134) and integrating all linearized effective source terms, we obtain the following estimation for $\delta\rho(\vec{x},t)$:

$$\begin{aligned}\delta\rho(\vec{x},t) \approx \delta\rho'(\vec{x},t) \propto\ & \frac{1}{4\pi c_0^2 r}\left[\frac{\partial}{\partial t}\left(\rho_V\phi_e\delta v_e\right)+\frac{1}{c_0}\frac{\partial}{\partial t}\left(2\rho_V\phi_e v_{e0}\delta v_e\right)\right] \\ =\ & \frac{\rho_V\phi_e}{4\pi c_0^2 r}\left[\delta\dot{v}_e+\frac{2v_{e0}}{c_0}\delta\dot{v}_e\right]\end{aligned} \tag{135}$$

Denoting $M = v_{e0}/c_0 \approx v_e/c_0$ as the Mach number of the vapor jet at $\partial\Omega_e(t)$, we can re-arrange Eq. (135) by substituting in Eq. (127):

$$\boxed{\delta p(\vec{x},t) \approx p_{\mathrm{acc}}^{(z)} \propto \frac{\rho_V\phi_e}{4\pi r}\left[2\left(\frac{v_e}{c_0}\right)+1\right]\dot{v}_e = \frac{G}{r}\left(2M+1\right)\frac{\partial v_e}{\partial t}} \tag{136}$$

where $G = \frac{\rho_V\phi_e}{4\pi}$. It's worth noting that the retardance between the source and the microphone still exists, *i.e.*, $v_e = v_e\left(t-\frac{r(\vec{x})}{c_0}\right)$, but we omitted it in the above equation for simplicity.

We call Eq. (136) the *Vapor-jet–Cavity Acoustic Equation* (VCAE). VCAE not only provides a linearized framework linking airborne acoustic emissions to vapor-jet cavity-associated quantities (particularly $v_e$), but also serves as a confirmation and extension of the previously investigated laser welding acoustic mechanism. Although derived under simplified perturbative assumptions, VCAE clearly indicates that the acoustic wave amplitudes of the monopole and dipole are driven by $\dot{v_e}$ and $M$ at $\partial\Omega_e(t)$, decaying proportionally to $1/r$ as the microphone distance from the vapor-jet cavity opening increases in the far field. Substituting $v_e = \delta v_e e^{i\omega_v t} + v_{e0}$ into VCAE, *i.e.*, Eq. (136), and recalling Eq. (93), we find that the integrated acoustic emission from monopole and dipole contributions is dominated by the frequency component of $v_e$ (which is the same as that of $L$, as shown in Sec. 1.3) with a phase-lag of $\frac{\pi}{2}$, while higher-frequency components are substantially attenuated. Assuming the characteristic frequency of $v_e$ and $L$ is $f_N$, VCAE further reveals that the emitted acoustic spectrum shifts toward higher-frequency components as $v_{e0}$ ($v_e$) decreases substantially. These behaviors align with the predictions in Sec. 1.3 and reinforce the potential of spectral acoustic analysis as a noninvasive approach for inferring the vapor-jet cavity depth oscillation frequency under small linear perturbations.

---

**Discussion 1.6.** *It is worth emphasizing that the analysis above targets acoustic emissions whose dominant source is the intrinsic unsteadiness of the vapor jet and its coupling to the cavity opening, rather than noise generated by fully developed turbulent mixing downstream. Accordingly, unless strong vortex shedding or intense shear-layer roll-up occurs at or near* $\partial\Omega_e(t)$*, we neglect contributions from local turbulence and shed eddies. Those mechanisms fall more naturally within the traditional "jet-noise" framework, where broadband radiation is governed by turbulent fluctuations, coherence scales, and mixing-layer dynamics, and where the relevant source terms introduce additional modeling complexity and stronger sensitivity to geometric and flow details. By making this separation explicit, we clarify what the airborne emissions measured near the vapor-jet exit are intended to represent in this chapter: a response primarily driven by coherent, cavity-coupled vaporization dynamics at the opening, rather than by far-field turbulence-dominated jet noise. This distinction also motivates the validity regime of the subsequent interpretation—namely, conditions* *under which the measured acoustic waveform can be meaningfully related to the cavity–jet forcing at the exit, with turbulent jet noise treated as a secondary contribution or a confounding background.*

---

## 1.5 Transient correspondence between acoustic and near-infrared signals

In this sub-section, we derive a theoretical relation between the vapor ejection velocity $v_e$ and the jet-driven cavity opening temperature $T_e$. This relation is important because it provides a theoretical basis for potentially using airborne acoustic emission amplitude to infer the jet-driven cavity temperature information, or vice versa—using the near-infrared sensing equipment to replace acoustic sensors for getting equivalent or even better measurement of the jet-driven cavity-affiliated quantities.

We assume that the equivalent recoil pressure and the vapor temperature at the jet-driven cavity opening $\partial\Omega_e(t)$, namely $p^{(z)}_{\text{recoil}}$ and $T_e$, satisfy the following relations:

$$p^{(z)}_{\text{recoil}} = \psi p_\infty \exp\left(\frac{L_V}{\eta T_0}\right) \exp\left(-\frac{L_V}{\eta T_e}\right) \tag{137}$$

Equation (137) mathematically parallels the *Clausius-Clapeyron relation*, where $\psi$ is a phenomenological constant; $p_\infty$ and $T_0$ are the ambient pressure and normal boiling temperature, respectively; $L_V$ is the latent heat of vaporization; and $\eta$ is the specific gas constant of the metal vapor. Next, recalling Eq. (27), we have:

$$p^{(z)}_{\text{recoil}} = (p_e - p_\infty) + \rho_V v_e^2 = p^{(z)}_{\text{acc}} - p_\infty = \psi p_\infty \exp\left(\frac{L_V}{\eta T_0}\right) \exp\left(-\frac{L_V}{\eta T_e}\right) \tag{138}$$

Since Lighthill's acoustic analogy suggests that a density wave is generated by the reactive pressure at $\partial\Omega_e\,(t)$, we should recognize that the incompressibility of the vapor phase is not a valid assumption anymore for the ambient gas. This means that, immediately outside the jet-driven cavity opening $\partial\Omega_e\,(t)$, the ambient gas becomes compressible and behaves like an ideal gas, and the gas pressure quickly adjusts from $p_e$ to the ambient pressure $p_\infty$ in an adiabatic way. Therefore, for the modeling of the airborne acoustic emission outside the jet-driven cavity, we shall assume $p_e \approx p_\infty$ and that the surrounding gas obeys the ideal gas law. Then, we have:

$$p_e \approx p_\infty = \eta \rho_V T_e \tag{139}$$

Substituting Eq. (139) into Eq. (138), we obtain:

$$p_{\text{recoil}}^{(z)} = p_{\text{acc}}^{(z)} - p_\infty \approx \frac{p_\infty}{\eta T_e} v_e^2 = \psi p_\infty \exp\left(\frac{L_V}{\eta T_0}\right) \exp\left(-\frac{L_V}{\eta T_e}\right) \tag{140}$$

$$v_e = \sqrt{\psi \eta T_e \exp\left(\frac{L_V}{\eta T_0}\right) \exp\left(-\frac{L_V}{\eta T_e}\right)} \tag{141}$$

Equation (140) indicates that $p_{acc}$ is positively correlated with $T_e$, explaining why the acoustic emission measurement is closely aligned with the near-infrared camera measurement. Equation (141) shows that $v_e$ and $T_e$ are positively correlated in a nonlinear manner, demonstrated by our experimental measurement in Fig. S5.

---

**Discussion 1.7.** *Equation* (141) *indicates that* $v_e$ *is a nonlinear function of* $T_e$ *and increases monotomically along with it. This relationship possesses a general mathematical structure as:*

$$v_e = v_e\left(T_e\right) \propto \sqrt{M_1 T_e \exp\left(-\frac{M_2}{T_e - M_3}\right)} \tag{142}$$

*where* $M_1$, $M_2$, $M_3$ *are tunable constants. In much of the previous literature, researchers have assumed or experimentally observed the following correlation between* $v_e$ *and* $T_e$*:*

$$v_e = v_e\left(T_e\right) \propto M_4 \sqrt{T_e - M_3} \tag{143}$$

*where* $M_4$ *is another tunable constant. Plotting these two functions together in one* $T_e$*–*$v_e$ *plane (shown in Fig. S6), we see that they are nearly identical to each other, demonstrating that our proposed Eq.* (141) *produces valid experimental observations on the correlation between* $v_e$ *and* $T_e$*.*

---

In summary, the investigated correlation between acoustics and near-infrared signals provides a link between airborne emissions and thermal quantities, offering alternative sensing modalities for microphones and paving the way toward thermal monitoring of the process (to be discussed in the next chapter).

## 1.6 Theory of LPBF vapor-jet cavity total vaporization with multiple laser reflections

For stable porosity-free vapor-jet cavitys, it is assumed that the mass rate of vaporization at $\partial\Omega_V(t)$, *i.e.*, $\dot{m}$, is a function of the input energy source, denoted as $Q(P)$, and the laser absorptance $\varepsilon$, both of which are functions of the fusion laser power $P$ and certain features of the vapor-jet cavity geometry, respectively. This relationship can be mathematically expressed as:

$$\dot{m} = \rho_L \varepsilon Q \tag{144}$$

To express this assumption in a more detailed formulation, we can write down the volume rate of liquid alloy vaporization ($\dot{V}_{\text{evapL}}$) as follows, ensuring both mathematical and physical consistency:

$$\boxed{\dot{V}_{\text{evapL}} = \varepsilon Q = \left[1 - \exp\left(-\frac{aL}{\sqrt{\phi_e}}\right)\right] \frac{bP}{\rho_L L_V}} \tag{145}$$

where $a,\ b$ are tunable coefficients. It is worth noting that $\varepsilon Q$ possesses a unit of $\text{m}^3/\text{s}$.

The assumed formulation of Eq. (144) and Eq. (145) can be derived from a new set of fundamental assumptions and a physical model describing multiple laser reflections within a vapor-jet cavity. In a vapor-jet cavity, where the fusion laser undergoes multiple reflections at the liquid–vapor interface $\partial\Omega_V(t)$, we assume that:

- The laser absorption rate at each instance when the laser beam impinges on a point of $\partial\Omega_V(t)$—denoted as $\varepsilon'$—is a material-dependent constant. This implies that the fraction of input laser energy utilized for vaporizing the liquid metal remains consistent each time the laser interacts with the liquid phase of the alloy.

- At any given time and location, the energy of the laser beam is either absorbed—primarily consumed by vaporization—or reflected. Consequently, potential losses due to dissipation, scattering, or radiation are not considered in this analysis.

- The total number of laser reflections within a vapor-jet cavity—denoted as $N$—is a linear function of the vapor-jet cavity's aspect ratio, *i.e.*, $L/\sqrt{\phi_e}$. Letting $k$ represent the dimension-less linear coefficient, this relationship can be expressed as:

$$\boxed{N = \frac{kL}{\sqrt{\phi_e}}, \quad k, N \in \mathbb{R}^+} \tag{146}$$

Although we acknowledge that $N$ is inherently a positive integer (*i.e.*, $N \in \mathbb{Z}^+$), we intentionally define both $k$ and $N$ as continuous variables within the positive real space in Eq. (146). This approach ensures mathematical consistency with the majority of other variables appearing in our derivation (For convenience, though, we still regard $N$ as a positive integer until Eq. (150)).

Based on these assumptions, the total absorbed laser energy at $\partial\Omega_V(t)$, denoted as $E_{\text{abs}}$, can be expressed as:

$$E_{\text{abs}} = \varepsilon' P \sum_{i=0}^{N} (1-\varepsilon')^i\,, \quad \varepsilon' \in [0,1] \tag{147}$$

The unit of $E_{\text{abs}}$ is $\text{W} = \text{J/s}$. Based on the key assumption regarding the utilization of fusion laser energy, we divide $E_{\text{abs}}$ by the enthalpy of vaporization $L_V$ to obtain the mass rate of liquid metal vaporization, *i.e.*, $\dot{m}$, expressed as:

$$\dot{m} = \frac{E_{\text{abs}}}{L_V} = \frac{\varepsilon' P}{L_V} \sum_{i=0}^{N} (1-\varepsilon')^i\,, \quad \varepsilon' \in [0,1] \tag{148}$$

Both Eq. (147) and Eq. (148) take the form of a geometric series with a common ratio of $1-\varepsilon'$. Therefore, the formula for a finite geometric series can be applied to obtain:

$$\frac{\varepsilon' P}{L_V} \sum_{i=0}^{N} (1-\varepsilon')^i = \frac{\varepsilon' P}{L_V} \frac{1-(1-\varepsilon')^N}{\varepsilon'} = \frac{P}{L_V}\left[1-(1-\varepsilon')^N\right], \quad \varepsilon' \in [0,1] \tag{149}$$

Now, defining $a = -k \log(1-\varepsilon')$ and substituting this expression along with Eq. (146) into Eq. (149), we obtain:

$$\dot{m} = \frac{P}{L_V}\left[1-(1-\varepsilon')^N\right] = \frac{P}{L_V}\left[1-\exp\left(-\frac{a}{k}N\right)\right] = \frac{P}{L_V}\left[1-\exp\left(-\frac{aL}{\sqrt{\phi_e}}\right)\right] \tag{150}$$

Moving back to our previous assumptions, if we substitute Eq. (145) with $b = 1$ into Eq. (144), we get:

$$\boxed{\dot{m} = \rho_L \varepsilon\, Q|_{b=1} = \rho_L\, \dot{V}_{\text{evapL}}\big|_{b=1} = \left[1-\exp\left(-\frac{aL}{\sqrt{\phi_e}}\right)\right]\frac{P}{L_V}} \tag{151}$$

It can be easily seen that Eq. (151) is identical to Eq. (150). This demonstrates that Eq. (144) will hold valid as long as the newly proposed assumptions and the physical model of laser reflections inside a vapor-jet cavity are corroborated (either computationally or experimentally).

Furthermore, the preceding derivation implies that the conservation of energy is inherently satisfied, as the energy input from the fusion laser effectively introduces a ”mass source term” within the model, provided that the previously stated assumptions remain valid.

## 1.7 A dimensionless number for LPBF vapor-jet cavity stability

From Eq. (78), we see that both $f_N = \frac{1}{2\pi}\sqrt{\frac{k_{\mathrm{eff}}}{m_{\mathrm{eff}}}}$, the damp-free natural frequency, and the phenomenological parameter $\varphi$, which represents vapor velocity gradient along $z$-axis and controls the damping coefficient, are key to the stability of vapor-jet cavity depth oscillations. Interestingly, both $f_N$ and $\varphi$ possess the same unit of $\mathrm{s}^{-1}$ and represent quantities related to certain timescales. This observation inspires us to define a new dimensionless number, namely the *vapor-jet cavity Oscillation Stability* (KOS) number, as the ratio constructed by the square of $f_N$ and $\varphi$:

$$\mathrm{KOS} = \frac{\varphi^2}{f_N^2} = \frac{2\pi\rho_V\phi_e\varphi^2 L}{C\gamma\cos\theta} = \frac{2\pi\rho_V\varphi^2\Omega}{C\gamma\cos\theta} \tag{152}$$

The KOS number can be interpreted as follows: the numerator $2\pi\rho_V\varphi^2\Omega$ represents the vaporization rate (*i.e.*, $\dot{m}$), which tends to destabilize the vapor-jet cavity depth oscillation when large. In contrast, the denominator $C\gamma\cos\theta$ captures the stabilizing effect of surface tension and the vapor-jet cavity front wall angle, which enhances stability when large. Thus, a larger KOS number corresponds to a more unstable vapor-jet cavity depth oscillation, while a smaller KOS number corresponds to a more stable one.

## 1.8 Fitting results for phenomenological constants

We report our calibrated phenomenological constants used in the theory validation and property inference in the paper:

- $C = 0.02$—constant for capillary force.

- $\varphi = 1 \times 10^6\ \mathrm{s}^{-1}$—constant for velocity gradient.
- $B = 1.5 \times 10^6\ \mathrm{kg \cdot s/m^3}$—constant for the experimental relationship between $\frac{\cos\theta}{L}$ and $\frac{P-P_0}{V^4}$.
- $b = 150\ \mathrm{m}^{-1}$—constant for the experimental relationship between $\frac{\cos\theta}{L}$ and $\frac{P-P_0}{V^4}$.

# Supplementary Figure S1

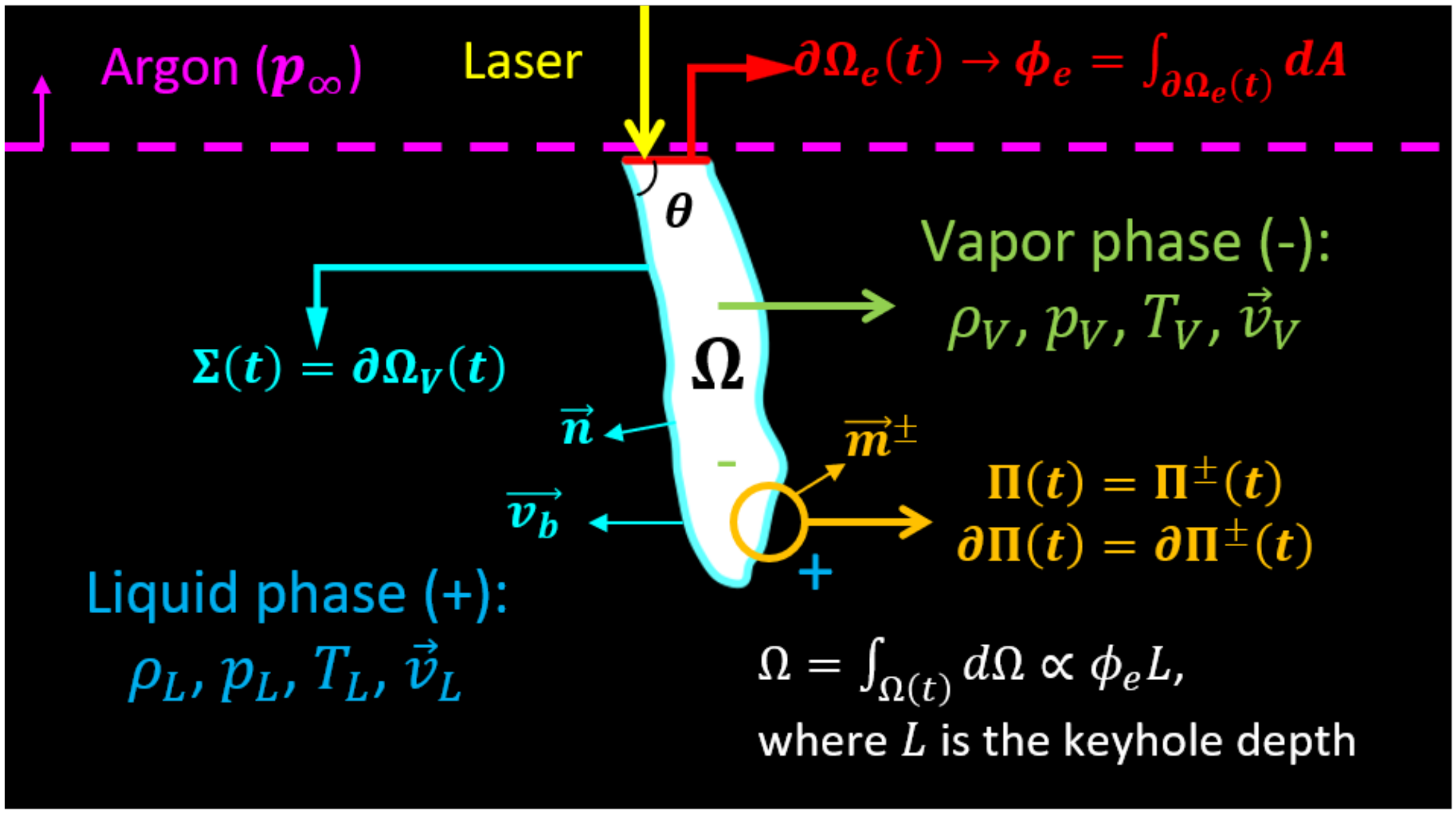

**Figure S1**: **Schematic of the physical picture of an LPBF keyhole (side-view).**

# Supplementary Figure S2

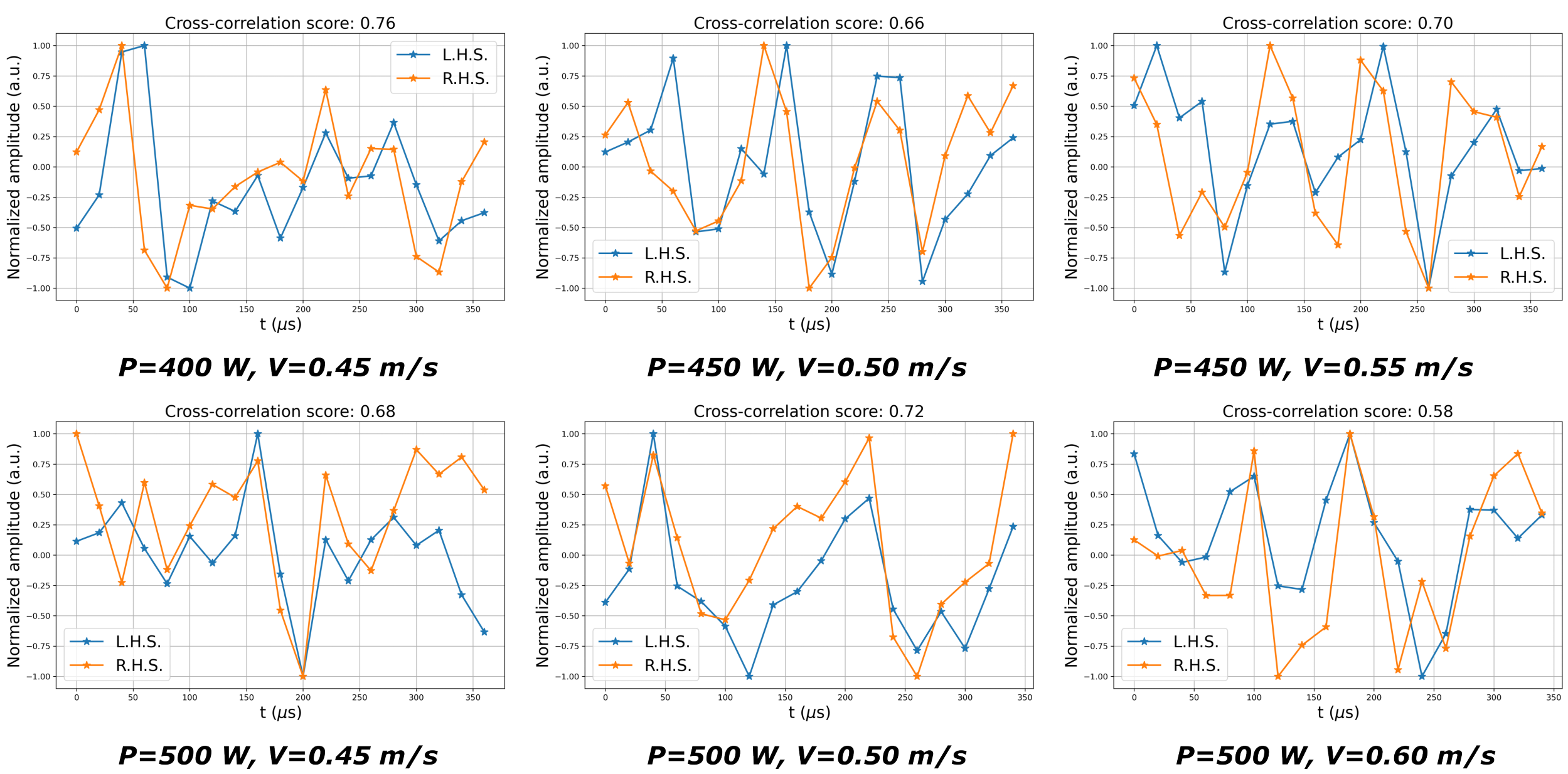

**Figure S2**: **More snippets demonstrate the L.H.S. and R.H.S. alignment of the JCODEs.** Each subplot corresponds to the same validation snippet with its $P$ and $V$ indicated at the bottom. The magnitude scale is normalized for better visualization.

# Supplementary Figure S3

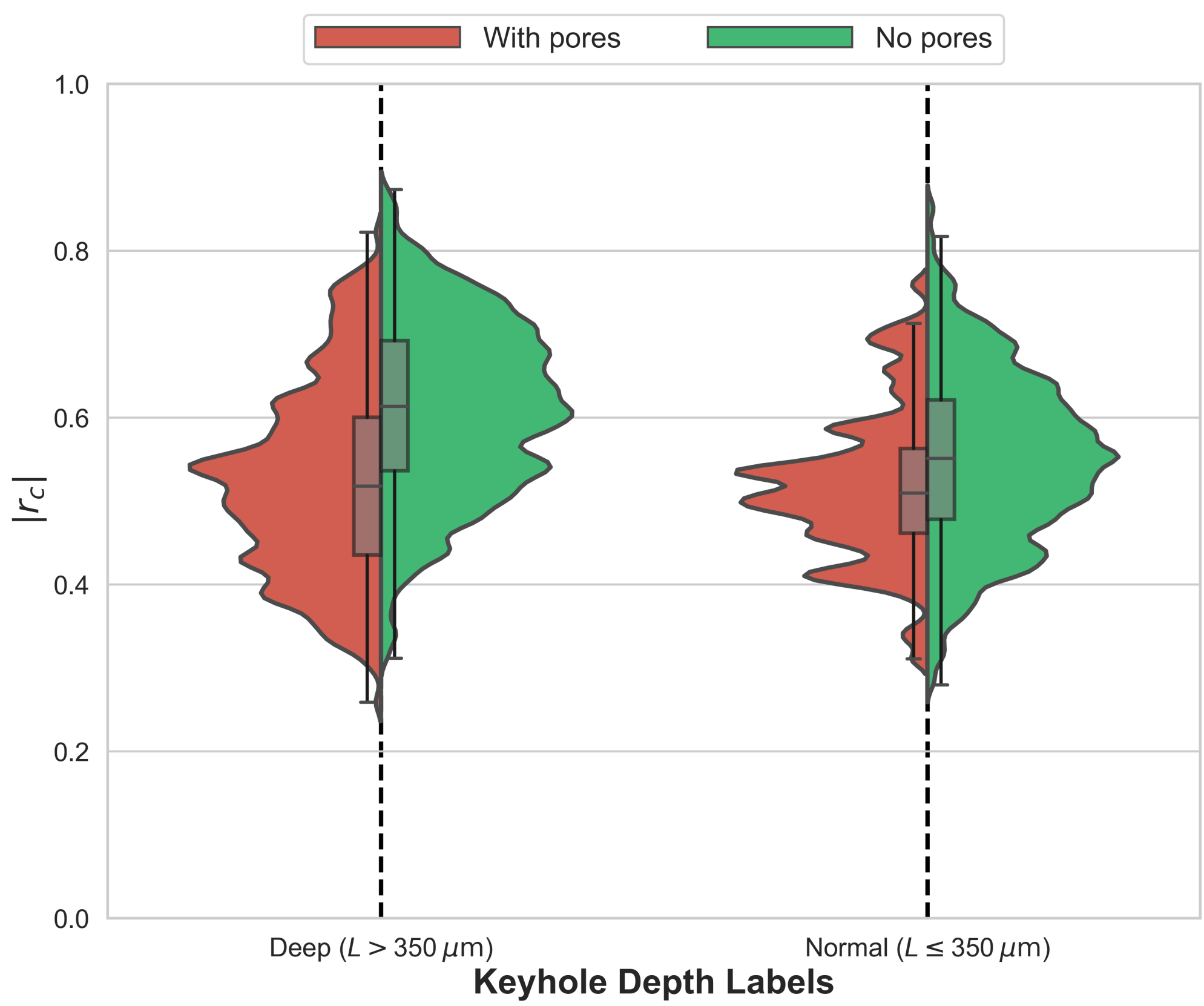

**Figure S3**: **Violin+box plots showing the absolute cross-correlation coefficient ($r_c$) distributions between the airborne acoustic amplitude ($\delta\dot{v}_e$) and the keyhole depth acceleration amplitude ($\delta\ddot{L}$) for all validation snippets.** The snippets are categorized based on two labels: (1) normal ($L_0 \leq 350\ \mu$m) or deep ($L_0 > 350\ \mu$m) keyhole, and (2) with pores (P-KH) or no pores (N-KH).

# Supplementary Figure S4

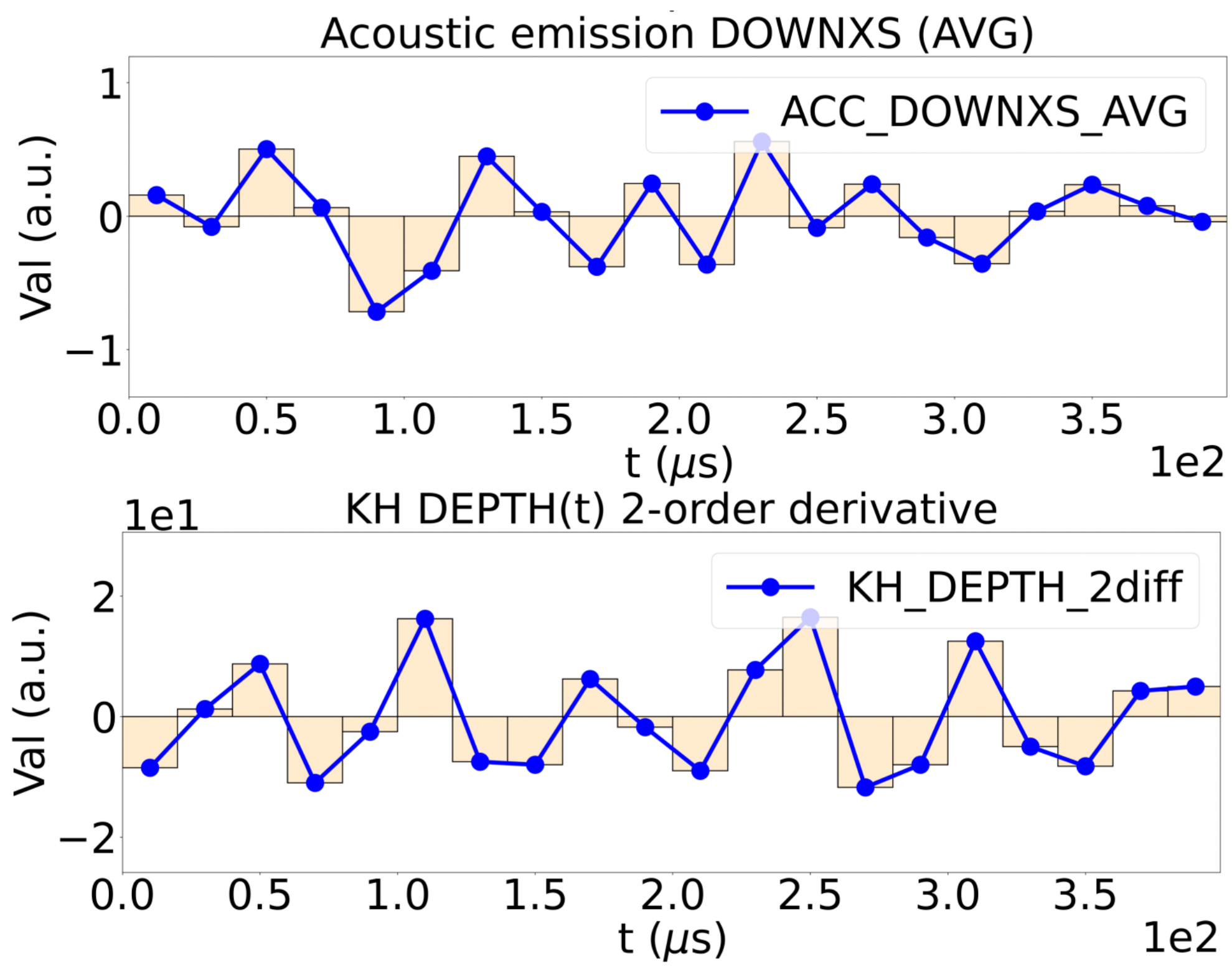

**Figure S4**: **Material generalizability: Time-domain comparison between the airborne acoustic emission (top) and the keyhole-depth oscillations (bottom) in an aluminum alloy (AlSi10Mg), demonstrating cross-material generalizability.** The top panel shows a representative normalized acoustic-emission channel ($\delta\dot{v}_e$; `ACC_DOWNXS_AVG`), and the bottom panel shows the corresponding normalized cavity-depth acceleration ($\delta\ddot{L}$; `KH_DEPTH_2diff`, *i.e.*, the second time derivative of $L(t)$), over the same time window $t$ ($\mu$s). Despite the change in material system, the two signals exhibit coherent oscillations with matching dominant periodicity, supporting the claim that airborne acoustic emission tracks the cavity-depth oscillation frequency beyond the baseline material (Ti-6Al-4V).

# Supplementary Figure S5

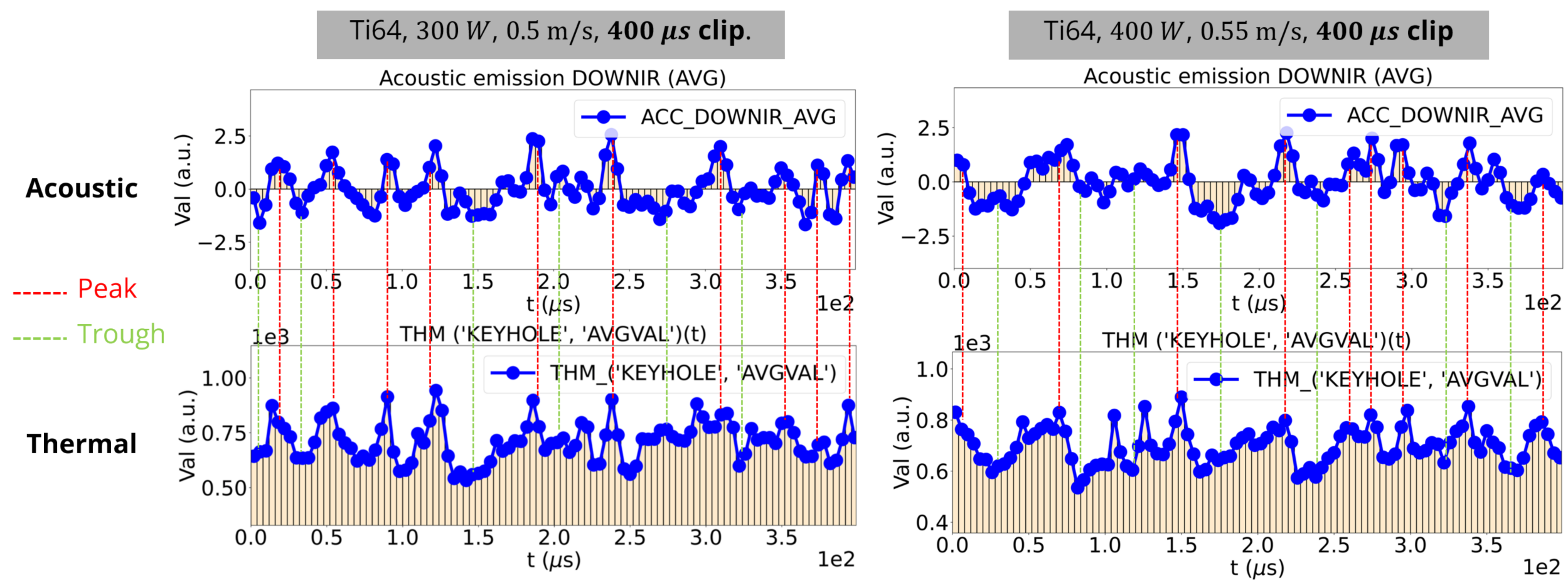

**Figure S5**: **Comparison between the airborne acoustic emission amplitude ($v_e$) and the keyhole opening near-infrared intensity (a quasi-representation of $T_e$) during LPBF experiments.** Two 400 $\mu$s clips were matched and compared with each other after data synchronization. The trends of these two quantities align well with each other, which is consistent with our theoretical analysis.

# Supplementary Figure S6

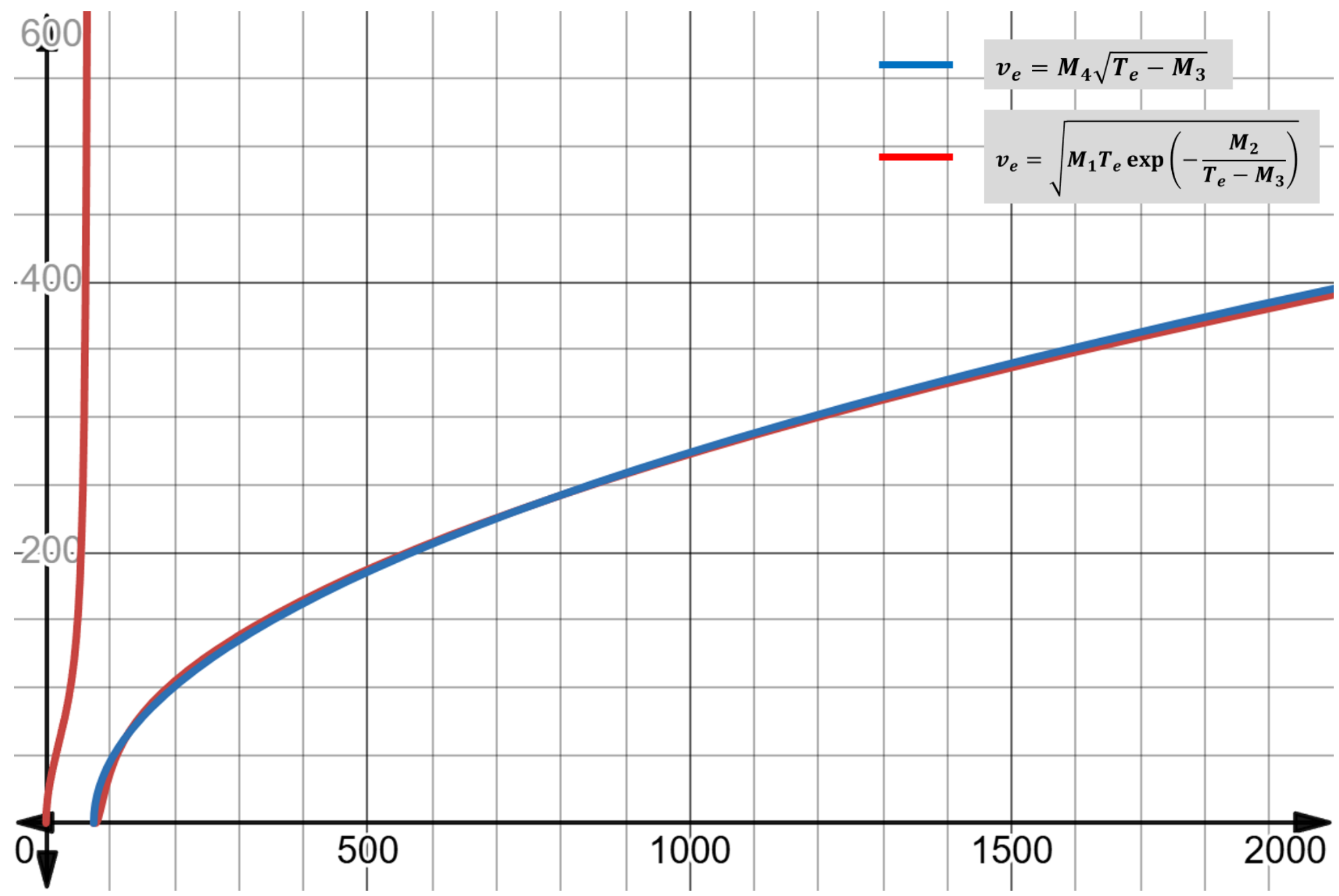

**Figure S6**: **Correlation between $v_e$ and $T_e$ in two mathematical structures.** The blue curve is the experimentally reported correlation between $v_e$ and $T_e$, while the red curve is our analytical curve, which is very close to the blue curve.

# Supplementary Table S1

| Symbol / Operator | Physical Description | Units |
|---|---|---|
| $\Omega(t)$ | Keyhole vapor region (total control volume) | $m^3$ |
| $\partial\Omega_e(t)$ | Boundary of $\Omega(t)$ at the keyhole opening | $m^2$ |
| $\partial\Omega_V(t)$ | Boundary of $\Omega(t)$ at the liquid–vapor interface (a.k.a. keyhole wall) | $m^2$ |
| $\partial\Omega(t)$ | Total boundary of keyhole region—$\partial\Omega(t) = \partial\Omega_e(t) \cup \partial\Omega_V(t)$ | $m^2$ |
| $\Xi(\Omega_e(t))$ | Differential volume around the keyhole opening | $m^3$ |
| $\vec{n}, \vec{e}_k$ | Outward normal vector of $\partial\Omega(t)$; $\vec{e}_k = \vec{n}\rvert_{\partial\Omega_e(t)}$ | — |
| $L$ | Keyhole depth | m |
| $R$ | Keyhole radius | m |
| $\phi_e$ | Keyhole opening area—$\phi_e = \pi R^2$ | $m^2$ |
| $\upsilon$ | Keyhole aspect ratio—$\upsilon = \frac{L}{2R}$ | (dimensionless) |
| $\dot{m}$ | Total mass excahnge rate through $\partial\Omega_V(t)$ | kg/s |
| $\rho_V$ & $\rho_L$ | Vapor & liquid density of the alloy | $kg/m^3$ |
| $\vec{v}_V$ & $\vec{v}_L$ | Vapor & liquid velocity of the alloy | m/s |
| $v_e$ & $v_B$ | Vapor ejection velocity at $\partial\Omega_e(t)$ (upward) and the bottom of the keyhole | m/s |
| $\vec{v}_b$ | Velocity of the liquid–vapor interface of the keyhole ($\partial\Omega_V(t)$) | m/s |
| $p_V$ & $p_L$ | Static pressure inside & outside the keyhole | Pa |
| $p_e$ | Static pressure at keyhole opening | Pa |
| $p_\infty$ | Ambient atmospheric pressure | Pa |
| $\tau_\mathbf{V}$ & $\tau_L$ | Vapor & liquid viscous stress tensor | Pa |
| $\sigma_\mathbf{V}$ & $\sigma_\mathbf{L}$ | Vapor & liquid Cauchy stress tensor | Pa |
| $\vec{f}$ | External surface force per unit area on $\partial\Omega_V(t)$ | Pa |
| $\mathbf{T}$ | Lighthill turbulence stress tensor | Pa |
| $S_m$ & $\vec{F}$ | Volumetric mass & momentum sources in $\Xi$ | Various |
| $\xi$ & $\varepsilon$ | Total laser reflectance & absorptance—$\varepsilon = 1 - \xi$ | (dimensionless) |
| $c_0$ & $M$ | Speed of sound & Mach number of the vapor ejection—$M = \frac{v_e}{c_0}$ | m/s, (dimensionless) |
| $P$ & $w$ | Laser power & scanning speed | W, m/s |
| $\gamma$ | Surface tension coefficient at $\partial\Omega_V(t)$ | N/m |
| $\theta$ | Characteristic angle of the keyhole geometry | rad / deg |
| $\alpha,\ C,\ \varphi,\ \varphi',\ \psi$ | Phenomenological/tunable constants | Various |
| $f_N$ | Keyhole depth oscillation natural frequency | Hz |
| $T_0$ | Alloy's normal (under $p_\infty$) boiling temperature | K |
| $T_e$ | Keyhole opening temperature | K |
| $T_\infty$ | Ambient atmospheric temperature | K |
| $\eta$ | Specific gas constant | J/(kg·K) |
| $L_V$ | Enthalpy of vaporization | J/kg |
| $\mathbf{I}$ | Identity tensor | — |
| $\otimes$ | Tensor product | — |
| $\langle\cdot\rangle$ | Volume average operator | — |
| $\cdot$ | Standard dot product | — |
| $\nabla$ & $\nabla\cdot$ | Gradient & divergence operator | 1/m |
| $\delta(\cdot)$ | Perturbation operator | (dimensionless) |
| $\frac{d}{dt}$ & $\frac{\partial}{\partial t}$, $\dot{(\ )}$ | Total & partial time derivative, dot notation | 1/s |

**Table S1**: **Glossary of key terms and operators in vapor-cavity modeling.**

# Supplementary Video S1

Example synchronous clip of Ti-6Al-4V multi-modality data.

# Supplementary Video S2

Example synchronous clip of Ti-6Al-4V multi-modality data.

# Supplementary Video S3

Example synchronous clip of AlSi10Mg multi-modality data.

# Supplementary Data S1

The attached file contains raw multimodal measurements collected across a range of process conditions, including X-ray synchrotron imaging, airborne ultrasonic emission, photodiode-based laser reflection signals, and thermal NIR imaging.